%% file: document_arxiv.tex
\documentclass[10pt,letterpaper]{article}

\input{macros/macros_arxiv}
\input{macros/math}

\usepackage{multirow}
\usepackage{rotating, graphicx}

\title{
Gaussianizing the Earth -- \\
Multidimensional Information Measures \\
for Earth Data Analysis
}

\author{
  J.~Emmanuel~Johnson\thanks{\url{https://jejjohnson.netlify.app}} \\
  Image Processing Laboratory \\
  Universitat de Val{\`e}ncia\\
  Val{\`e}ncia, Spain\\
  \texttt{juan.johnson@uv.es} \\
  \And
  Valero Laparra\thanks{\url{https://www.uv.es/lapeva/}} \\
  Image Processing Laboratory \\
  Universitat de Val{\`e}ncia\\
  Val{\`e}ncia, Spain\\
  \texttt{valero.laparra@uv.es} \\
  \And
  Maria Piles\thanks{\url{https://sites.google.com/site/mariapiles/}} \\
  Image Processing Laboratory \\
  Universitat de Val{\`e}ncia\\
  Val{\`e}ncia, Spain\\
  \texttt{maria.piles@uv.es} \\
  \And
  Gustau Camps-Valls\thanks{\url{https://www.uv.es/gcamps/}} \\
  Image Processing Laboratory \\
  Universitat de Val{\`e}ncia\\
  Val{\`e}ncia, Spain\\
  \texttt{gcamps@uv.es} \\
}

\begin{document}

\maketitle

% As a general rule, do not put math, special symbols or citations
% in the abstract or keywords.
\input{sections/0_abstract}

% make the title area

\keywords{\textbf{Density estimation, Information theory, entropy, mutual information, Gaussianization, multivariate data, droughts, climate extreme, anomaly, spatio-temporal Earth data.}}

\section{Introduction}
    \label{sec:intro}
    \input{sections/1_introduction}

\section{Multivariate Gaussianization}
    \label{sec:methods}
    \input{sections/3_methods}

\section{Experiments}
    \label{sec:experiments}
    \input{sections/4_experiments}

\section{Conclusions}
    \label{sec:concl}
    \input{sections/6_conclusions}

\section{Acknowledgements}

This research was funded by the European Research Council (ERC) under the ERC-Consolidator Grant 2014 Statistical Learning for Earth Observation Data Analysis. project (grant agreement 647423). J.E.J. thanks the European Space Agency (ESA) for support via the Early Adopter Call of the Earth System Data Lab project; M.D.M. thanks the ESA for the long-term support of this initiative. Additional support was provided by the Project RTI2018-096765-A-100 (MCIU/AEI/FEDER, UE).

% trigger a \newpage just before the given reference
% number - used to balance the columns on the last page
% adjust value as needed - may need to be readjusted if
% the document is modified later
%\IEEEtriggeratref{8}
% The "triggered" command can be changed if desired:
%\IEEEtriggercmd{\enlargethispage{-5in}}

% references section
\bibliographystyle{IEEEtran}
\bibliography{bibtex/biblio,bibtex/ci,bibtex/ESDL}
\end{document}

%% file: macros/macros_arxiv.tex
\usepackage{macros/arxiv}

\usepackage[utf8]{inputenc} % allow utf-8 input
\usepackage[T1]{fontenc}    % use 8-bit T1 fonts
\usepackage{url}            % simple URL typesetting
\usepackage{booktabs}       % professional-quality tables
\usepackage{amsfonts}       % blackboard math symbols
\usepackage{nicefrac}       % compact symbols for 1/2, etc.
\usepackage{microtype}      % microtypography
\usepackage{lipsum}
\usepackage{helvet}
\usepackage{amsmath}
\usepackage{array}

%% file: macros/math.tex
\usepackage{wrapfig}

\usepackage{mdframed}% http://ctan.org/pkg/mdframed
\usepackage{fancybox}

\usepackage{pgffor}
\usepackage{lscape}

\usepackage[normalem]{ulem}
\usepackage{color}
\usepackage[dvipsnames]{xcolor}
\newcommand{\commentE}[1]{\textbf{\textcolor{Plum}{E: #1}}}

\definecolor{amber}{rgb}{1.0, 0.75, 0.0}

\usepackage{tikz}
\usetikzlibrary{shapes.geometric, arrows}

\usepackage[colorlinks=true,citecolor=blue]{hyperref}

\usepackage{amsmath,amssymb}
\usepackage{amsfonts}
\usepackage{amsbsy} % Per a \boldsymbol
\usepackage{bbm} % for the \ii
\usepackage{mathtools}
\usepackage{footnotehyper}

\newcommand{\ra}[1]{\renewcommand{\arraystretch}{#1}}

\newcommand{\TODO}[1]{\textcolor{red}{#1}}

\newcommand{\Real}{\mathbb{R}}

\newcommand{\cL}{\mathcal{L}}

\newcommand{\cN}{\mathcal{N}}

\newcommand{\cX}{\mathcal{X}}
\newcommand{\cY}{\mathcal{Y}}
\newcommand{\cZ}{\mathcal{Z}}

\newcommand{\bE}{\mathbf{E}}

\newcommand{\bG}{\mathbf{G}}

\newcommand{\bR}{\mathbf{R}}

\newcommand{\bX}{\mathbf{X}}
\newcommand{\bY}{\mathbf{Y}}

\newcommand{\bPsi}{\boldsymbol \Psi }

\newcommand{\btheta}{\boldsymbol \theta }

\newcommand{\dR}{\mathbb{R}}

\newcommand{\vect}[1]{{\boldsymbol{\mathbf{#1}}}} % vector
\newcommand{\x}{\vect{x}}
\newcommand{\y}{\vect{y}}
\newcommand{\z}{\vect{z}}

\newcommand{\kld}{\text{D}_{\text{KL}}}

\newcommand{\remove}[1]{}

%% file: sections/0_abstract.tex
\begin{abstract}

\textcolor{black}{Information theory is an excellent framework for analyzing Earth system data because it allows us to characterize uncertainty and redundancy, and is universally interpretable. However, accurately estimating information content is challenging because spatio-temporal data is high-dimensional, heterogeneous and has non-linear characteristics.}
\textcolor{black}{In this paper, we apply multivariate Gaussianization for probability density estimation which is robust to dimensionality, comes with statistical guarantees, and is easy to apply. In addition, this methodology allows us to estimate information-theoretic measures to characterize multivariate densities: information, entropy, total correlation, and mutual information.}
% 
% In this paper, we apply multivariate Gaussianization for probability density function and information-theory measure estimation for analyzing Earth data. This methodology allows us to estimate information-theoretic measures to characterize multivariate densities: information, entropy, total correlation, %Kullback-Leibler (KL) divergence 
% and mutual information. The methodology is robust to dimensionality and sample size, comes with statistical guarantees and is easy to apply. % in spatio-temporal domains. 
% 
% We show empirical evidence of performance in several Earth system data analysis problems. 
We demonstrate how information theory measures can be applied in various Earth system data analysis problems.
First we show how the method can be used to jointly Gaussianize radar backscattering intensities, synthesize hyperspectral data,  and quantify of information content in aerial optical images. 
We also quantify the information content of several variables describing the soil-vegetation status in agro-ecosystems, and investigate the temporal scales that maximize their shared information under extreme events such as droughts. 
Finally, we measure the relative information content of space and time dimensions in remote sensing products and model simulations involving long records of key variables such as precipitation, sensible heat and evaporation. 
Results confirm the validity of the method, for which we anticipate a wide use and adoption. Code and demos of the implemented algorithms and information-theory measures are provided. %, and evaluate climate model differences and shared information.

\end{abstract}

%% file: sections/1_introduction.tex
%% Earth data is complex and hetero
%%      * RS + geo + climate models
%%      * multivariate data, multisource, nonlinearities,
%%      * typically spatio-temporal structures, where's the info, man???

%% Importance of Spatial-Temporal
Earth system models and observational data are fundamental to monitor our planet and to understand climate change~\cite{Buermann2001,Moss10,Overpeck11,eyring_taking_2018}. We now face a data deluge which comes from remote sensing platforms that continuously increase the spatial, temporal and spectral resolution of data sources.
% from in-situ measurements with more and better instrumentation, and lately from citizen sensors too. 
In recent years, Earth system data comes in high volume, heterogeneity, and uncertainty \cite{reichstein19nat} which poses important challenges in analysis, modeling and understanding. The statistical analysis of remote sensing data and model simulations requires dealing with large amounts of heterogeneous, multivariate, and spatio-temporal data. 
%% Data volume vs information: What are the overarching goal and the challenges?
The volume of data from high-resolution models and observations have substantially increased to petabyte scales. Yet, we are well aware that copious amounts of data does not necessarily mean large amounts of {\em information}. For example, it is now widely acknowledged that models are often correlated and share common traits, features and information content. Which is the most appropriate and representative model? How can we best quantify their information content in meaningful units? Essential Earth variables and data products exhibit high levels of redundancy in space and time. What is the appropriate space, time or spatio-temporal scales one should look at? The same questions arise when trying to assess and choose the most adequate observational variable or bio-geo-physical parameter for Earth monitoring.

%% ML and statistics as the right framework to study this
\begin{wrapfigure}{r}{4.4cm}
\vspace{-0.5cm}
\begin{mdframed}[backgroundcolor=gray!20] 
Density estimation is at the core of all problems in statistics and machine learning, but is still a challenging and unresolved problem.
\end{mdframed}
\vspace{-0.5cm}
\end{wrapfigure}
From a pure statistical standpoint, the problem of {\em information quantification} for Earth and climate data is challenging. Information theory (IT) is the appropriate framework to study information content, uncertainty and redundancy \cite{Cover06}. The estimation of entropy and mutual information for discrete and continuous random variables has been addressed under different approaches in the statistics literature~\cite{Darbellay2006,Kraskov04,Wang06,leonenko2008}. %, yet most of them are based on estimating the densities first. %%% GCV: don't want to introduce the pdf yet!
An important problem is that IT measure estimation of multivariate data is difficult, and very often only unidimensional/marginal measures of information are computed in practice. Many are largely based on histogram estimates, which is a very limiting factor~\cite{Cover06,scott2015multivariate}. Many multivariate estimates based on nearest neighbors \cite{Kraskov04,Wang06,leonenko2008} either do not scale well, do not converge to the true measure, or show high estimation bias~\cite{PerezCruz08nips}. Measures such as entropy or mutual information have been used in remote sensing and the geosciences to study feature redundancy in image classifiers \cite{paul2018spectral}, to assess the maximum number of parameters that can be estimated given a set of observations \cite{Konings2015}, for remote sensing feature extraction and weighting~\cite{marinoni2017unsupervised,zhang2018mutual}, data fusion~\cite{prasad2007hyperspectral}, image registration \cite{zhao2015multi,chen2018medium,xu2016multimodal}, and to quantify uncertainty in models and observations~\cite{ruddell2013applying}. 
%The concepts of entropy and mutual information are widespread and used in practice \cite{IT4RStutorial}. 

Information quantification requires estimating multivariate densities which is a challenging and unresolved problem. This is especially problematic in Earth observation data with moderate-to-high dimensional problems with nonlinear feature relations.
%% But IT is challenging because we need PDF estimation, which sucks in highdim data
%Density estimation is at the core of all statistical/machine problems, from supervised regression and classification to unsupervised clustering and learning representations. 
%This is because IT measure estimation boils down to estimating multivariate densities, which is particularly difficult in high dimensions. % where the space readily becomes empty and one needs an exponential increase in sampling. 
These issues affect the classic parametric density estimators based on the exponential family of solutions or mixture distributions as well as non-parametric methods based on histograms, kernel density estimation, and $k$-nearest neighbors. As an alternative to these traditional methods, there is a new class of methods called neural density estimators \cite{Papamakarios2019NeuralDE} which are parameterized neural networks that estimate densities. They use the `change of variables' formula to estimate densities of inputs and also allow one to draw samples of your input data. They have promise as they have been successfully used in applications related to Earth system sciences including inverse problems~\cite{Ardizzone2019INNF} and density estimation~\cite{Rezende2020NFTS}. 

\begin{wrapfigure}{r}{4.4cm}
\vspace{-0.5cm}
\begin{mdframed}[backgroundcolor=gray!20] 
Gaussianization is effective in multivariate density estimation, and allows one to estimate multi\-dimensional information measures
\end{mdframed}
\vspace{-0.5cm}
\end{wrapfigure}
In this paper, we look at a particular class of models inside the neural density estimation family. In particular, we introduce the Gaussianization method \cite{Chen2000Gaussianization} and in particular a generalized algorithm called Rotation-Based Iterative Gaussianization (RBIG) \cite{Laparra2011RBIG}. This method uses a repeated sequence of simpler feature-wise Gaussian transformations and orthogonal rotations until convergence. It can be shown that in each iteration the total correlation and the non-Gaussianity are reduced and converge towards zero, that is, towards full independence.
The learned transformation towards the Gaussian domain is invertible, which allows us to synthesize data easily by inverting samples drawn from the Gaussian domain. 
The method is also advantageous because it %it has connections to IT measures and 
allows us to estimate IT measures such as entropy, total correlation, non-Gaussianity % divergences 
and mutual information effectively in high dimensional data. The method is easy to apply, fast, and has links to deep neural networks ~\cite{Laparra2011RBIG, Ball2016GDN, Meng2020GaussianizationF}. Section \S2 will review the theoretical properties of RBIG and their practical use for information theoretic measures estimation. 
%% Our proposal for multivariate PDF estimation
%% We introduce/review a method for PDF estim in multivariate
%% datasets, and show performance in RS+geo problems

%We here take an indirect pathway to probability density function (PDF) estimation via multivariate Gaussianization, which was originally introduced in~\cite{Chen2000Gauss} and further generalized in~\cite{Laparra2011RBIG}. The method allows one to map any multivariate data distribution to a multivariate Gaussian distribution. Through an invertible transformation, it allows us to estimate PDFs and compute information theory measures, such as entropy, total correlation or mutual information. The method is easy to run and apply, fast, and interesting links to deep learning emerge~\cite{Ball2016GDN, Emman2019}. 

%% Our contributions, both theoretical and experimental
%%    * Theory: RBIG is cool: addresses the problem, splits
%%      multiv into univar ---> curse of dim, connection to IT,
%%      connection to nnets/NF, comput efficient (sic), 
%%      code available!
%%    * Applied: show performance in sevearal problems. 
%%      This allows us to: 1) characterize spatio vs temp 
%%      compon in information theoretic terms}, 2) PDF estim
%%      allows charact. multivar relations (droughts)

In \S3 we take advantage of RBIG to estimate information theoretic measures in different Earth system science problems of interest. %In particular, we estimate the information content in Earth and climate data, and pay attention to the evaluation of information in three different environments: 
Three settings of increasing scale and sophistication are given (cf.~Table~\ref{tab:expsummary}): from working at pixel and patch level (fully spectral and spatio-spectral domains) to studying information in time series (fully temporal domain), and finally to quantifying  redundancy and probability in Earth data cubes (spatio-temporal domain). 
We first illustrate the use of RBIG with three standard remote sensing data modalities and for three different illustrative applications: in Gaussianizing radar backscattering intensity data, synthesizing hyperspectral spectra and quantifying information in RGB aerial images. 
Our second application is concerned with assessing the information content conveyed by a selection of remotely-sensed variables widely used in vegetation/land monitoring -temperature, moisture and vegetation indices-, and with investigating the temporal scales that maximize their shared information under extreme events such as droughts. %, and 3) climate models intercomparison of sea surface temperature and their differences and shared information.
Finally, we focus on quantifying the information content in spatio-temporal data cubes of selected climate variables (precipitation, sensible heat, evaporation) over a decade of global data. We are interested in quantifying and contrasting the information content of the space versus time dimensions as a means to understand the scales of the underlying physical processes. 
%
% To end with the boring, probably unread and useless outline paragraph
%The remainder of the paper is organized as follows. After summarizing the main aspects of the method and how to use it to estimate information theory measures in \S2, we will illustrate applicability in Earth and climate data problems in \S3.
We conclude with some remarks and outline further work in \S4.

%% file: sections/3_methods.tex
\iffalse
\begin{itemize}
 \item \sout{PDF estim in general}
 \item \sout{Family of methods: histo, knn, kde}
 \item \sout{Neural Density estimators (hot topic)}
 \item Gaussianization because its clever
 \item RBIG in particular
 \item How RBIG works with
 \item remarks: comput, code, toy example...
\end{itemize}
\fi

\subsection{PDF Estimation}\label{sec:pdf}

\iffalse
\commentE{
\begin{itemize}
 \item I introduce PDF estimation, what is it?
 \item Why it's important
 \item define the PDF estimation problem somewhat explicitly
\end{itemize} 
}
\fi 

Most problems in signal and image processing, information theory and machine learning involve the challenging task of multidimensional probability density function (PDF) estimation. %It is well-known that accurately estimating the PDF of a data distribution is important for many machine learning problems e.g. regression, classification and data representation. 
A probability density function or simply a density $p(\cdot)$ takes an input $\x\in{\mathcal X}$ and outputs a density, which follows the properties that 1) $p(\x) \geq 0, \forall \x\in\mathbb{R}^{D}$, and 2) it has to sum to one, $\int_{\mathcal X} p(\x)\text{d}\x=1$. 
In practice, we usually do not have access to the PDF $p(\cdot)$ but we do have a set of (multivariate) samples drawn from the generating process $\x = \left\{ \x_1, \x_2, \ldots, \x_N \right\}$ to estimate the PDF from. An accurate PDF estimation is important because it allows us to: 1) calculate the probability of any arbitrary input data point, which accounts for the relative likelihood that the value of the random variable (r.v.) would equal the sample; 
2) generate samples $\x'\sim p(\x)$ from this distribution thus allowing data synthesis, background and support estimation, as well as anomaly detection; and 
3) calculate expectations for functions (or transformations) of arbitrary form $f(\x)$ given $p(\x)$, i.e. ${\mathbb E}_\x[f(\x)]$, which allows us to e.g. characterize the system. %and to assess how well it performs. %, as well as to compare datasets, systems and processes too through some sort of distance measure between probabilities,  $D(p(\x)|p(\y))$.

Having access to all of these properties gives us the ability to tackle long-standing problems in machine learning and statistics. With accurate PDF estimates, one can model conditional densities of data generated from a prior distribution, develop accurate and efficient compression schemes, and use principled objective functions such as maximum likelihood. % or the Kullback-Leibler (KL) divergence criteria. 
In addition, having access to an accurate density estimator can be useful in many hybrid applications to deal with out-of-sample or out-of-distribution problems too \cite{Nalisnick2019HybridMW}. The problem is therefore to estimate the density $p(\x)$ given a set of samples from $\cX$.

\iffalse
\commentE{
\begin{itemize}
 \item I want to briefly mention parametric models
 \item I want to stress that these methods are good on univariate data
\end{itemize} 
}
\fi

% The problem has been tackled using a plethora of methods and approaches, yet always stumbled on the curse of dimensionality problem, which is omnipresent in all fields of science and engineering. 
The simplest approach to PDF estimation assumes the density has a {\em parametric} functional form defined by a fixed number of tuneable parameters. The Gaussian assumption is the most widely adopted for unimodal distributions, which comes parameterized by a mean $\boldsymbol{\mu}$ and a covariance function $\boldsymbol{\Sigma}$. 
%Further restrictions on the form of $\boldsymbol{\Sigma}$ can lead to other related methods such as PCA, MCA, or ICA. 
If more than one mode is assumed, then a mixture of Gaussians %or logistics 
generally leads to better fits. % as with sufficiently enough components $\pi$, it should approximate most univariate densities arbitrarily well. 
% The main advantage of parametric methods is their simplicity that involve closed-form solutions and can be solved either exactly or with methods like expectation-maximization. 
%
%
%
\iffalse
\commentE{
\begin{itemize}
 \item I want to mention the core PDF estimators
 \item how histograms have a \textbf{fixed} bandwidth, kernels are \textbf{smooth} and kNN are \textbf{adaptive}...because I like the way it sounds.
 \item I want to stress that they suck with high-dimensional data
 \item I want to stress that they're good on univariate data
\end{itemize} 
}
\fi
%
%Sometimes a specific functional form is not good enough and we want to estimate the PDF by actually modeling the data. 
However, finding a parametric form for the distribution that fits properly to a particular data is really difficult in most cases.
%assuming a specific, often Gaussian, distribution may be unrealistic in many data problems. 

The alternative approach comes from {\em non-parametric models}, which do not assume a specific form for the distribution and are learned from data. The simplest non-parametric method estimates the PDF %an empirical distribution via histograms 
by partitioning the data space in non-overlapping bins whereby the density is estimated as the fraction of data points in the bin divided by the volume of the bin. This histogram-based PDF estimation method poorly copes with dimensionality, typically leads to either overfitting or underfitting, and selecting an appropriate number of bins per dimension is a challenge in itself. %is typically non-parametric with a bias-variance trade-off of too many bins leading to the method to overfit and too little bins will cause the method to overfit. 
Alternative parametric estimates for these methods following likelihood-estimation schemes for the optimal bin width determined by the maximum likelihood have been introduced \cite{Papamakarios2019NeuralDE}. However, they are very {\em rigid} %and {\em fixed} 
approaches %to density estimation 
and lead to very rough density functions. To achieve {\em smoother} PDF estimates, the kernel density estimation (KDE) method is another popular non-parametric method. % that is a smoothed version of the empirical distribution. 
KDE places a non-linear kernel function with a varying bandwidth parameter to control the degree of smoothness on top of each example. Unfortunately, a bias-variance trade-off %where too low 
will result in over/underfitting the PDF, especially in moderate-to-high dimensional problems. 
In the previous approaches, the bandwidth is typically fixed {\em a priori} following heuristics in the literature~\cite{Bishop2007PatternRA}, and rarely take into account the concentration of points, i.e. that smaller bins should be placed in regions with a higher concentration of points, in a form of adaptive bit-allocation scheme. This can be addressed by using $k$-nearest neighbors (kNN), which has an {\em adaptive} bandwidth per location and depends on the number of training points available. However, all of the density estimators above suffer from the curse of dimensionality: as the dimensionality increases, the space becomes sparser and density estimates are unreliable. %For example, popular density estimators such as histograms, KDE estimators and kNN approximations. 

\begin{table*}[t!]
% \begin{minipage*}{\textwidth}
    \caption{Summary of all components of the Gaussianization algorithm.}
    \vspace{-0.0cm}
    \label{tab:rbigparts}
    \centering
\begin{tabular}{|c|c|m{3.5cm}|p{1.5cm}|m{1.1cm}|m{1.1cm}|}
\hline \hline
\multicolumn{1}{c}{\textbf{Description}}               
& \multicolumn{1}{c}{\textbf{Notation}}                              
& \multicolumn{1}{c}{\textbf{Transformation}} 
& \multicolumn{1}{c}{\textbf{Domain}}  
& \multicolumn{1}{c}{\textbf{Before}}  
& \multicolumn{1}{c}{\textbf{After}}  \\ \hline \hline
     Marginal Uniformization  & $\text{U}$                                                  & Histogram \cite{Laparra2011RBIG}, \newline Kernel Density Estimation \cite{Meng2020GaussianizationF},\newline Lambert\cite{goerg2012lambert}, Splines\cite{NSF}, Box-Cox \cite{boxcox}      
     & $\mathbb{R} \rightarrow \mathbb{R}^{[0,1]}$       &  \includegraphics[width=1.2cm,height=1.2cm]{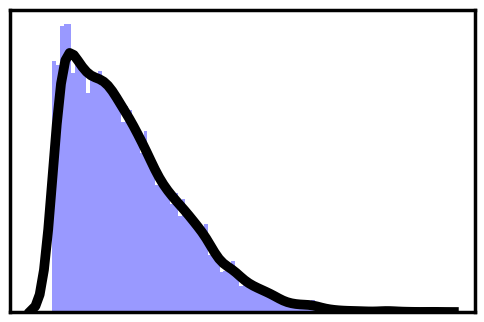}            &      \includegraphics[width=1.2cm,height=1.2cm]{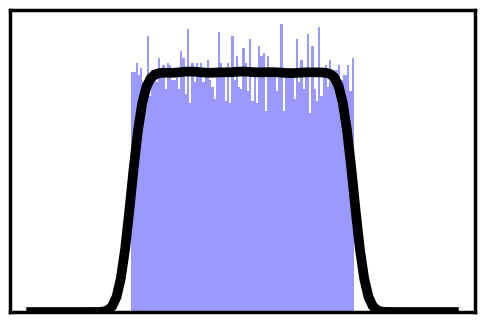}        \\ \hline
     Inverse CDF              & $\text{CDF}^{-1}$
     & Inverse Gaussian CDF, Logit,\newline Inverse Cauchy CDF
     & $\mathbb{R}^{[0,1]} \rightarrow \mathbb{R}$
      & 
     \includegraphics[width=1.2cm,height=1.2cm]{figures/rbig_components/hist_x_u.png}            &      \includegraphics[width=1.2cm,height=1.2cm]{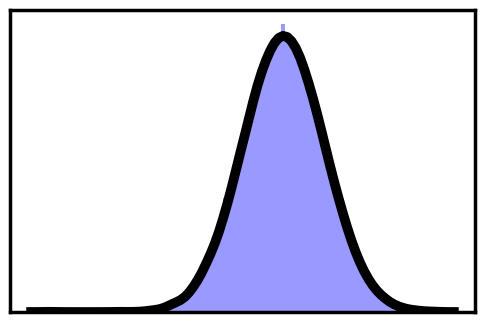}  \\ \hline
     Marginal Gaussianization & $\Psi = \text{CDF}^{-1}\circ \text{U}$                      & Marginal Uniformization + Inverse CDF                                                                                         &
     $\mathbb{R} \rightarrow \mathbb{R}$ &
     \includegraphics[width=1.2cm,height=1.2cm]{figures/rbig_components/hist_x.png}            &      \includegraphics[width=1.2cm,height=1.2cm]{figures/rbig_components/hist_x_g.png}\\ \hline
     Rotation                 & $\bR$
     & Principal Components Analysis \cite{Laparra2011RBIG},\newline Independent Components Analysis \cite{Chen2000Gaussianization}, \newline Random Rotations \cite{Laparra2011RBIG}
     & 
     $\mathbb{R}^d \rightarrow \mathbb{R}^d$ &
     \includegraphics[width=1.2cm,height=1.2cm]{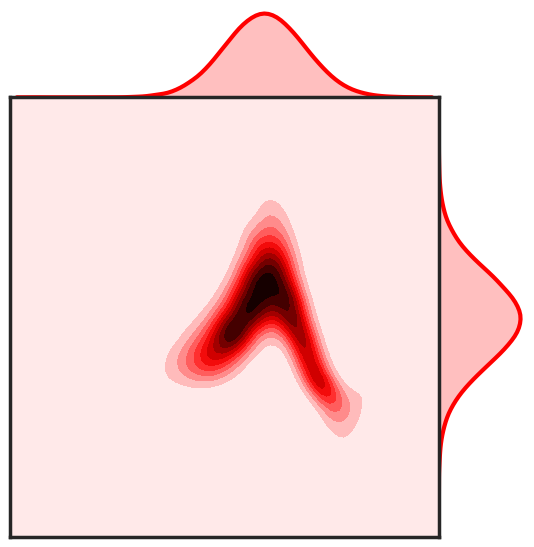}            &      \includegraphics[width=1.2cm,height=1.2cm]{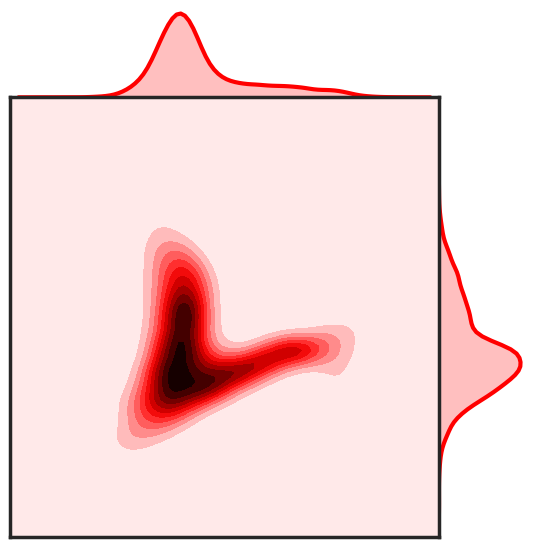}\\ \hline
     Gaussianization Block    & $G_\ell = \mathbf{R} \left[ \Psi^1 \cdots \Psi^d  \right]$ 
     & Composition of \newline Rotation + Marginal Gaussianization
     & $\mathbb{R}^d \rightarrow \mathbb{R}^d$
     &
     \includegraphics[width=1.2cm,height=1.2cm]{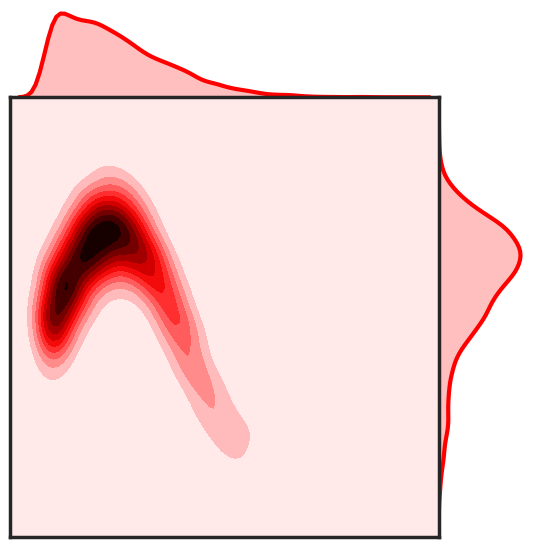}            &      \includegraphics[width=1.2cm,height=1.2cm]{figures/rbig_components/joint_x_l1.png}\\ \hline \hline
     Gaussianization transform                     
     & $G= \left[G_{1} \circ  \cdots \circ G_{L}  \right]$         
     & Composition of \newline Gaussianization Blocks                               
     &       
     $\mathbb{R}^d \rightarrow \mathbb{R}^d$
     &\includegraphics[width=1.2cm,height=1.2cm]{figures/rbig_components/joint_x.png}            &      \includegraphics[width=1.2cm,height=1.2cm]{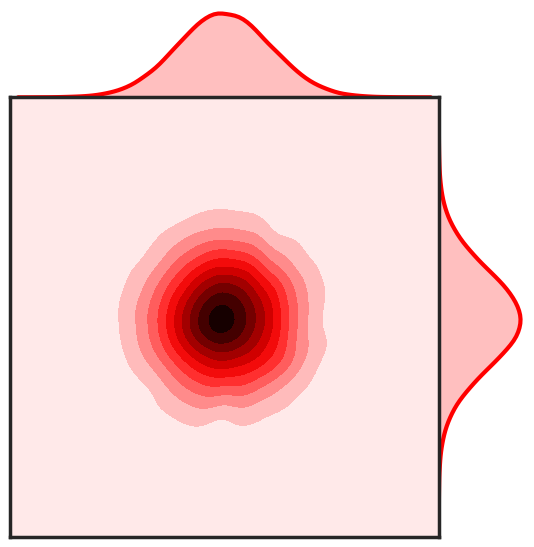} \\ \hline\hline
\end{tabular}
\end{table*}
\begin{figure*}
    \centering
    \vspace{3mm} \includegraphics[width = 16cm]{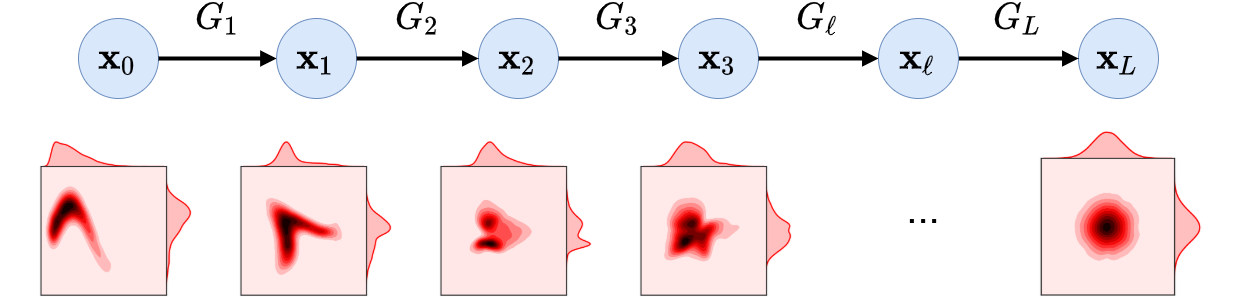}
    \caption{A demonstration of a complete Gaussianization of a noisy sine wave to a marginally and jointly Gaussian distributed. We use {PCA} for the rotation matrix and  {histogram} CDF estimator for the marginal transformation.}
    \label{fig:rbigflow}
\end{figure*}

\subsection{Gaussianization for PDF estimation}

\iffalse
\commentE{
\begin{itemize}
 \item Introduce Gaussianization - original (chen), random rotations (laparra), fully parameterized (meng)
 \item redo the cost function with KL divergence related to the above section
 \item stress on how that representation relates to the IT measures
 \item talk about some of the details (rotations, marginal gaussianization options, computational cost)
\end{itemize}}
\fi

An alternative way to estimate a PDF from observational data is through a data transformation to a {\em convenient} domain, instead of working explicitly in the high-dimensional input domain. The question of what is a convenient domain is a long-standing one, yet ideally this should be a domain with independent components so one can work in each dimension independently to get rid of the curse of dimensionality, one that allows to perform operations and compute quantities therein, and one that is invertible so that one can express these quantities in meaningful units of the input domain. 

The Gaussian distribution has desirable properties of showing independent components and being mathematically tractable and is thus a good candidate for density estimation. A class of Gaussianization methods \cite{Meng2020GaussianizationF, Laparra2011RBIG} look for transforms to a multivariate Gaussian domain.  
These transforms are related to projection pursuit transformations originally introduced in \cite{Friedman1987ExploratoryPP} and seek to transform a multivariate distribution $p(\x)$, where $\x\in\Real^d$, into a standardized multivariate Gaussian distribution  \cite{Chen2000Gaussianization,Laparra2011RBIG}: 
\begin{eqnarray}
\begin{array}{llll}
      G_{\btheta}: & \x\in \Real^d & \mapsto & \z\in \Real^d \\ 
      &   \sim p(\x) & &  \sim {\mathcal N}({\bf 0},{\bf I}_d),
\end{array}
\end{eqnarray}
where $\btheta$ are the parameters learned to Gaussianize the data $\x$, $\bf{0}$ is a vector of zeros (for the means) and $\bf{I}_d$ is the identity matrix (for the covariance). By construction, the Gaussianization transform is a parameterized function $G_{\btheta}$ consisting of a sequence of $L$ iterations (or layers), each one of them performing an orthogonal rotation of the data and a marginal Gaussianization transformation to each feature.
% 
% \commentE{
% So the reviewer was super confused about this $\theta$. They wanted to be clearer about what $\theta$ is. In our case, it's a few things: 
% \begin{itemize}
%     \item The marginal transformation used
%     \item Monotonically increasing functions that convert the marginal distribution into a Gaussian. (Point to the table!)
%     \item Rotation matrix (stress multidimensional)
%     \item The number of layers $L$
% \end{itemize}
% }

% =[x_1,\ldots,x_D]^\top with PDF $p(\x^{(0)})$, 
The transformation $G_{\btheta}$ in each iteration $\ell$ is defined as:
\begin{equation*}
G_{\btheta}: \x_{\ell+1} = {\bf R}_\ell \boldsymbol{\Psi}_\ell(\x_{\ell}),~~~\ell=1,\ldots,L
\end{equation*}
where $\x_{0}$ corresponds to the original data $\x$, $\boldsymbol{\Psi}_\ell$ is the marginal Gaussianization of each dimension of $\x^\ell$ for the iteration $\ell$, and ${\bf R}_\ell$ is a rotation matrix for the marginally Gaussianized variable $\boldsymbol{\Psi}_\ell(\x^{\ell})$. After convergence in $L$ iterations, the transformation contains all the needed information to transform data coming from the original density into a multivariate Gaussian. \textcolor{black}{$\btheta$ collectively group all parameters in the method: those from the rotation matrix $\bR$ and the marginal transformation $\bPsi$. For example, one could use a principal components analysis (PCA) transformation for the rotation matrix $\bR$ and a histogram transformation for the marginal Gaussianization transformation $\Psi$. Then, the eigenvectors obtained from PCA describing $\bR$ and the parameterizations of $\bPsi$ would define $\btheta$. See Table \ref{tab:rbigparts} for more details on the decomposition of this formula and Fig. \ref{fig:rbigflow} for a full decomposition of a toy dataset.}

We can use the change of variables formula to calculate the PDF of $\x$ as
\begin{equation}\label{eq:changeofvariables}
     p_{\x}(\x) = p_{\z}\left(G_{\btheta}(\x) \right)\left| \nabla_\x G_{\btheta(\x)} \right|,
\end{equation}
where $\left| \nabla_{\x} G_{\btheta}(\x) \right|$ is the determinant of the Jacobian of $G_{\btheta}$ w.r.t. $\x$. Generally, any unknown PDF of $\x$ can be estimated as long as we have the transformation $G_{\btheta}$ along with its Jacobian. 
Intuitively, this transformation essentially destroys the density of $\cX$ into unstructured noise (often Gaussian) \cite{Inouye2018DeepDD}. There is no limit to the amount of composite transformations $G_{\mathbf{\Theta}}=G_{\btheta_1} \circ G_{\btheta_2} \circ \cdots \circ G_{\btheta_L}$ that can be used in order to sufficiently converge to the Gaussian distribution. In addition, because $G_{\mathbf{\Theta}}$ is invertible, we can sample points in the original domain $\x'\in\cX$ by generating samples in the transformed Gaussian domain and propagating this through the inverse transformation $G_{\mathbf{\Theta}}^{-1}$. Because the transform is a product of linear and marginal operations, both the Jacobian and the inverse transform can be computed easily~\cite{Laparra2011RBIG, sosjaini19a}.

\begin{wrapfigure}{r}{4.4cm}
\vspace{-0.5cm}
\begin{mdframed}[backgroundcolor=gray!20] 
The transformation is invertible and allows for density estimation, synthesis and information quantification in high-dimensional Earth data problems
\end{mdframed}
\vspace{-0.5cm}
\end{wrapfigure}

\textcolor{black}{
The original Gaussianization algorithm \cite{Chen2000Gaussianization} worked by applying an orthogonal rotation matrix via independent components analysis (ICA) and then a mixture of Gaussians (MOGs) for the marginal Gaussian transformation. After enough repetitions $L$, it was shown that this converged to a multivariate Gaussian distribution \cite{Chen2000Gaussianization}. In \cite{Laparra2011RBIG} we extended Gaussianization by realizing that the method will converge with any orthogonal rotation matrix $\bR$ and we named the algorithm \textit{Rotation-Based Iterative Gaussianization} (RBIG). This allowed for more simpler and faster algorithms such as the Principal Components Analysis (PCA) and even randomly generated orthogonal rotation matrices. In addition, much simpler univariate estimators like the histogram was used to speed up the algorithm significantly.
%They also introduced a stopping criteria based on the difference in total correlation between layers (\commentE{is this true?}).
%
Meng et. al. \cite{Meng2020GaussianizationF} coined the term {\em Gaussianization Flows} and extended the iterative algorithm to be fully parameterized and trainable by incorporating a mixture of logistics as the marginal Gaussianization layer and a sequence of Householder Flows \cite{Liu2019GMMHF, Tomczak2016HF} as the rotation layer. They also proved this is a universal approximator and showed convincing results that Gaussianization is comparable to some other classes of methods specifically designed for density estimation or sampling \cite{Meng2020GaussianizationF}. All transformations and example variants can be found in table \ref{tab:rbigparts}.}
% state-of-the-art (SOTA) algorithms  
% , such as the RealNVP, FFJord and GLOW in applications involving density estimation, sampling and robustness to out-of-distribution samples.
% \commentE{Rewrite - make it clear its PDF estimation.}

Irregardless of the method chose, in order to find the parameters $\btheta$ for the transformation $G_{\btheta}$, we minimize the following cost function w.r.t. $\btheta$:
\begin{equation}\label{eq:loss_ng}
 \cL(\btheta) = \kld \left[ p_\z\left(G_{\btheta}(\x) \right) || \cN({\bf 0},{\bf I}_D) \right],
\end{equation}
which is the Kullback-Leibler (KL) divergence between the estimated Gaussian distribution and the true multivariate Gaussian distribution of mean $\mathbf{0}$ and covariance $\mathbf{I}$; in other words, this is a measure of how much {\em non-Gaussian} our distribution is after transformation. This reveals a direct relationship with information theoretic concepts and measures. 
% While it is equivalent to maximizing the likelihood as shown with eq. \ref{eq:loss_ll} this representation . 
Chen \cite{Chen2000Gaussianization,Cardoso03} showed that \eqref{eq:loss_ng} can be decomposed as
\begin{equation}\label{eq:loss_it}
 \cL(\theta) = T(\x) + J_m(\x),
\end{equation}
where $T(\x)$ is the Total Correlation (T) (a.k.a. Multi-Information, Multivariate Mutual Information) between all of the marginal distributions, and $J_m(\x)$ is the KL divergence between the marginal distributions and the standard Gaussian normal distribution. Intuitively, this cost function is trying to minimize the shared information between each of the marginal distributions and ensuring that they follow a standard normal Gaussian distribution. 
% See \cite{Laparra11,Meng2020GaussianizationF} for more details on how this cost function can be decomposed even further. 
We want to highlight here that RBIG vastly transforms and simplifies the PDF estimation problem: from estimating the density of the high-dimensional multivariate distribution in $\cX$ directly, to doing it indirectly through a transformation to a Gaussian domain. All this by using a series of marginal transformations, which are straightforward and fast. 

\begin{figure}[t!]
\small
\begin{center}
\setlength{\tabcolsep}{2pt}
\begin{tabular}{ccc}
\Large$\mathcal X$ & \Large$\hat \z = \bG_\theta(\x)$ & \Large$\x = \bG_\theta^{-1} (\hat \z)$\\[0mm]
\includegraphics[width=4cm,height=4cm]{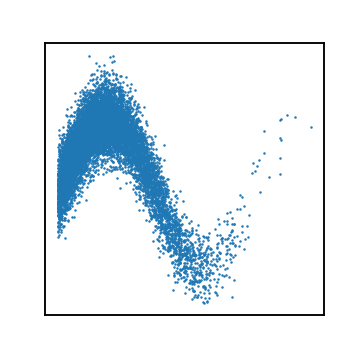} &
\includegraphics[width=4cm,height=4cm]{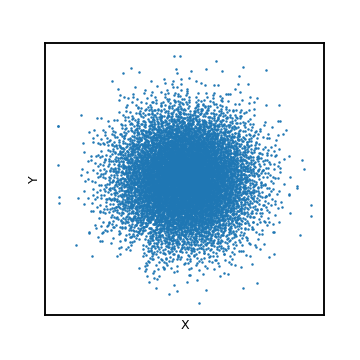} &
\includegraphics[width=4cm,height=4cm]{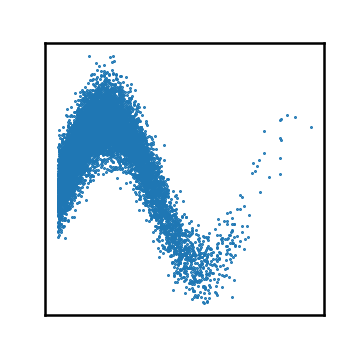} \\
% \hspace{-1.5cm}
\Large$\mathcal Z$ & \Large$\hat \x = \bG_\theta^{-1} (\z)$ & \Large$\Delta T$, Non-Gaussianity \\[0mm]
\includegraphics[width=4cm,height=4cm]{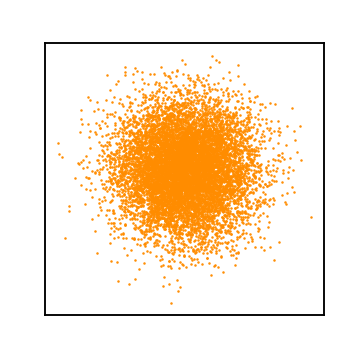} &
\includegraphics[width=4cm,height=4cm]{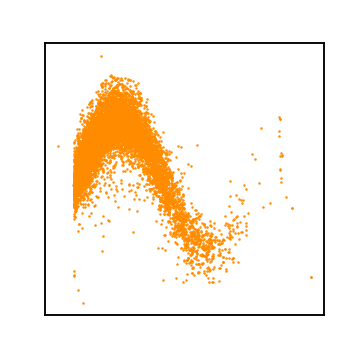} & 
\hspace{-3mm}
\includegraphics[width=4cm,height=4cm]{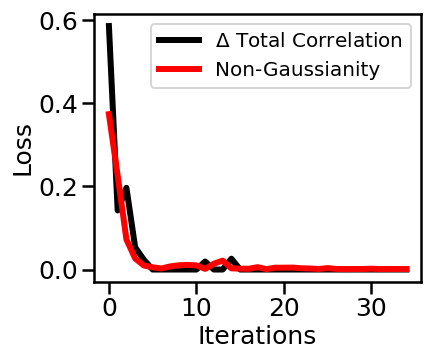}
\end{tabular}
\vspace{-0.0cm}
\caption{Density estimation of sinusoid with heteroscedastic noise using RBIG. 
Top: The original data distribution $\cX$ is mapped to a Gaussian domain $\cZ$ with transform $\bG_\theta$ parameterized by a set of rotations and marginal Gaussianizations collectively denoted as $\theta$, which has an analytic inverse transformation, $\x = \mathbf D_\theta^{-1} (\hat z)$ to recover the original data.
Bottom: One can sample random data from the Gaussian in domain $\cZ$ and use the inverse transformation of $\z$ to $\hat \x$ for data synthesis. We also demonstrate the losses; the equivalence of the change in total correlation between layers $\Delta T$ and the KL-Divergence between transformed data and a multivariate Gaussian (Non-Gaussianity)} % used as stopping criterion with the 
\label{fig:rbigtoy}
\end{center}
\end{figure}

An illustrative example of how RBIG works on a simple 2D toy dataset is shown in Figure~\ref{fig:rbigtoy}. We transform a non-Gaussian 2D dataset into a 2D marginal and jointly Gaussian distribution along with the inverse transformation (first row). The second row showcases how we can use RBIG to synthesize points in the data domain using the inverse transformation. The bottom right figure shows evolution through iterations of the final total correlation (as a measure of redundancy) and the Non-Gaussianity (as a measure of distance to a Gaussian). Please see

\centerline{\fbox{RBIG site: \url{https://ipl-uv.github.io/rbig/}}}
\vspace{0.25cm}

\noindent for a working implementation of the RBIG algorithm in Python and MATLAB. %So we can use many standard univariate density estimators such as the ones mentioned in \ref{sec:pdf}. 
% This is the Kullback-Leibler (KL) Divergence between the 

% \begin{equation}
%  p_\theta(\x) = p_z(\bF_\theta(\x))\left| \text{det} \nabla_\x \bT_\theta(\x) \right|
% \end{equation}
% %
% \begin{equation}
%  \bD_\text{KL}(p(\x)|| \cN(\mathbf{0,I})) = J(\x) = I(\x) + J_m(\x)
% \end{equation}

% So, we have effectively transformed our problem into one where we can use the plethora of methods mentioned in section \ref{sec:methods}A that were designed for univariate density estimation.

%
\subsection{Information Theory Measures using the RBIG Transform}

\iffalse
\commentE{
\begin{itemize}
 \item Talk about how RBIG calculates the IT measures we mentioned 
 \item Information - density estimation
 \item Entropy - for free through the density estimation
 \item Mutual Info - 2 RBIGs and 1 RBIG with the residual entropy of the transformed variables
 \item Probably need a diagram for how RBIG does all of these similar to the ICML poster. It's \textbf{much} easier to understand that way.
 \item Demo Toy Dataset - something similar to the circles, IT measures demo from Valero compared to RV Coefficient.
\end{itemize}}
\fi

RBIG was designed for density estimation, but was inspired by and had connections to information theory~\cite{Cover06}. The series of transformations learned by RBIG lead to a Gaussian domain so features are statistically  independent. This reduction in redundancy is achieved iteratively and can be explicitly computed by summing up all the layer redundancy reductions. This metric is known as the total correlation and computing this metric subsequently allows us to derive {\em some information-theoretic measures} from the data. %, such as information, entropy, total correlation and multivariate mutual information.

\subsubsection{Information} 
\label{sec:shannon_info}

Shannon information $I$~\cite{Shannon1948TheMT} is based on the idea that a sample, $\x_i$, is more interesting (carries more information) if it is less probable. The formal definition of information is:
\begin{equation}
    I(\x_i) = -\log(p_\x(\x_i)).
    \label{eq:information}
\end{equation}
It can be used for instance to highlight regions of more interest in a dataset. Information can be computed for each sample in our dataset by using RBIG and Eq.~\eqref{eq:changeofvariables}.

The expected value of the provided information by a complete dataset, $\x$, is called entropy: 
\begin{equation}
    H(\x) = \mathbb{E}_{\x} [-\log(p_\x(\x))].
    \label{eq:entropy}
\end{equation}
While the entropy could be computed by estimating the information of each sample in the dataset using Eq.~\eqref{eq:information} and averaging, it is more convenient to compute it by using the ability of RBIG to compute Total Correlation as we will see in the following section.
%where we have used the definition of information given by Shannon. 
%This definition can be extended to variables of multiple dimensions. In this case, the \emph{joint differential entropy} of a multidimensional random variable, $H(\x) = H([x_1,\ldots,x_{D}])$, is given by the union of the univariate sets. 

\subsubsection{Total Correlation} 

The total correlation, $T$, accounts for the information shared among the dimensions of a multidimensional random variable \cite{Watanabe1960TC,Studeny1998MI}. Details on how to compute $T$ using RBIG can be found in \cite{Laparra2011RBIG}, here we sketch the main idea. Given data  $\x \in \Real^D$, we first learn the Gaussianization transform with $L$ iterations, and compute the cumulative reduction in total correlation in each iteration as:
\begin{equation}\label{eq:T}
 T(\x) = \sum_{\ell=1}^{L} \bigg(D~H(\mathcal{N}(0,1)) - \sum_{d=1}^D H(\x^{\ell}_d) \bigg).
\end{equation}
The number of layers $L$ will be determined by the reduction in total correlation with each transformation. If there is no change in total correlation after some threshold number of layers, we can assume that $x_d$ are completely independent. It is important to note that all entropy calculations only involved marginal operations which are simple and fast which allows RBIG to be used on large datasets with a high number of dimensions. 

\subsubsection{Joint entropy} 

While the concept of information is attached to a a particular sample, entropy is used in different fields to characterize how unpredictable a complete process is. 
%The joint entropy $H$ is a function that generalizes the concept of entropy for multidimensional data, and reduces to the expected (average) amount of information from an outcome. 
Entropy can be easily computed from the learned RBIG transformation by 
\begin{equation}
 H(\x) = \sum_{d=1}^D H(\x_i) - T(\x),
\end{equation}
where $\sum_{d=1}^D H(\x_i)$ are marginal entropy estimations and $T(\x)$ also involves marginal estimations, cf.~\eqref{eq:T}.

\subsubsection{Multivariate Mutual Information} 

The multivariate mutual information (MI) accounts for the information shared by two datasets \cite{Cover06}. Estimating MI can be very challenging when working with high dimensional data. Our approach is based on the invariance property of mutual information to reparameterize the space of each variable \cite{Kraskov2004}. Therefore, we essentially Gaussianize the two datasets, $\bX$ and $\bY$, with corresponding transforms that remove their total correlations. Then, the total correlation remaining among both Gaussianized datasets is equivalent to the mutual information between the original datasets: 
\begin{equation}
MI(\bX, \bY) = T([\bG_{\theta_\x}(\bX), \bG_{\theta_\y}(\bY)]),
\end{equation}
which again implies only marginal operations, cf.
eq.\eqref{eq:T}. 

\begin{figure}[h!]
% \small
\begin{center}
\vspace{0.5cm}
\begin{tabular}{c}
\includegraphics[width=10.0cm]{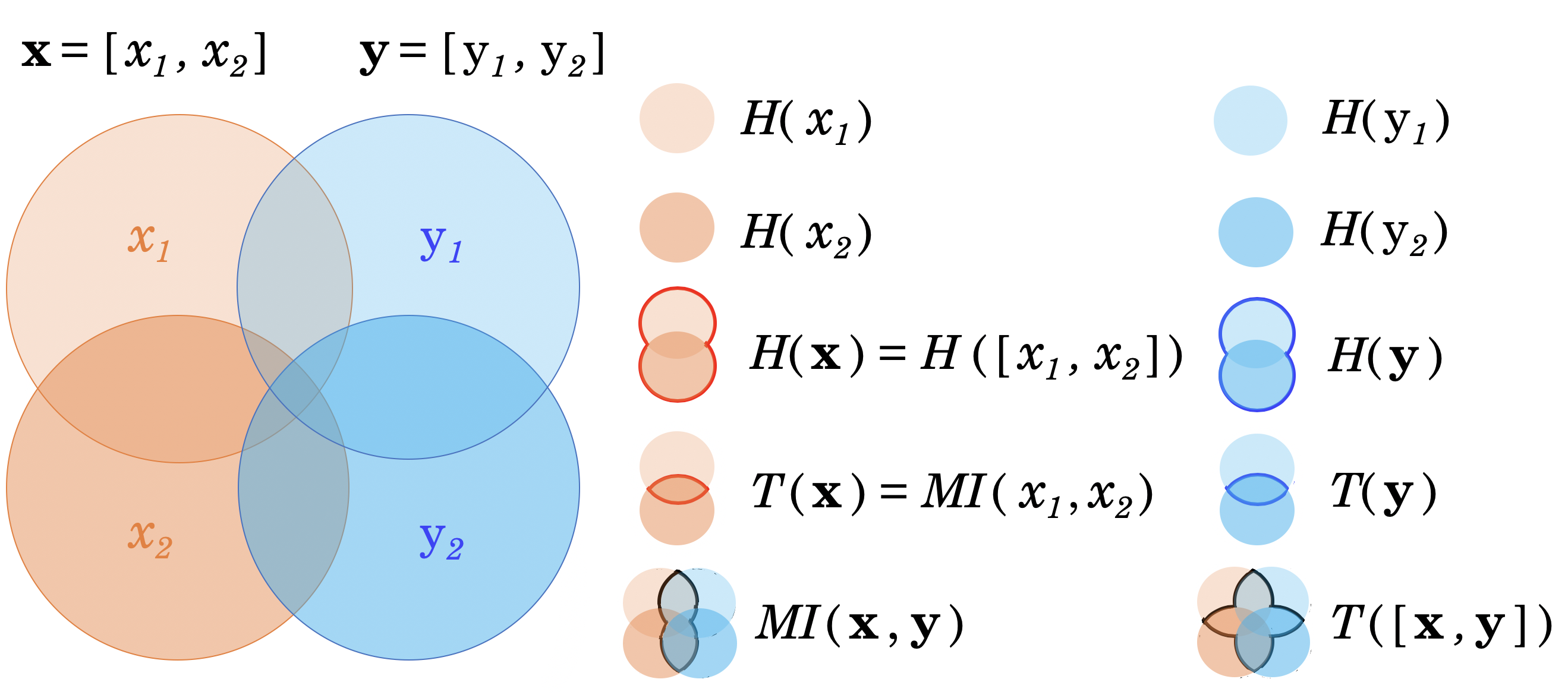}
\end{tabular}
\vspace{-0.15cm}
\caption{A venn diagram showing the relationships of all information theory measures used in this paper. The solid coloured circles represent marginal variables and the intersection regions with bold lines represent regions for information theory measures like mutual information, $MI$ and total correlation, $T$} 
\label{fig:it_measures3}
\end{center}
\end{figure}

Figure~\ref{fig:it_measures3} shows a venn diagram illustrating the different information theory measures used in this paper and Table~\ref{tab:itms} shows a visual demonstration of how they compare to the popular Pearson correlation coefficient for different toy datasets. 

%Table~\ref{tab:itms} shows a more complete example on different information theory measures applied to different toy datasets and figure~\ref{fig:it_measures3} and a complete venn diagram with how they're all related, 
%$I(\x,\y) = T([{\mathcal G}_x(\x),{\mathcal G}_y(\y)])$

%All of these %information theory measures can be calculated using RBIG and further explanations can be found in~\cite{Emman2019,Laparra2011RBIG}.

%Coming back to Gaussianization, the primary reason why we use Gaussianization is because it can calculate information theoretic measures (ITM) by design. In other words, not only do we get a density estimator that allows us to calculate probability estimates and synthesize data, we also get measures like entropy, mutual information and total correlation.

\ra{1.2}

\begin{table}[h!]
\caption{\textcolor{black}{Demonstration showing how different information theory measures discussed compare to the popular Pearson correlation coefficient, $\rho$. This table is also a visual demonstration of how to interpret Mutual Information and how it's related to marginal entropy and the joint entropy; $MI(\mathbf{x}, \mathbf{y})=H(\mathbf{x}) + H(\mathbf{y}) - H(\mathbf{x}, \mathbf{y})$}}
\label{tab:itms}
\begin{tabular}{m{1.5cm}ccccccc}
&
& \includegraphics[width=1.8cm, height=1.9cm]{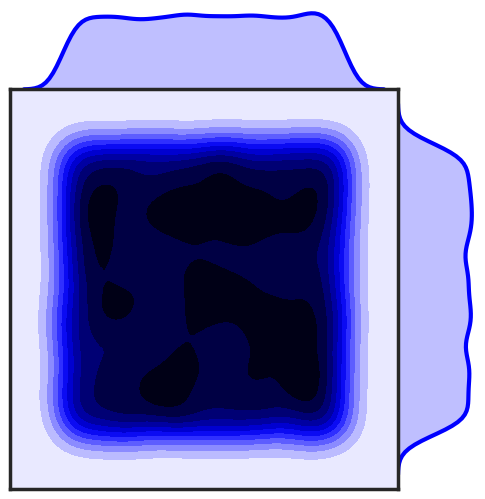}
& \includegraphics[width=1.8cm, height=1.9cm]{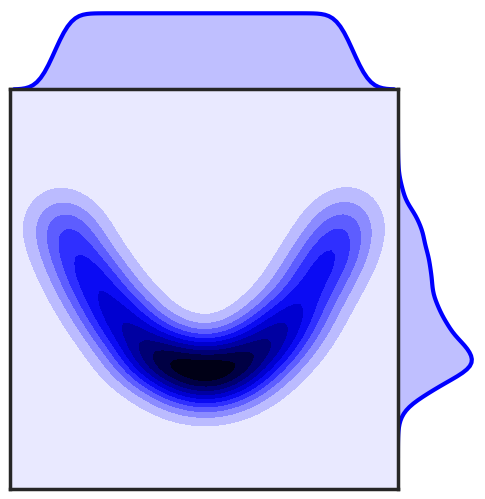}      
& \includegraphics[width=1.8cm, height=1.9cm]{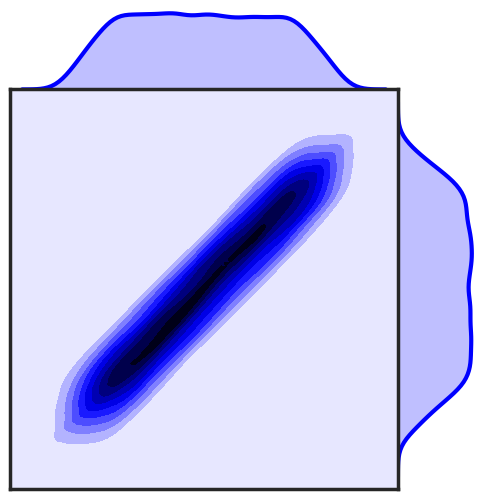} 
& \includegraphics[width=1.8cm, height=1.9cm]{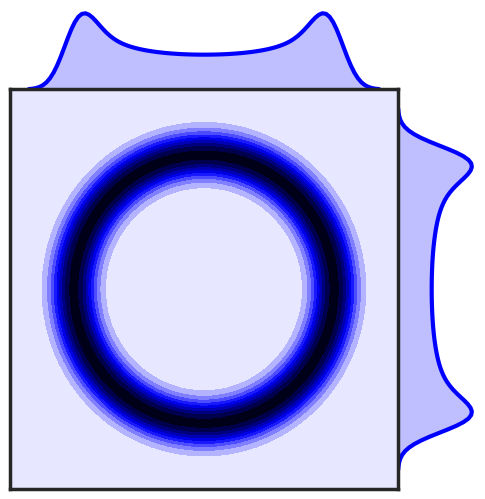}
& \includegraphics[width=1.8cm, height=1.9cm]{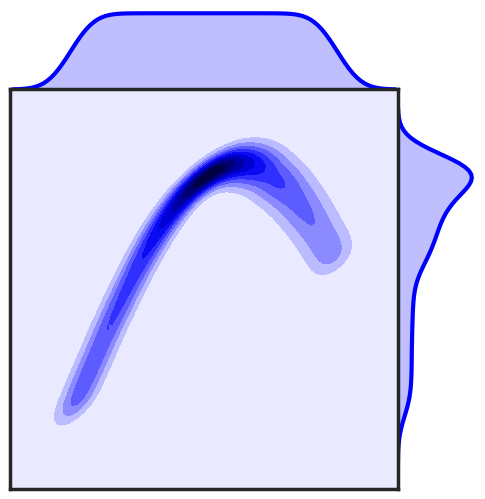}
& \includegraphics[width=1.8cm, height=1.9cm]{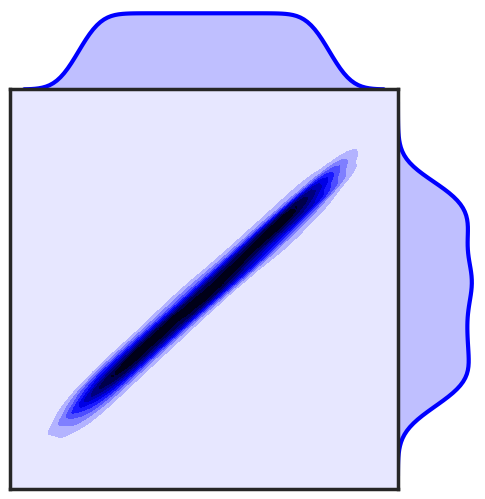} \\ \hline\hline
\textbf{Correlation}  & $\rho(\mathbf{x}, \mathbf{y})$                          
& Low  & Medium & Low    & Low  & Medium & High \\
\textbf{Mutual Information} & $MI(\mathbf{x}, \mathbf{y})$
& Low  & Medium & High   & High & High   & High \\
\textbf{Marginal Entropy} & $H(\mathbf{x}), H(\mathbf{y})$                    
& High & High   & High   & High & High   & High \\
\textbf{Joint Entropy} & $H(\mathbf{x}, \mathbf{y})$                        
& High & Medium & Medium & Low  & Low    & Low \\ \hline\hline
\end{tabular}
\end{table}

%% file: sections/4_experiments.tex
In this section, we explore the information content, the redundancy and the relation in a selection of Earth data analysis problems, involving both remote sensing data and models, using RBIG. 
First we illustrate the method ability to analyze standard remote sensing settings involving total correlation estimation in hyperspectral, radar and very high resolution imagery. 
Second, we quantify the information content of several variables describing the soil-vegetation status,  and investigate the temporal scales leading to maximum shared information for the detection and precursors of anomalies such as droughts.
Finally, we explore the challenging problems of IT measure estimates and the quantification of the spatio-temporal information tradeoff in global Earth products.  
Table~\ref{tab:expsummary} summarizes the experiments in terms of measures, applications and data/simulations used. 

\begin{table}[t!]
    \caption{Summary of experiments with details on the data sets, configurations, application and measures employed.}
    \vspace{-0.0cm}
    \label{tab:expsummary}
    \centering
    \renewcommand{\arraystretch}{1.2}
    \begin{tabular}{c|p{3.2cm}p{4cm}p{1cm}p{1.9cm}p{3cm}p{1.cm}}
\hline 
\hline 
Exp.  &  Data sets  &  Characteristics  & Ref. &  Configuration &  Application  &  Measures\\
\hline 
\hline 
  &  SAR: ERS-2  &  26m, backscatter intensity  & \cite{Chova06} & Pixel-wise  &  Gaussianization  &  T\\
\cline{2-7}
1  &  Hyperspectral: AVIRIS  &  30m, 224 channels  &  \cite{PURR1947} & Pixel-wise  &  Synthesis  &  T\\
\cline{2-7}
 &  Airborne camera: RGB images  &  10cm, 21 classes, 100 images/class   &  \cite{yang2010bag} & Spatial  &  I quantification  &  T\\
\hline 
2  & Optical: MODIS LST, NDVI  &  0.05º, 5.5 years, 14-day  & \cite{Sanchez2018} &  Temporal  &  I quant., PDF comparison  &  H, MI\\
\cline{2-7}
& Passive MW: SMOS SM, VOD  &  25km, 5.5 years, daily  & \cite{Fernandez_Moran_2017} &  Temporal  &  I quant., PDF comparison  &  H, MI\\
\hline
3  &  Obs. \& Sim.: E, SH, Precip  &  0.083º, 10 years, monthly, global  & \cite{Mahecha19esdc} &  Spatio-Temporal  &  I quantification  &  I, H\\
\hline 
\hline 
\end{tabular}
%\begin{flushleft} *SMADI includes optical (LST, NDVI) as well as passive MW (SM) in its computation \cite{Sanchez2016}.
%\newline\commentE{Im a bit confused of the measures column. Is it things that we do or actual things one could do? For example the first two applications (SAR, Hyperspectral), we don't actually do anything related to mutual information. In fact, there's no mention of IT measures at all.}
%\end{flushleft} 
\end{table}

\input{sections/41_remote_sensing}
\input{sections/42_droughts}

\input{sections/43_world_info}

%% file: sections/41_remote_sensing.tex
\subsection{Experiment 1: Gaussianization in remote sensing data} 

This first set of experiments considers the use of RBIG for standard remote sensing image processing. We will show the performance of RBIG in hyperspectral, very high resolution and radar imagery, and for several applications: joint (multivariate) Gaussianization, data synthesis and information estimation.

\subsubsection{Gaussianization of radar images}

\begin{figure*}[t!]
\begin{center}
\setlength{\tabcolsep}{0pt}
\begin{tabular}{cc}
\includegraphics[height=3.5cm]{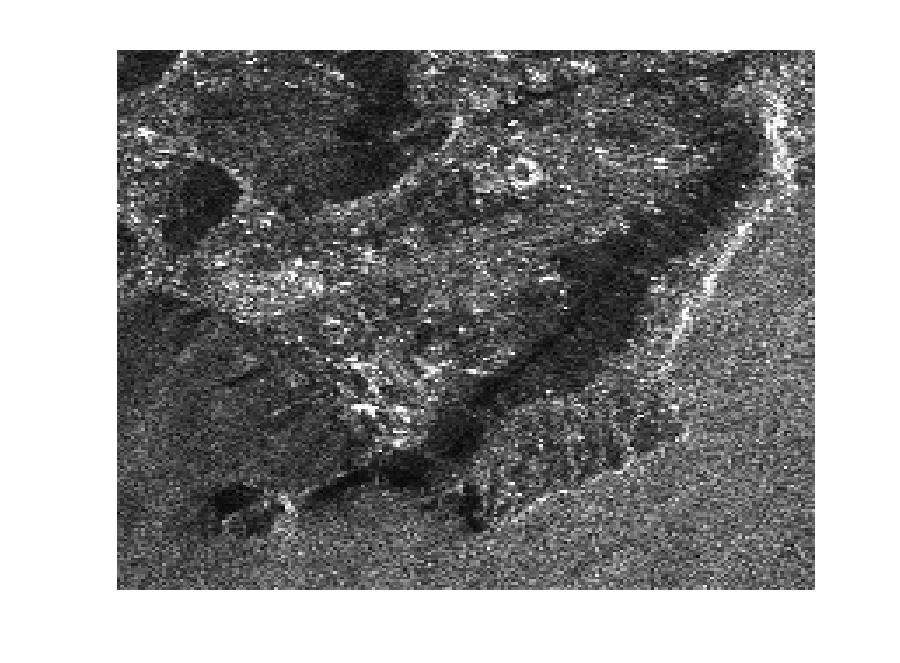} & 
\includegraphics[height=3.5cm]{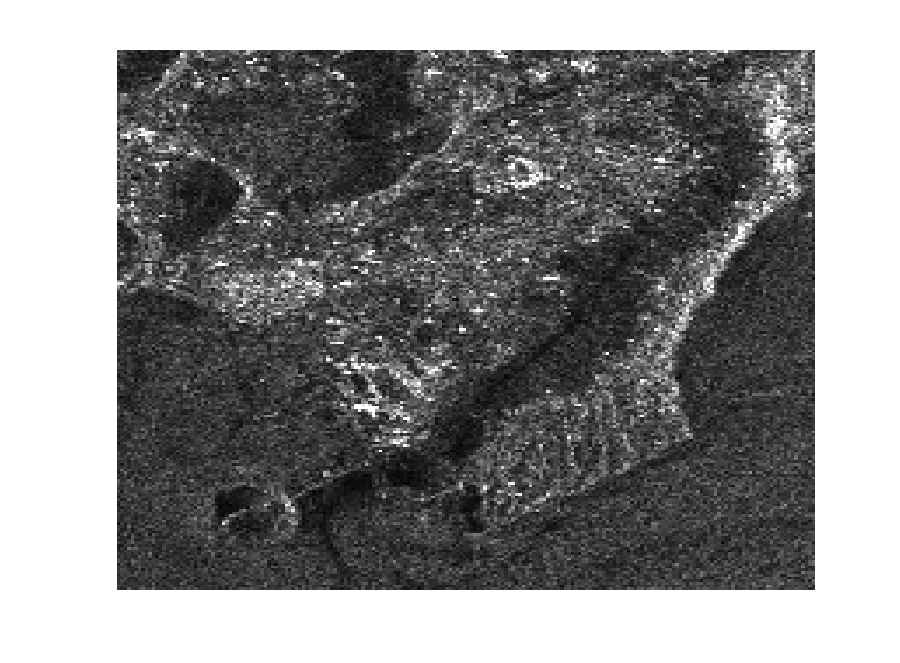}\\
% \begin{tikzpicture}
%     \draw (0, 0) node[inner sep=0] {\includegraphics[height=.8cm]{figures/experiments/remote_sensing/sarimage1.png} };
%     \draw (0.6, -0.4) node {\color{yellow}{$I_1$}};
% \end{tikzpicture}
% \begin{tikzpicture}
%     \draw (0, 0) node[inner sep=0] {\includegraphics[height=1.8cm]{figures/experiments/remote_sensing/sarimage2.png} };
%     \draw (0.6, -0.4) node {\color{yellow}{$I_2$}};
% \end{tikzpicture}
\end{tabular}
\qquad
\begin{tabular}{ccc}
(a) & (b) & (c)\\
\includegraphics[height=5cm]{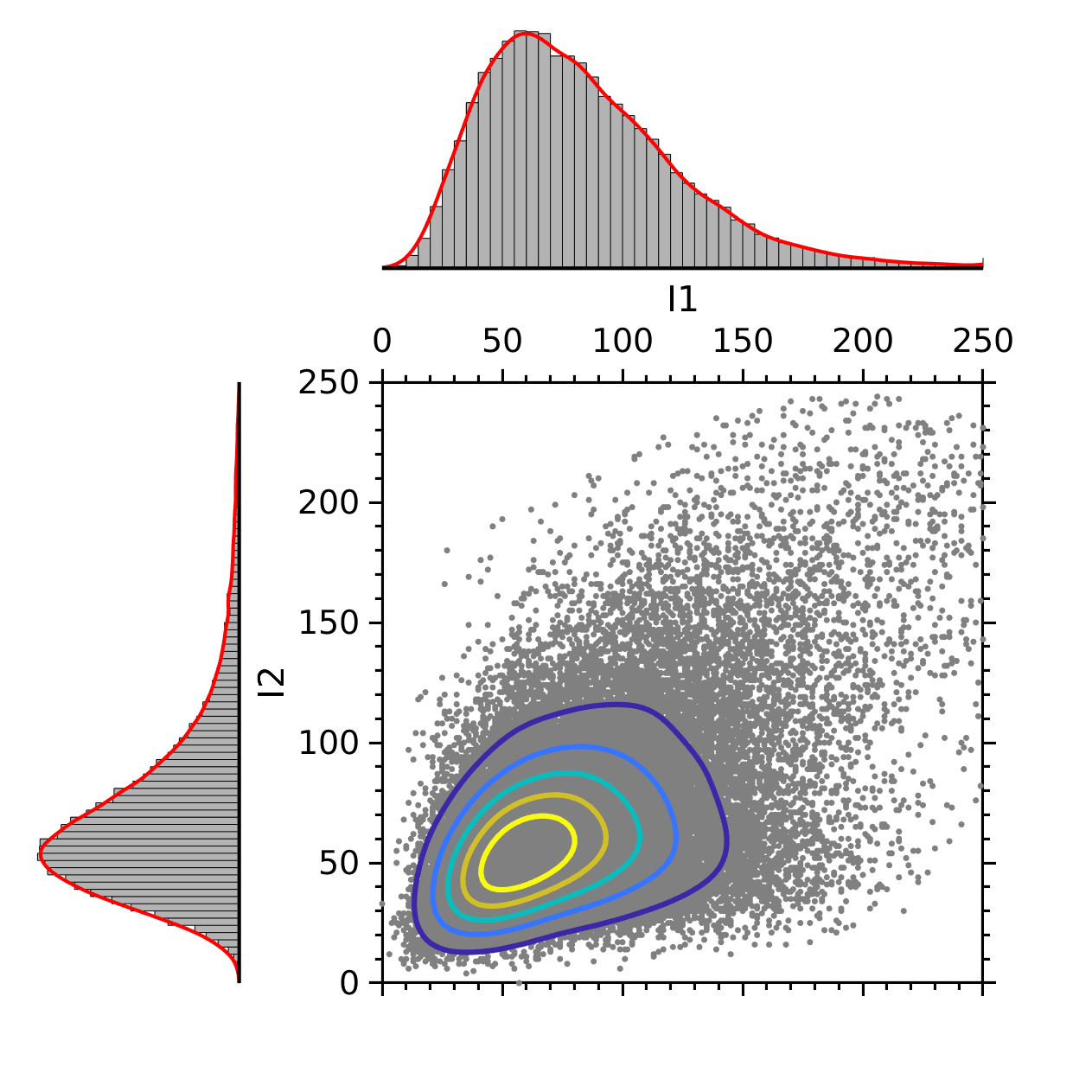} & 
\includegraphics[height=5cm]{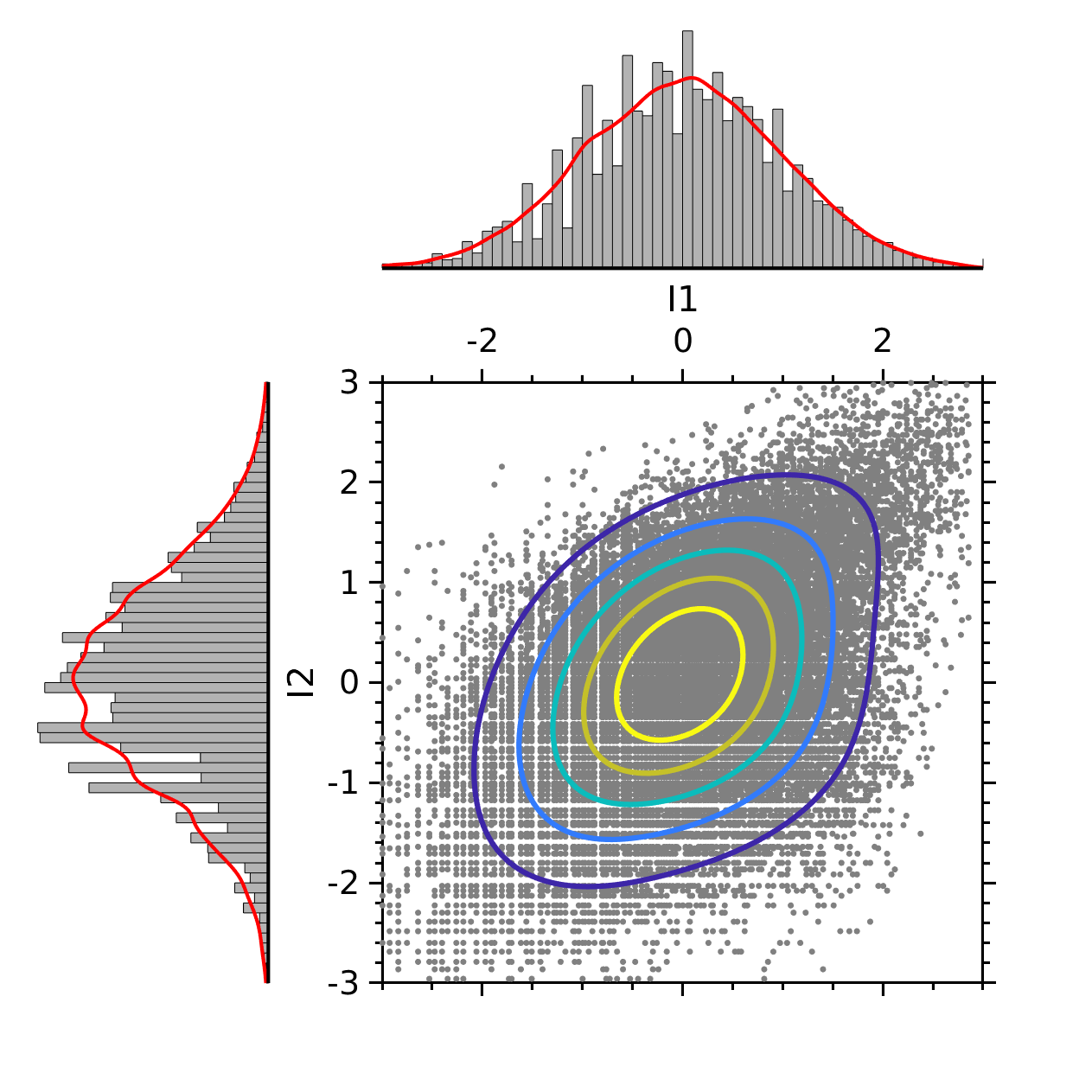}&
\includegraphics[height=5cm]{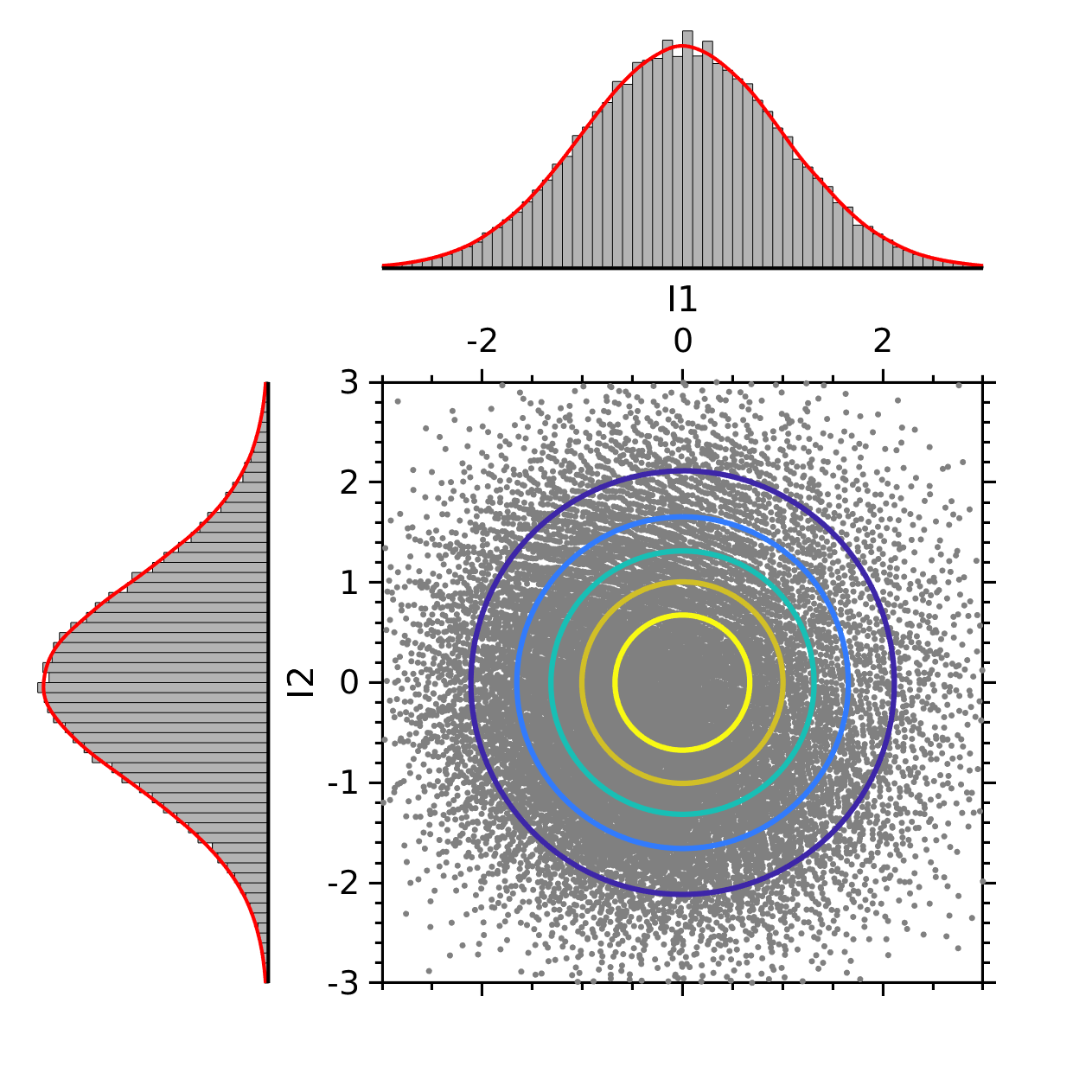}  \\
\end{tabular} 
\end{center}
\vspace{-0.5cm}
   \caption{Radar image processing. We illustrate the Gaussinization of 2D radar data comprised of a pair (I1, I2) of ERS-2 SAR backscattering intensities. The joint distribution is non-Gaussian (a) and preprocessing before applying any algorithm is generally convenient. The standard marginal Gaussiaization (b) does not achieve a full spherical (joint) Gaussian, unlike the RBIG transformation (c). %The computed total correlation, $T$, is indicated for all domains.
   }
\label{fig:sar}
\end{figure*}

The first part of the experiment focuses on analysis of radar imagery. Data used here was collected in the Urban Expansion Monitoring (UrbEx) \href{http://dup.esrin.esa.int/ionia/projects/summaryp30.asp)}{ESA-ESRIN DUP} project \cite{Chova06}. Results from UrbEx project were used to perform the analysis of the selected test sites and for validation purposes. We consider an ERS-2 SAR pair selected with perpendicular baselines between 20 and 150 m in order to obtain the interferometric coherence from each complex SAR image pair. The corresponding pair ($I_1$, $I_2$) of the SAR backscattering intensities (0-35 days) were stacked for analysis, Figure~\ref{fig:sar}[left]. The relation between the intensity features is strongly nonlinear and non-Gaussian and shows a large dispersion, see Figure~\ref{fig:sar}[a]. The total correlation, $T$ is computed with RBIG for the original domain is $T=0.0929$ bits. 
A standard approach in SAR image (pre)processing consists of noise removal and marginal Gaussianization, which can address these problems only partially. This marginal Gaussianization cannot deal with the saturation for high and low signal values, Figure~\ref{fig:sar}[b]. %Here, $T$ reduces to $T=0.0409$ bits. 
A multivariate Gaussianization leads to a fully Gaussian density, Figure~\ref{fig:sar}[c]. This is confirmed by the estimated total correlation $T=0.0095$ bits as it is less than the marginally Gaussianized data.

\begin{wrapfigure}{r}{4.4cm}
\vspace{-0.5cm}
\begin{mdframed}[backgroundcolor=gray!20] 
Multispectral, radar and hyperspectral data exhibit complex nonlinear relations in high dimensional spaces. RBIG maps the data to a convenient Gaussian domain that allows to synthesize new data and quantify information
\end{mdframed}
\vspace{-0.5cm}
\end{wrapfigure}

\subsubsection{Synthesizing hyperspectral images}

\begin{figure*}[h!]
    \ra{1.2}
\begin{center}    
\begin{tabular}{ccc}
(a) Generated spectra & (b) Real & (c) RBIG \\
\includegraphics[height=4.5cm]{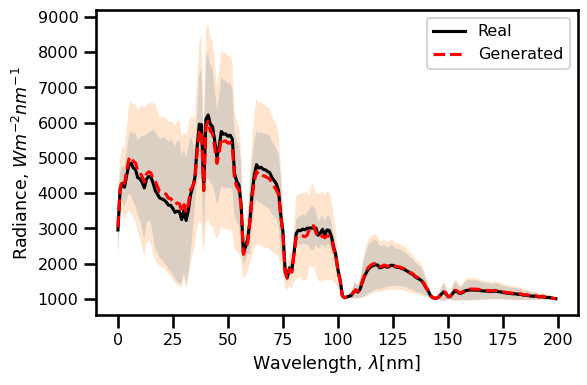}&
\includegraphics[height=4.5cm]{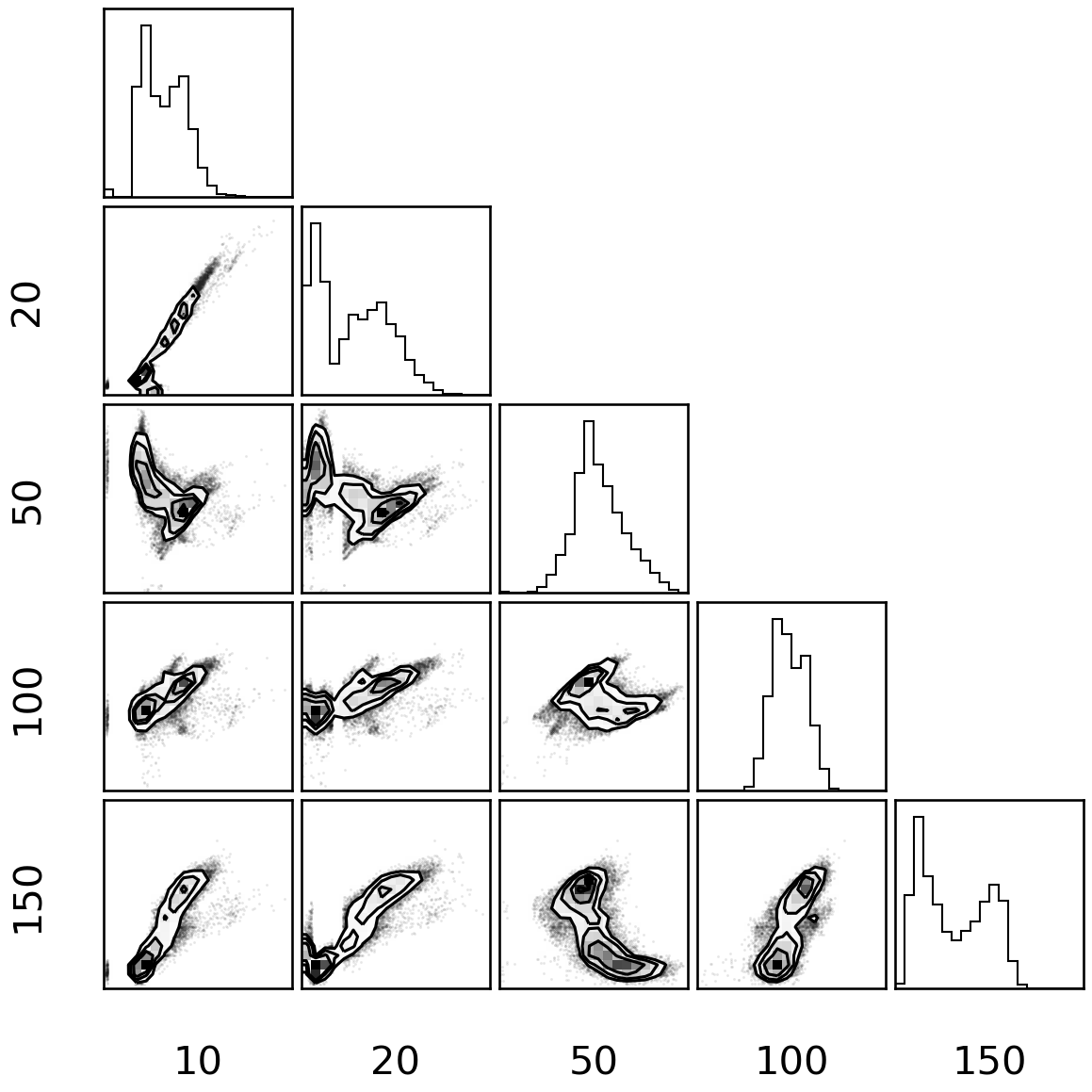}&
\includegraphics[height=4.5cm]{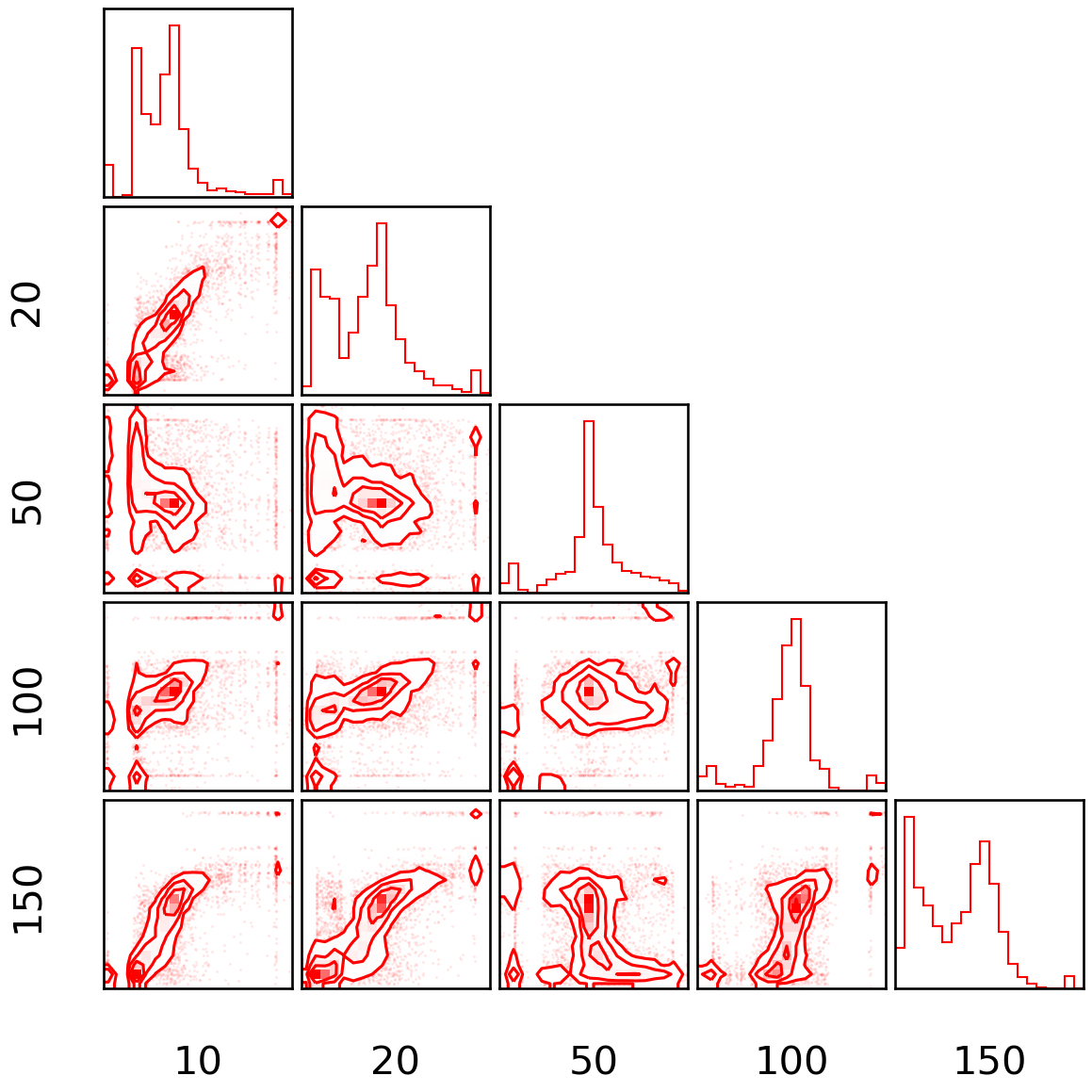} 
\end{tabular}
\end{center}
% \begin{tabular}{c}
% \includegraphics[height=5cm]{figures/experiments/hsi/gen_data.png} \\
% \includegraphics[height=5cm]{figures/experiments/hsi/gen_data_skew.png} \\
% \includegraphics[height=5cm]{figures/experiments/hsi/gen_data_kurtosis.png}
% \end{tabular}
%\\
%\end{tabular} 
\caption{\textcolor{black}{Gaussianization and synthesis of hyperspectral data using RBIG. In (a) we show the mean and standard deviation spectrum for the 21000 real pixels (mean = black, standard deviation = darker shade) and the 1 million pixels generated synthetically (mean = red, standard deviation = lighter shade) using RBIG.  In (b) and (c) we show the marginal and joint distributions of 10, 20, 50, 100, 150 spectral bands for the real data and data generated with RBIG respectively.}}
\label{fig:indianpines}
\end{figure*}

To show the capabilities of the method to deal with high dimensional data here we consider hyperspectral image processing. We took the standard AVIRIS Indian Pines data set~\cite{PURR1947}, % depicted in Fig.~\ref{fig:aviris}[left], 
where the data has
spectrally redundancy and complex joint distrubtions. The image contains 200 spectral channels, which are considered here as the (very high) input dimensionality. We learned a Gaussianization transform leading to a multivariate Gaussian domain of 200 dimensions spectral bands. Then we sampled from a multivariate Gaussian $n=10^6$ samples of 200 dimensions, and inverted them back to the spectral domain. RBIG can be used this way to generate synthetic spectra easily. Figure~\ref{fig:indianpines} (a)
shows the original and the synthesized spectra. This shows how the proposed method allows us to generate/synthesize seemingly spectral distributions even in such a high-dimensional setting. 

\textcolor{black}{Figure~\ref{fig:indianpines} (a) shows the original and the synthesized spectra. This shows how the proposed method allows us to generate/synthesize seemingly spectral distributions even in such a high-dimensional setting. In addition, figure~\ref{fig:indianpines} (b) and (c) show corner plots illustrating the joint distributions between various spectral bands (10, 20, 50, 100, and 150). We see that the marginal and joint distributions for the generated spectra by RBIG in (c) are very similar to the real data in (b) across all pairwise band combinations. This is important to highlight that some of the most widely used methods such as PCA would be able to replicate figure~\ref{fig:indianpines}(a) with a good approximate mean and standard deviation but they would not be able to replicate figure~\ref{fig:indianpines}(d) where {\em all joint distributions} are approximately Gaussian.}

\subsubsection{Information and redundancy in high resolution images}\label{sec:ucmerced}

Very high resolution images are constantly acquired with the new generation of sensors, both on airborne and spaceborne platforms. A systematic analysis of the images is necessary. Machine learning, and deep learning in particular, has led to an important leap in classification accuracy. However, owing to the wealth of data and the diversity, it becomes necessary to design algorithms that exploit most of the information content of images, both in terms of relevant features and examples. 

\iffalse
\begin{figure*}[h!]
\small
\begin{center}
\centerline{\includegraphics[width=18cm]{figures/experiments/remote_sensing/ucmercedpanel.png}}
\setlength{\tabcolsep}{0pt}
\vspace{0.15cm}
\begin{tabular}{cc}
(a) Averaged convergence & (b) Class-specific $T$ \\%& (c) KLD \\
\includegraphics[height=7cm]{figures/experiments/remote_sensing/ucmerced_mi_convergence.png} &  \includegraphics[height=7cm]{figures/experiments/remote_sensing/ucmerced_mi.pdf} 
%& \includegraphics[height=5cm]{figures/experiments/remote_sensing/ucmerced_kld.png}\\
\end{tabular}
\end{center}
\caption{
\textcolor{blue}{Estimation of total correlation, $T$ in aerial imagery. Top: We show four illustrative images for each of the 21 classes in the database. 
Bottom: (a) Average total correlation computed iteratively for the different 21 class-specific RBIG models over 50 iterations, with the mean $T$ (solid) and the $T$ standard deviation (shaded) over all models. Convergence is achieved very rapidly for all classes (note the log-scale). (b) ranked $T$ per class computed from the RBIG models.}
}
%(c) Symmetrized KLD between the 21 classes in the UC Merced database computed with RBIG.} 
\label{fig:ucmerced}
\end{figure*}
\fi

%%%%%%%%%%%%%   new one
%%%%%%%%%%%%%
\begin{figure*}[h!]
\small
\begin{center}
%\centerline{\includegraphics[width=18cm]{figures/experiments/remote_sensing/ucmercedpanel.png}}
\scriptsize
\setlength{\tabcolsep}{1pt}
\begin{tabular}{ccccccccccccccccccccc}
forest & chap. & agric.& park & sparse & golf & airplane & beach & river & harbor & tennis & parking & medium & tanks & dense & baseball & inters. & build & freeway & pass & runway\\
% idx = 8     6     1    14    19    10     2     4    17    11    21    16    13    20     7     3    12 5 9  15    18
\includegraphics[height=0.80cm]{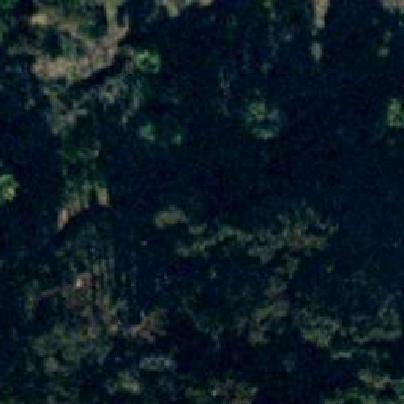} & 
\includegraphics[height=0.80cm]{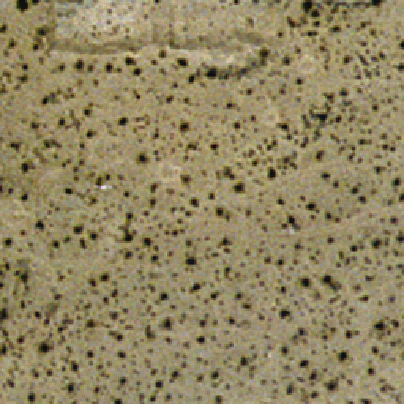} & 
\includegraphics[height=0.80cm]{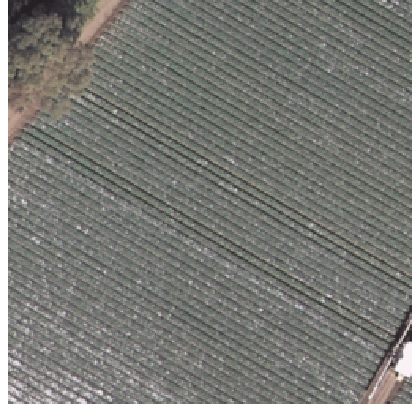} & 
\includegraphics[height=0.80cm]{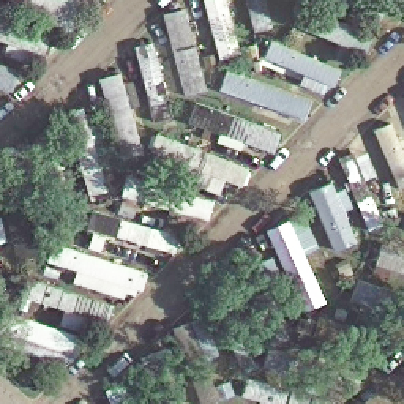} & 
\includegraphics[height=0.80cm]{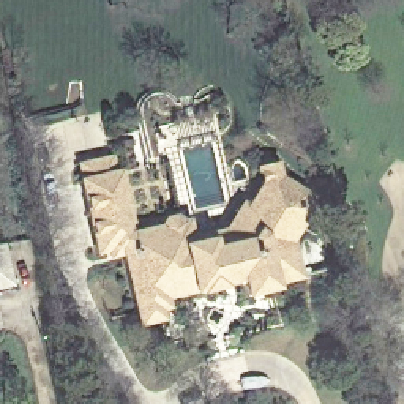} & 
\includegraphics[height=0.80cm]{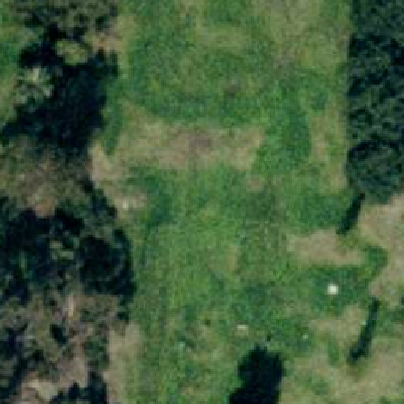} & 
\includegraphics[height=0.80cm]{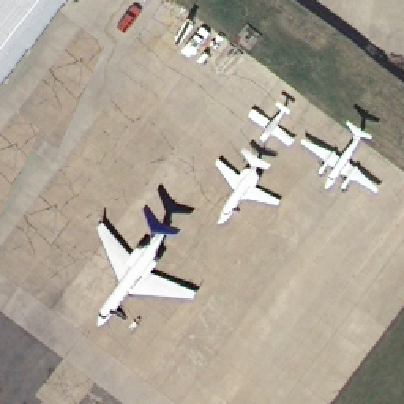} & 
\includegraphics[height=0.80cm]{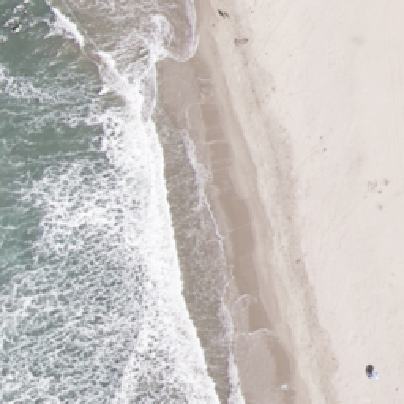} & 
\includegraphics[height=0.80cm]{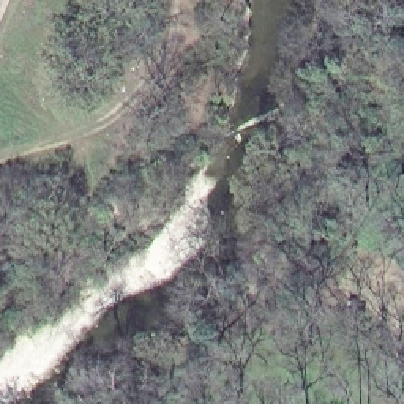} & 
\includegraphics[height=0.80cm]{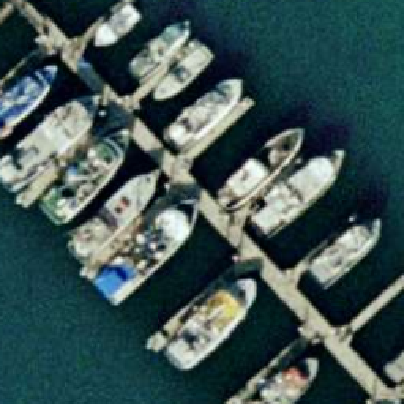} & 
\includegraphics[height=0.80cm]{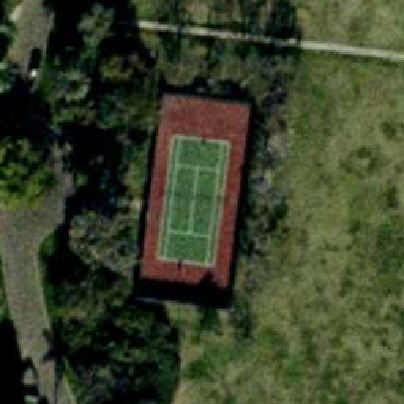} &
\includegraphics[height=0.80cm]{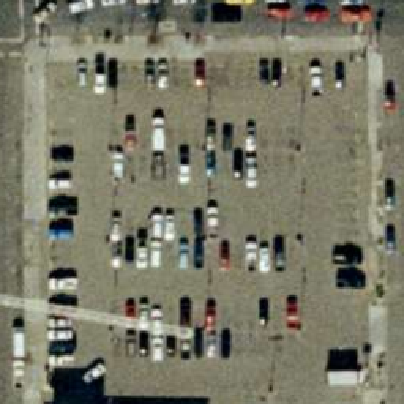} & 
\includegraphics[height=0.80cm]{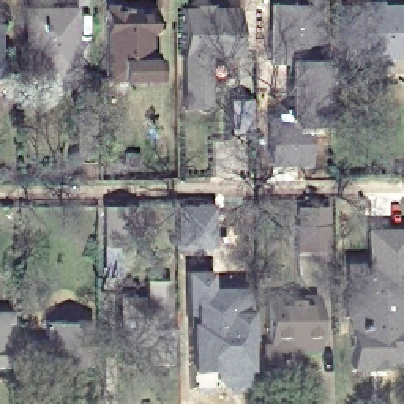} & 
\includegraphics[height=0.80cm]{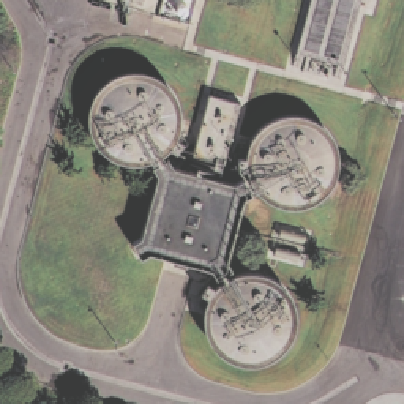} & 
\includegraphics[height=0.80cm]{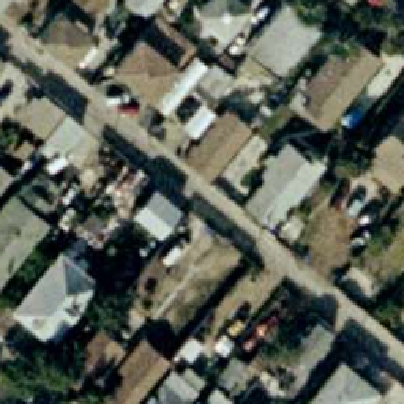} & 
\includegraphics[height=0.80cm]{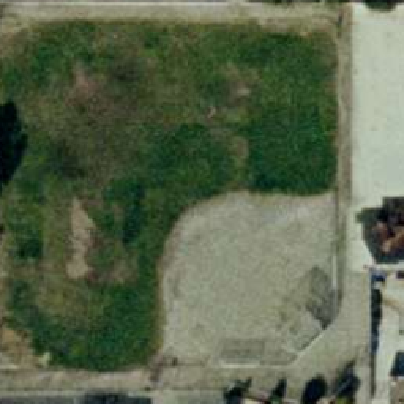} & 
\includegraphics[height=0.80cm]{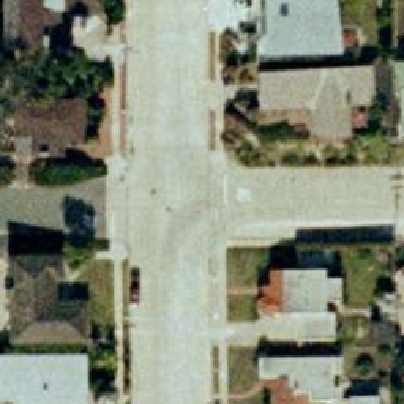} & 
\includegraphics[height=0.80cm]{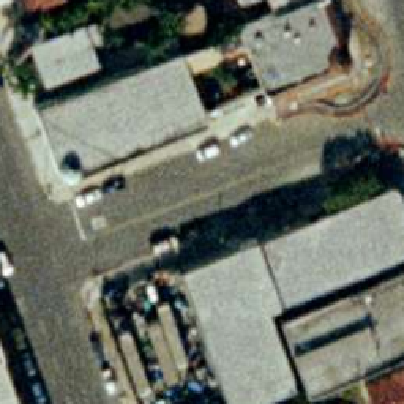} & 
\includegraphics[height=0.80cm]{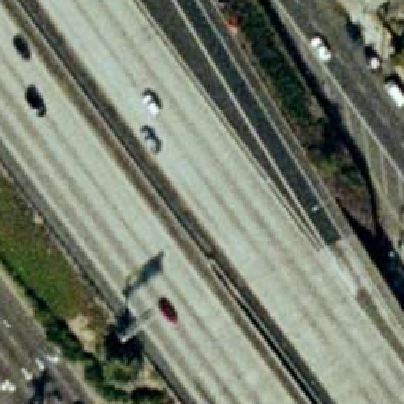} & 
\includegraphics[height=0.80cm]{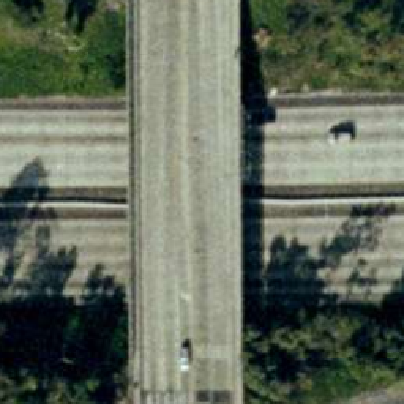} & 
\includegraphics[height=0.80cm]{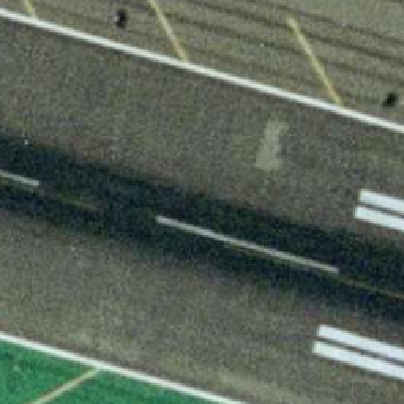} \\
\includegraphics[height=0.80cm]{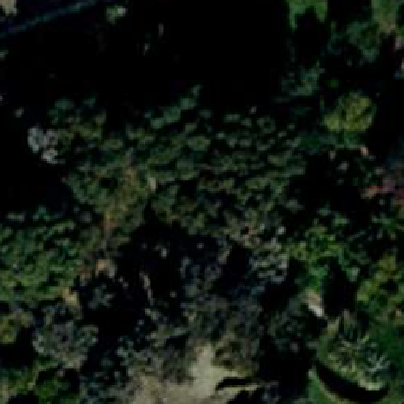} & 
\includegraphics[height=0.80cm]{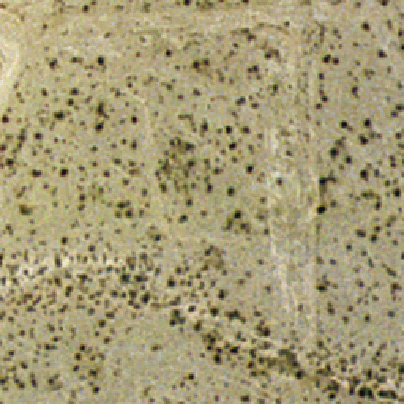} & 
\includegraphics[height=0.80cm]{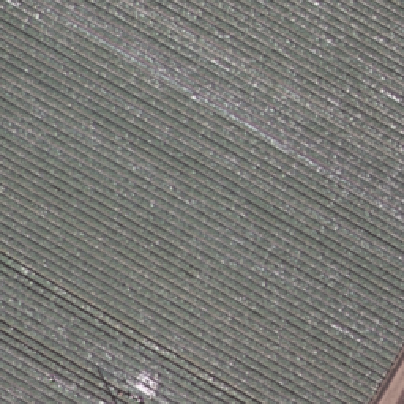} & 
\includegraphics[height=0.80cm]{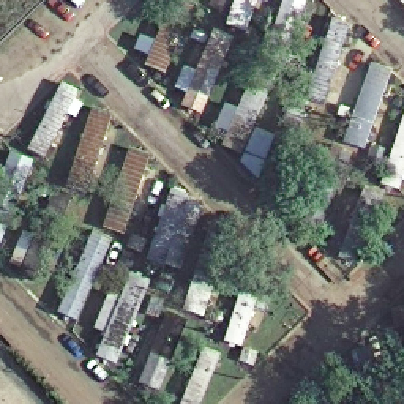} & 
\includegraphics[height=0.80cm]{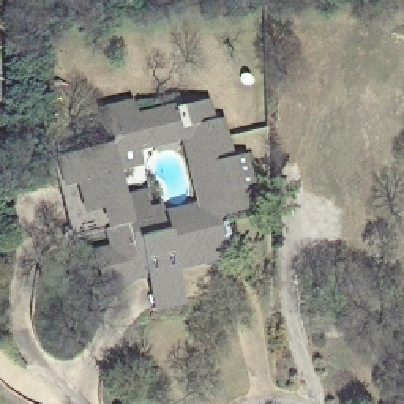} & 
\includegraphics[height=0.80cm]{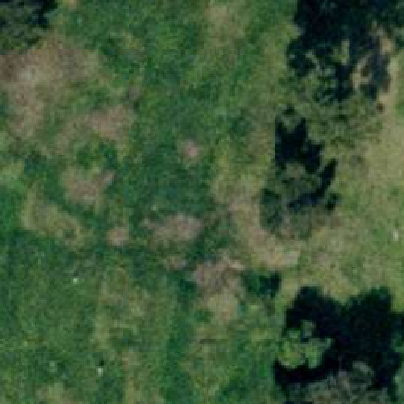} & 
\includegraphics[height=0.80cm]{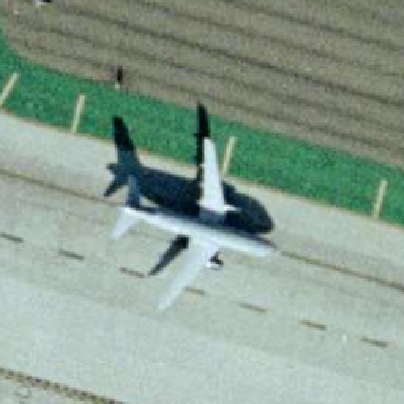} & 
\includegraphics[height=0.80cm]{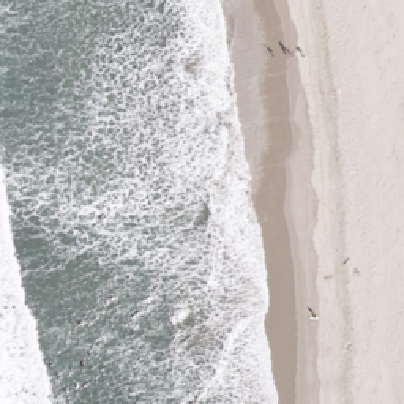} & 
\includegraphics[height=0.80cm]{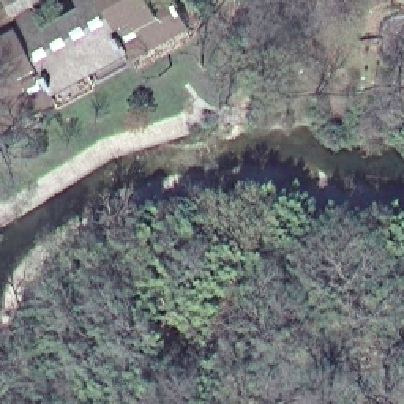} & 
\includegraphics[height=0.80cm]{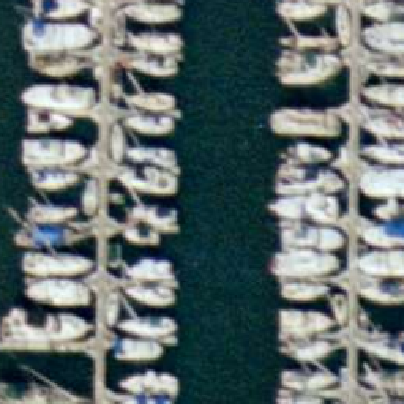} & 
\includegraphics[height=0.80cm]{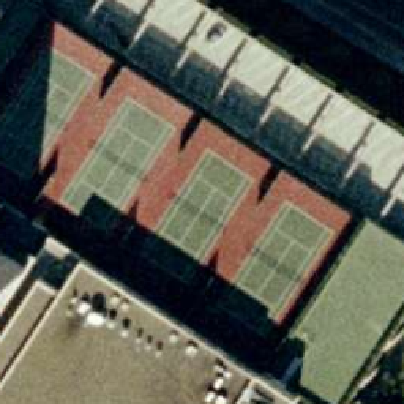} &
\includegraphics[height=0.80cm]{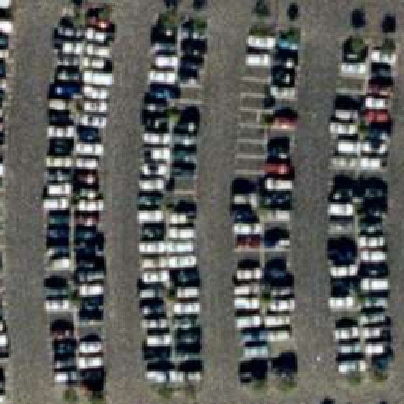} & 
\includegraphics[height=0.80cm]{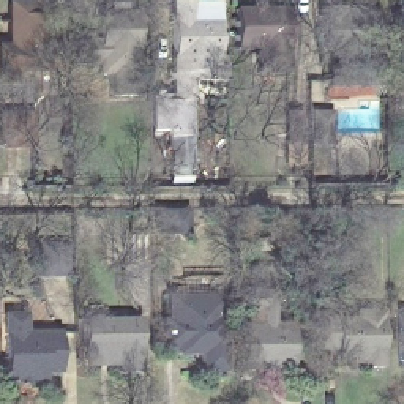} & 
\includegraphics[height=0.80cm]{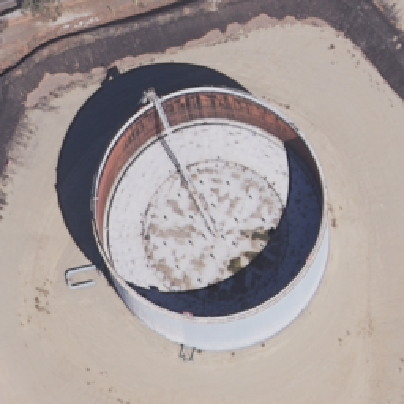} & 
\includegraphics[height=0.80cm]{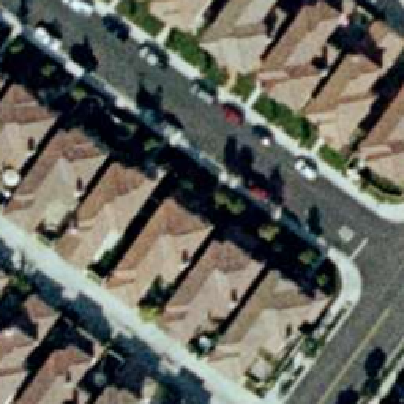} & 
\includegraphics[height=0.80cm]{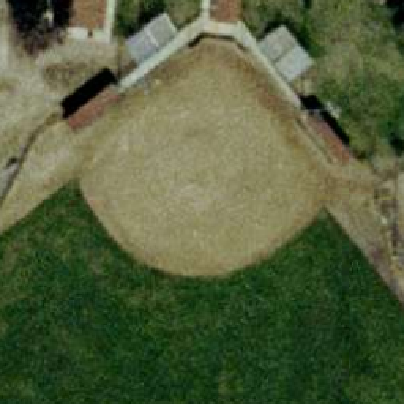} & 
\includegraphics[height=0.80cm]{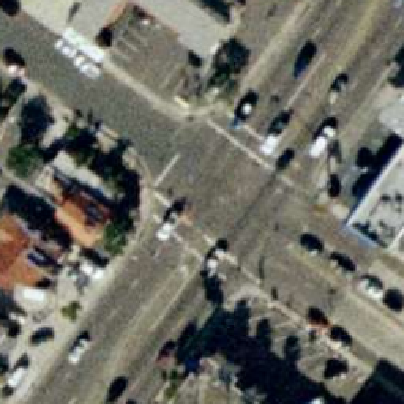} & 
\includegraphics[height=0.80cm]{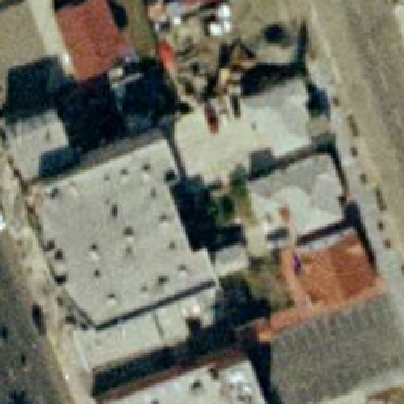} & 
\includegraphics[height=0.80cm]{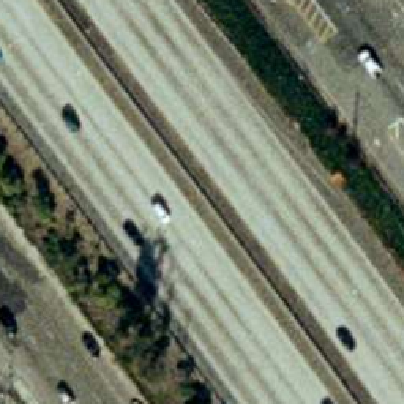} & 
\includegraphics[height=0.80cm]{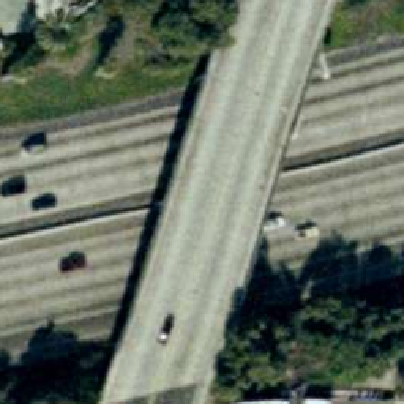} & 
\includegraphics[height=0.80cm]{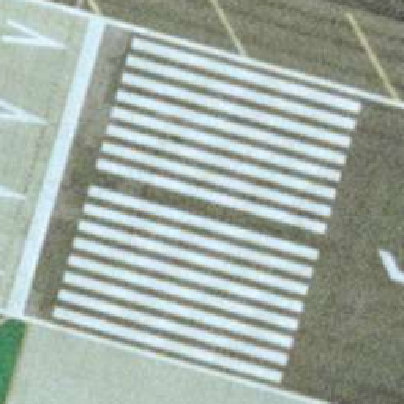} \\
\includegraphics[height=0.80cm]{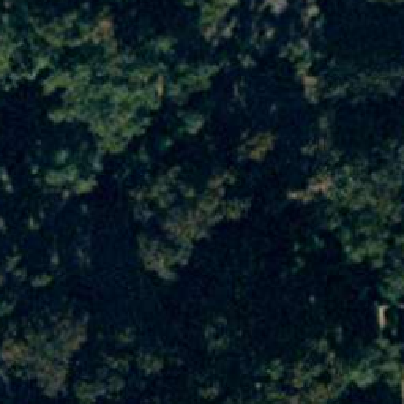} & 
\includegraphics[height=0.80cm]{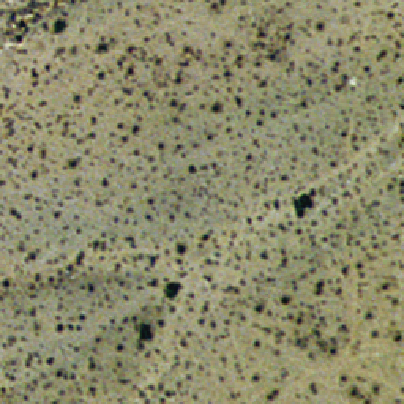} & 
\includegraphics[height=0.80cm]{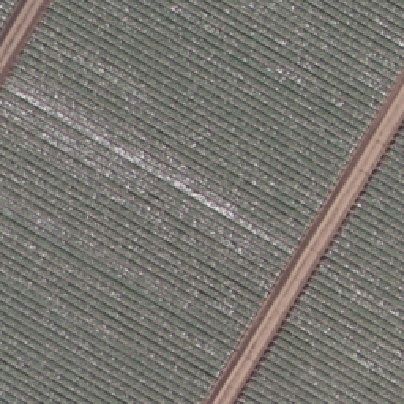} & 
\includegraphics[height=0.80cm]{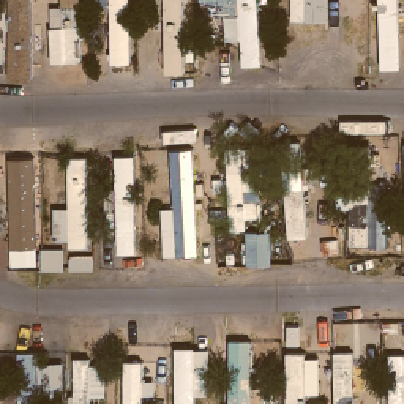} & 
\includegraphics[height=0.80cm]{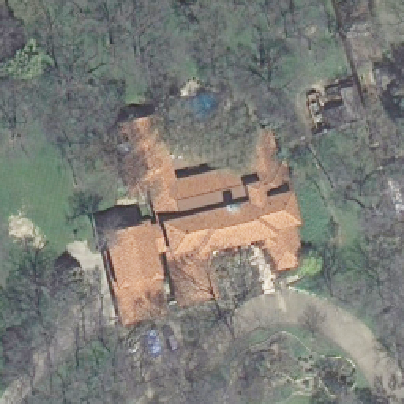} & 
\includegraphics[height=0.80cm]{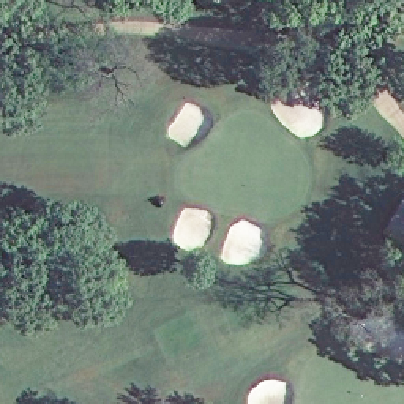} & 
\includegraphics[height=0.80cm]{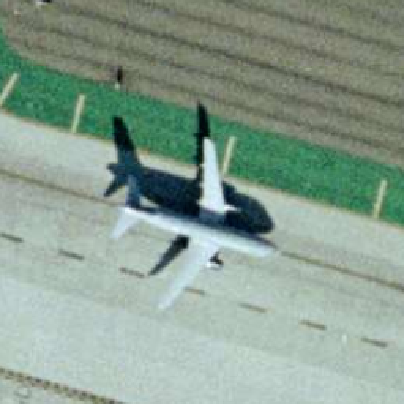} & 
\includegraphics[height=0.80cm]{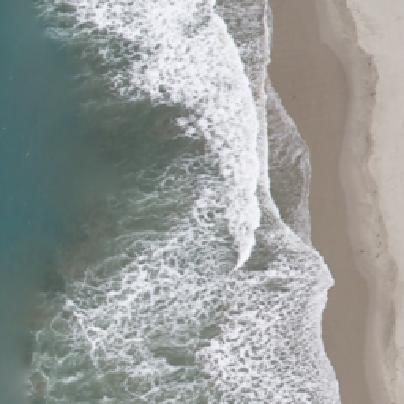} & 
\includegraphics[height=0.80cm]{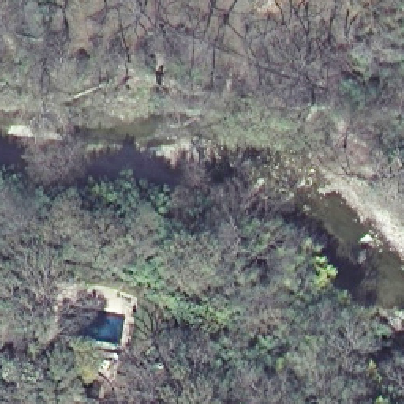} & 
\includegraphics[height=0.80cm]{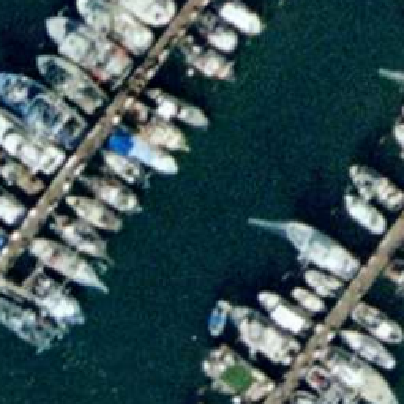} & 
\includegraphics[height=0.80cm]{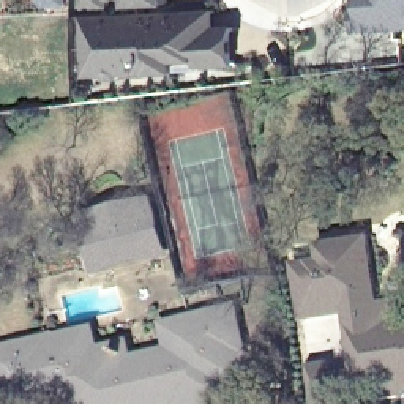} &
\includegraphics[height=0.80cm]{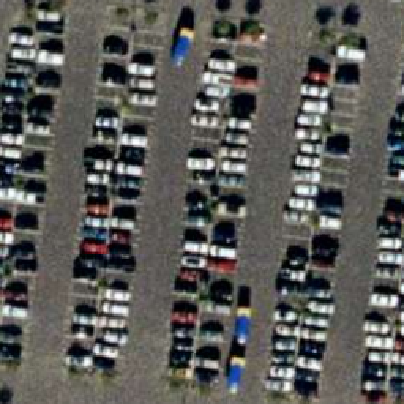} & 
\includegraphics[height=0.80cm]{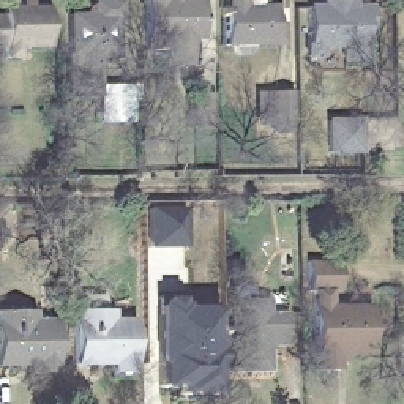} & 
\includegraphics[height=0.80cm]{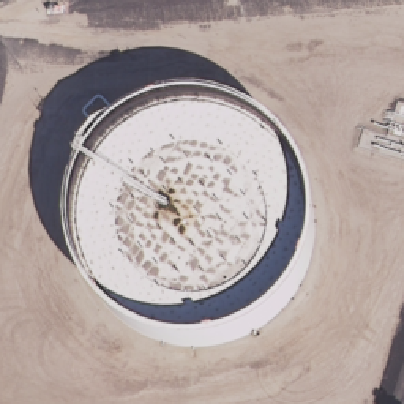} & 
\includegraphics[height=0.80cm]{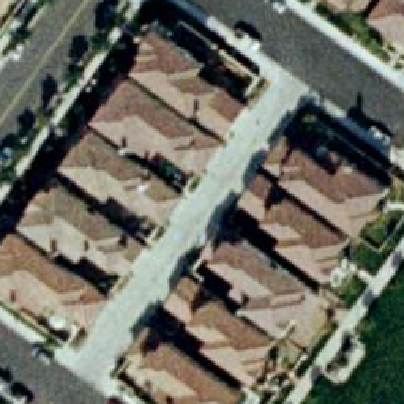} & 
\includegraphics[height=0.80cm]{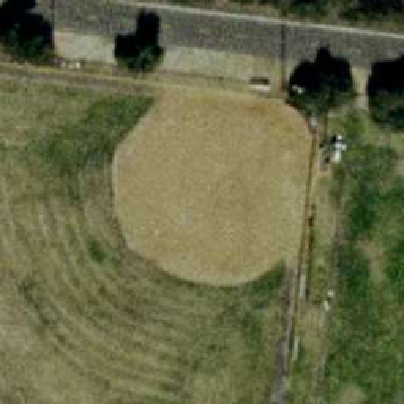} & 
\includegraphics[height=0.80cm]{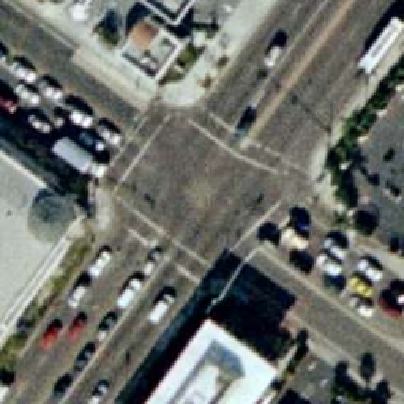} & 
\includegraphics[height=0.80cm]{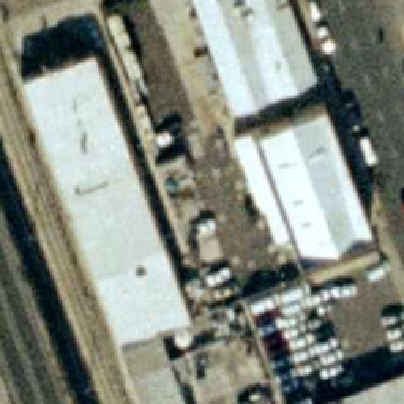} & 
\includegraphics[height=0.80cm]{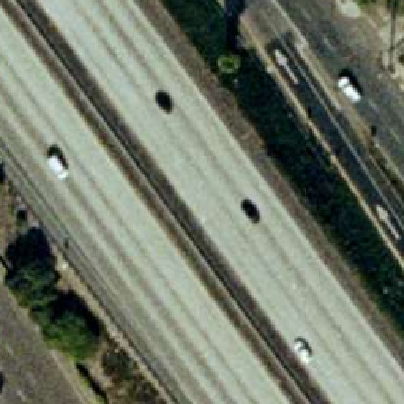} & 
\includegraphics[height=0.80cm]{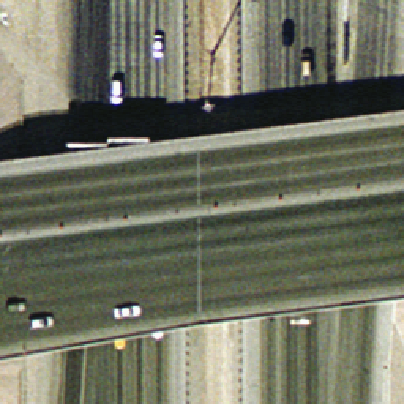} & 
\includegraphics[height=0.80cm]{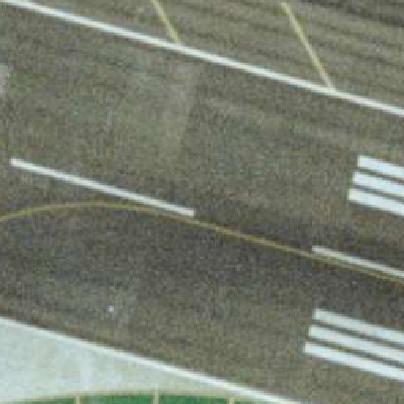} \\
% classes sorted by T
% idx = 8     6     1    14    19    10     2     4    17    11    21    16    13    20     7     3    12 5 9  15    18

% sorted already:
%\def\names{{forest},{chaparral},{agricult},{park},{sparse~resid.},{golf},{plane},{beach},{river},{harbor},{tennis},{parking},{medium~resid.},{tanks},{dense~resid.}{baseball},{intersec.},{building},{freeway},{overpass},{runway}}
 
%{\scriptsize forest chap. agric. homepark sparse~resid. golf airplane beach river harbor tennis parkinglot medium~resid. tanks dense~resid. baseball inters. build freeway overpass runway}\\

%forest&chap.&agric.&homepark&sparse~resid.&golf&airplane&beach&river&harbor&tennis&parkinglot&medium~resid.&tanks&dense~resid.&baseball&inters.&build&freeway&overpass&runway\\

%\foreach \name in \names
%{\scriptsize \name\hspace{0.15cm}}\\

%\foreach \c in {8,6,1,14,19,10,2,4,17,11,21,16,13,20,7,3,12,5,9,15,18} 
%{\includegraphics[height=0.80cm]{figures/experiments/remote_sensing/ucmerced_5_examples/ucmerced_c\c_im1.png}\hspace{0.02cm}}  \\

%\foreach \c in {8,6,1,14,19,10,2,4,17,11,21,16,13,20,7,3,12,5,9,15,18} {\includegraphics[height=0.80cm]{figures/experiments/remote_sensing/ucmerced_5_examples/ucmerced_c\c_im2.png}\hspace{0.02cm}} \\

%\foreach \c in {8,6,1,14,19,10,2,4,17,11,21,16,13,20,7,3,12,5,9,15,18} %{\includegraphics[height=0.80cm]{figures/experiments/remote_sensing/ucmerced_5_examples/ucmerced_c\c_im3.png}\hspace{0.02cm}} \\
\end{tabular}

\setlength{\tabcolsep}{0pt}
\vspace{0.5cm}
\begin{tabular}{ccc}
(a) Data partition & (c) Averaged convergence & (d) Class-specific $T$ \\ 
 \includegraphics[height=2.5cm]{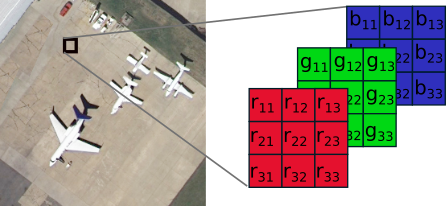} & &
 % \multirow{1}{*}{\includegraphics[height=5cm]{figures/experiments/remote_sensing/ucmerced_mi_convergence.pdf}} & 
 %\multirow{1}{*}{\includegraphics[height=5cm]{figures/experiments/remote_sensing/ucmerced_mi.pdf}} 
 \\
 (b) Measured magnitude & &  \\    
 \includegraphics[height=2.5cm]{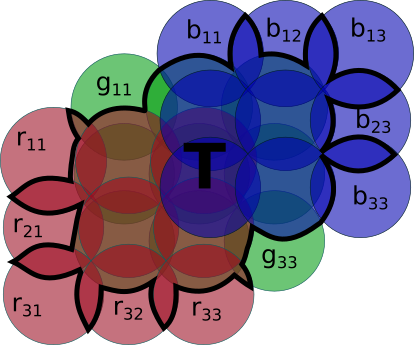} & \multirow{-1}{*}[5cm]{\includegraphics[height=5cm]{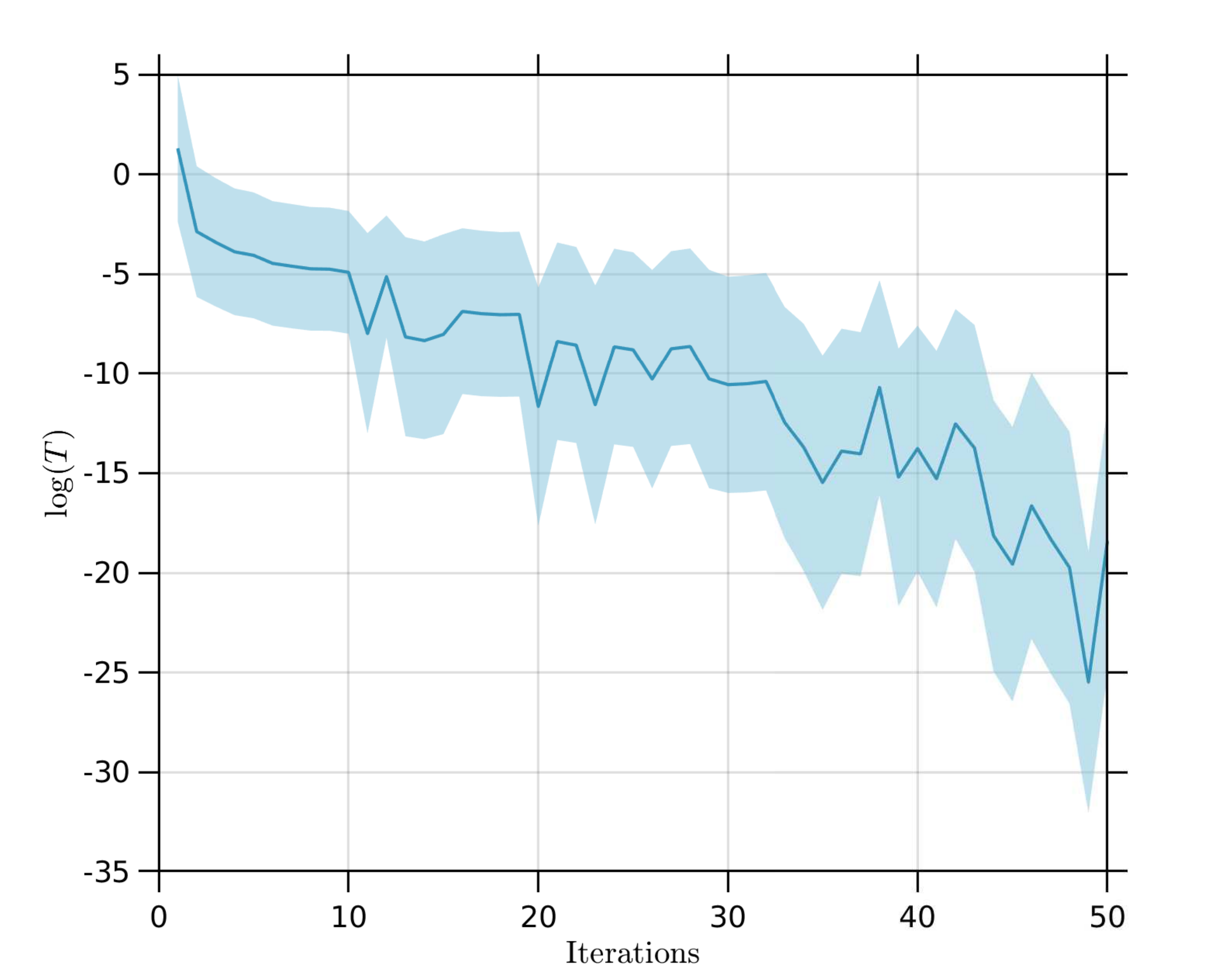}} & \multirow{-1}{*}[5cm]{\includegraphics[height=5cm]{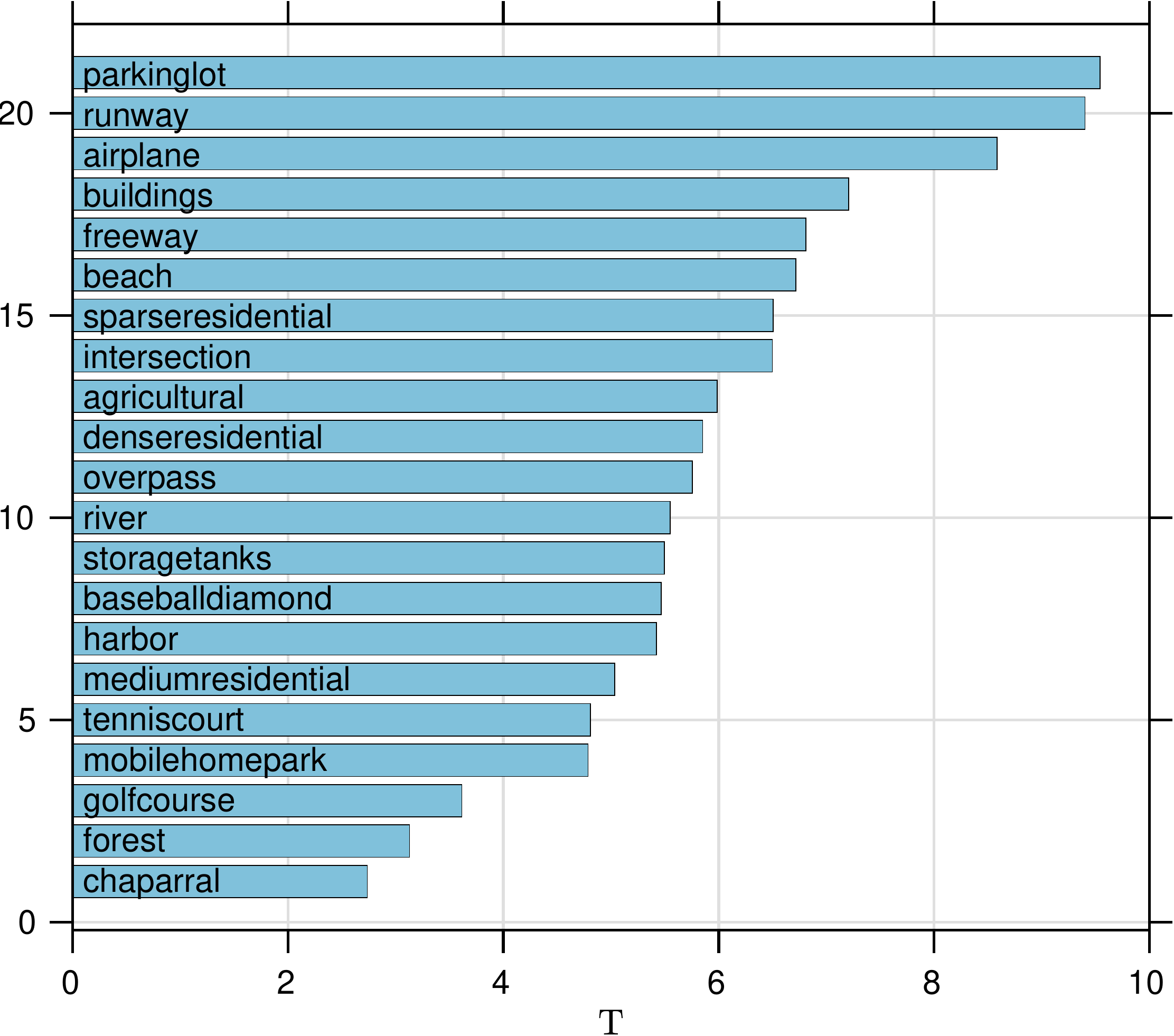}}
\end{tabular}
\end{center}
\caption{Estimation of total correlation, $T$ in very high resolution aerial imagery. Top: Three illustrative images for each of the 21 classes in the database, ranked according to their estimated $T$. 
Bottom: \textcolor{black}{(a) Each image is decomposed in 3 $\times$ 3 patches with three channels (rgb), making samples of 27 dimensions. (b) We measure how much overlapping there is between the information content (i.e. the total correlation) of these 27 dimensions for each class. We show a Venn diagram to illustrate the measured information following the same criteria as in Fig.\ref{fig:it_measures3}.} (c) Average total correlation computed iteratively for the different 21 class-specific RBIG models over 50 iterations, with the mean $T$ (solid) and the $T$ standard deviation (shaded) over all models. Convergence is achieved very rapidly for all classes (note the log-scale). (d) ranked $T$ per class computed from the RBIG models.
}
%(c) Symmetrized KLD between the 21 classes in the UC Merced database computed with RBIG.} 
\label{fig:ucmerced}
\end{figure*}
%%%%%%%%%%%%%
%%%%%%%%%%%%%

Here we validate RBIG to estimate total correlation (multi-information) %and the Kullback-Leibler divergence in and between 
in a set of aerial scenes collected in the \href{http://vision.ucmerced.edu/datasets/landuse.html}{UCMerced} data set~\cite{yang2010bag}. The data set contains manually extracted images from the USGS National Map Urban Area Imagery collection from 21 aerial scene categories, with 1-ft/pixel resolution. \textcolor{black}{The data set contains highly overlapping classes and has 100 images per class, see some examples per class in Fig.~\ref{fig:ucmerced}[top]. 
We extracted color patches of size $3\times 3 \times 3$ from each image, which yielded a total of 6499950 27-dimensional feature vectors per class. Then, we developed a Gaussianization transformation for each class individually and computed the (spatio-spectral) $T$ using RBIG, see Fig.~\ref{fig:ucmerced}[bottom left]. We show in Fig.~\ref{fig:ucmerced}[bottom right] the average and standard deviation of the $T$ evolution through 50 iterations for the 21 classes (note the log-scale) and the total correlation per class. More textured classes like runaways, freeway, buildings and intersections lead to higher $T$, while rather homogeneous/flat classes like chaparral, agricultural or forests reveal low information content.}

%% file: sections/42_droughts.tex
\subsection{Experiment 2: Information Quantification of Terrestrial Biosphere Variables in Time} % for Drought Analysis}
\label{sec:exps}

%Intro (from CI2019) 

\begin{figure}[h!]
\begin{center}
\setlength{\tabcolsep}{0pt}
\begin{tabular}{lc}
\multirow{2}{*}[4cm]{\includegraphics[width=0.34\textwidth]{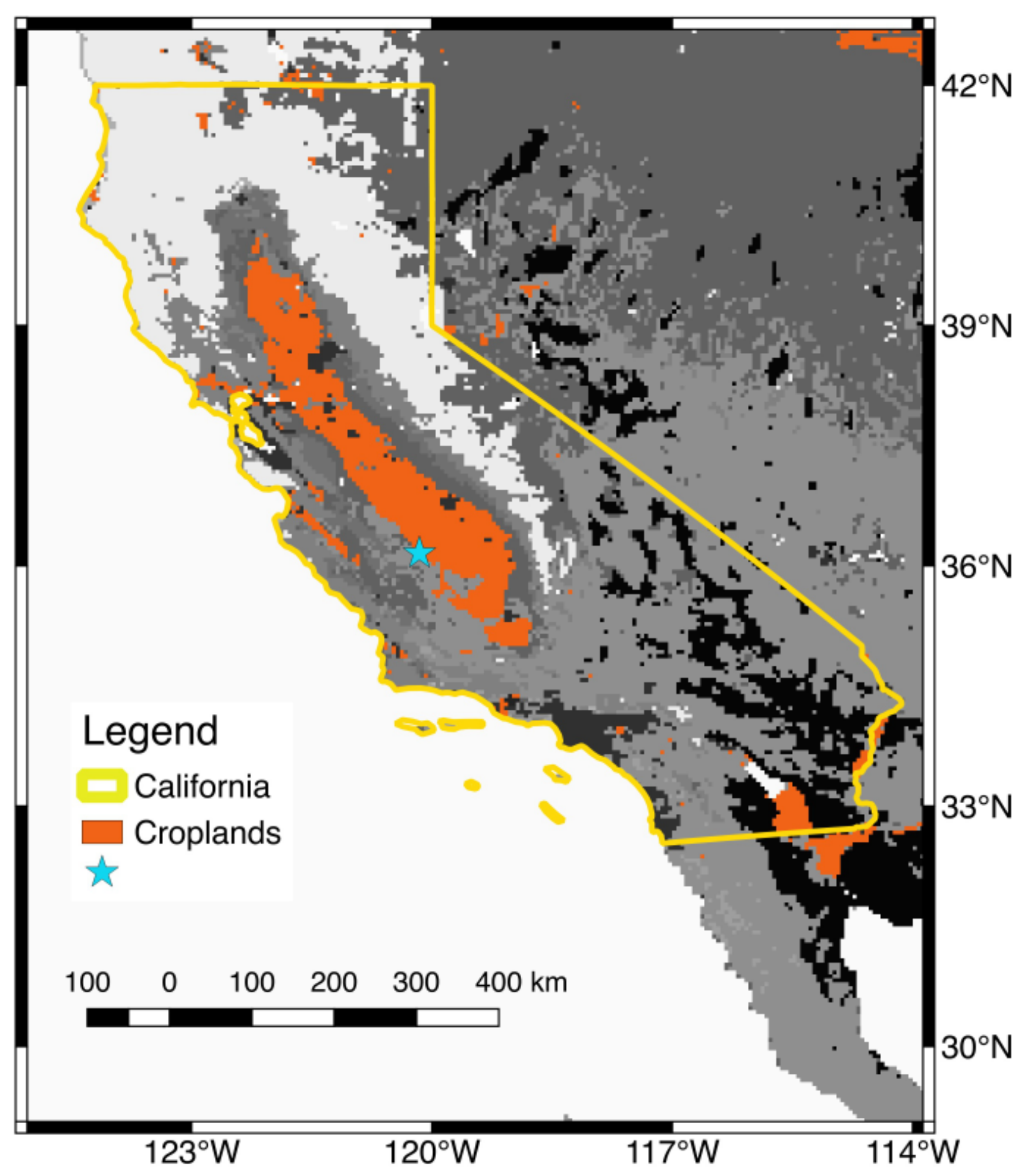}} &
\includegraphics[width=0.45\textwidth]{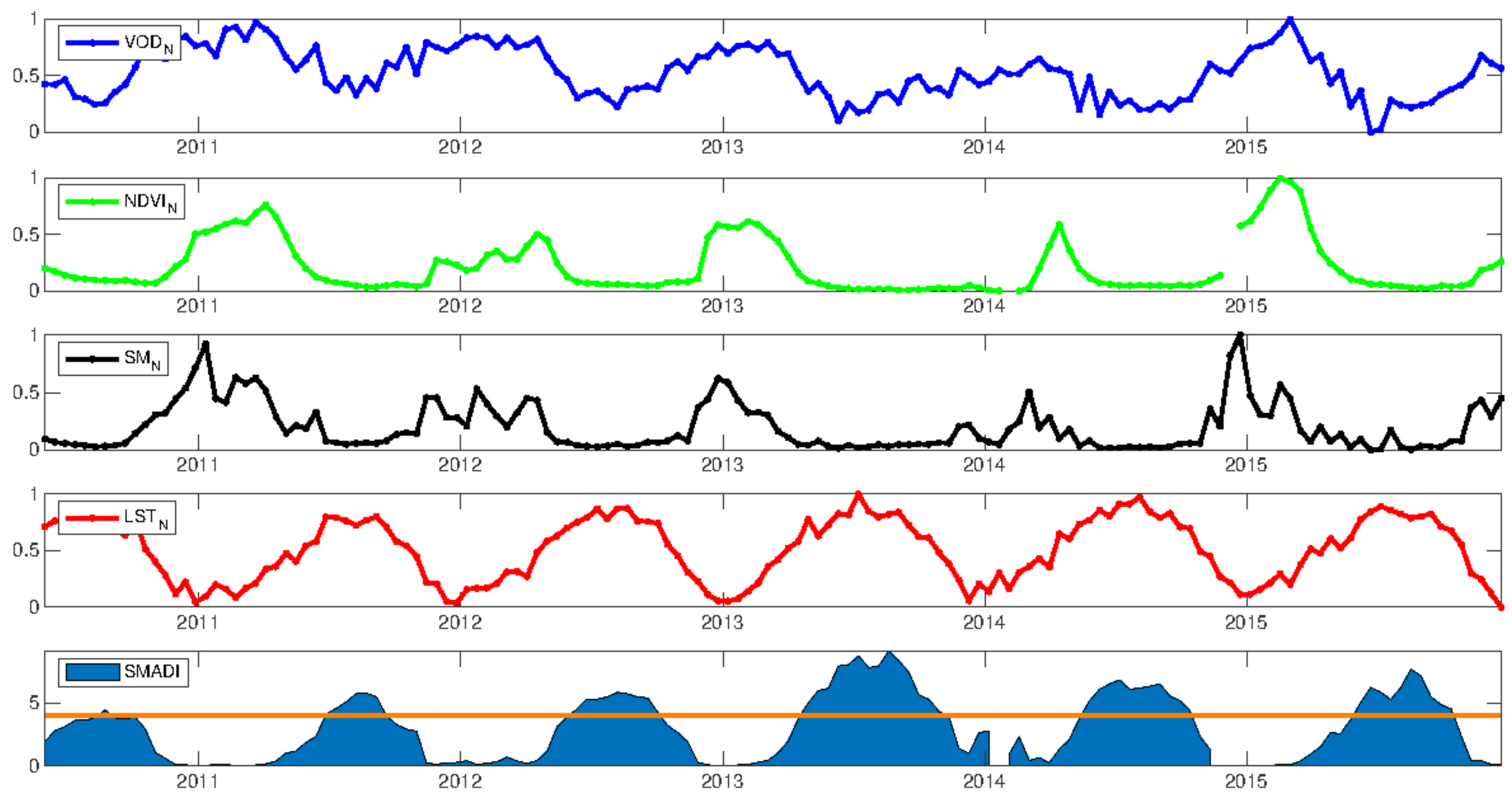}\\
 & \includegraphics[width=0.46\textwidth]{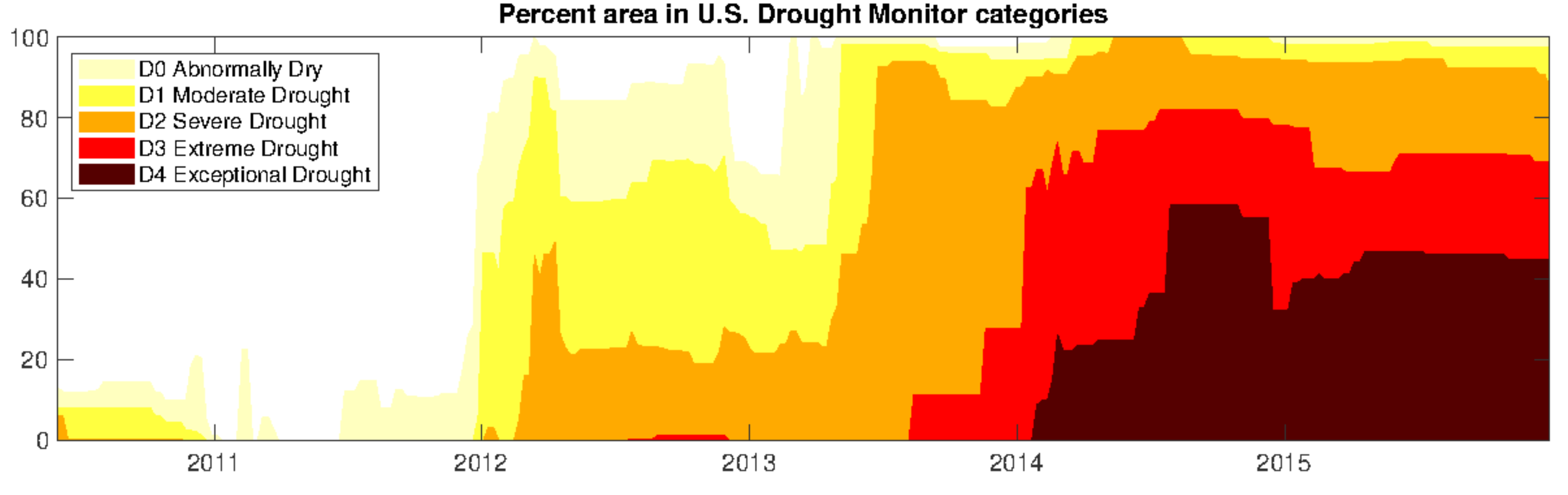} 
\end{tabular} 
\vspace{-0.15cm}
\caption{Left: distribution of croplands within California, according to the MODIS IGBP land cover classification. Top right: time series of normalized VOD, NDVI, SM, and LST, as well as the SMADI index \cite{smadidata} obtained at the selected pixel (blue star). SMADI extreme drought category is marked with an orange horizontal line. Bottom right: percent area of California in U.S. drought monitor~\cite{usdrought} categories
.} 
\label{fig:cali}
\end{center}
\end{figure}

According to climate projections, extreme events are likely to intensify and become more frequent over the coming years \cite{Zscheischler2013}. The effects of extreme events (such as droughts) are prevalent, not only in the biosphere and atmosphere, but in the anthroposphere too. Drought is a major cause of limited agricultural productivity which accounts for a large proportion of crop losses and annual yield variations throughout the world \cite{Boyer1982}. Droughts are also being currently placed as direct contributors to social conflicts, migration, and political unrest (e.g. \cite{Kelley2015}). 

There are many studies showing the value of incorporating Earth observation (EO) data for global agricultural systems and applications \cite{Fritz2019,Weiss2020}. Variables such as land surface temperature (LST) and the normalized difference vegetation index (NDVI) derived from optical satellites and, more recently, soil moisture (SM) and vegetation optical depth (VOD) derived from passive microwave sensors are just a few of the many features that can potentially be key for the early detection of drought events \cite{Sanchez2018,Sadri2018,Fernandez_Moran_2017}. The Soil Moisture Agricultural Drought Index (SMADI) was proposed in \cite{Sanchez2016} to integrate SM with LST and NDVI, showing good agreement with other drought indices and with documented events of drought world-wide~\cite{Sanchez2018}. 

In this experiment, we quantify the information in and between LST, NDVI, SM and VOD variables for the study area of California (only agricultural fields), see figure~\ref{fig:cali}. LST and NDVI are descriptors of the surface temperature and vegetation chlorophyll content, whereas SM and VOD characterize the water content in soils and vegetation \cite{Sanchez2016,Fernandez_Moran_2017}. We will also use information measures as a means to evaluate whether it would be worthwhile to include VOD as an additional variable in the SMADI ensemble to characterize droughts. Prior to the analyses, variables were resampled into a common 0.05$^\circ$ grid and biweekly temporal resolution. Details on the data sets are provided in Table \ref{tab:expsummary}. Measures are conducted for years 2010-2011 and years 2014-2016 separately, which are representative of non-drought and drought conditions in the study region (see Figure~\ref{fig:cali}).

We focus here in computing multivariate information theory measures in a temporal feature setting, in which previous time steps are included as input features. So for example, 1 input feature includes the current time stamp, 2 input features includes the current time stamp and the time stamp 14 days previously, and so on. This allows us to investigate the temporal scales that maximize the shared information among the remotely-sensed variables. This is particularly relevant for droughts, since there is a time lag between soil/climatic conditions (e.g. represented here by SM, LST) and the plant response (e.g. described by NDVI and VOD), which varies in the literature from two or three weeks up to three months \cite{Petropoulos2017}.

%Here, SM and LST represent the soil/climatic conditions and NDVI and VOD describe the vegetation status.
%In this experiment, LST and SM represent the soil/climatic conditions and NDVI and VOD describe the plant response. 

%Figure \ref{fig:cali} provides a summary of the database. 
%We explore how information theory can help with the analysis of these variables by looking at: 1) the comparison between each of the variables computing entropy and mutual information in a instantaneous setup (no temporal information is taken into account), and 2) the combination of NDVI, LST, and SM compared to VOD in a temporal feature setting. This works by including previous temporal time steps as input features. So for example, 1 input feature includes the current time stamp, 2 input features includes the current time stamp and the time stamp 14 days previously, 3 input features would include 

\begin{figure}[h!]
\small
\begin{center}
\setlength{\tabcolsep}{-0pt}
\begin{tabular}{cc}
(a)  \\
\includegraphics[height=5.5cm]{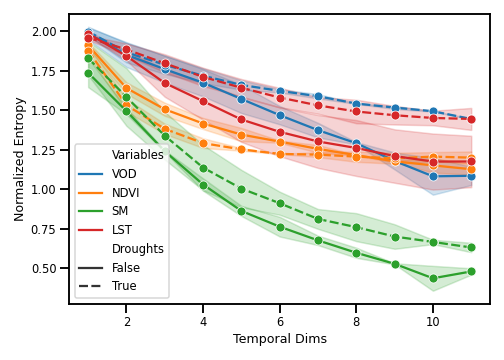} & \multirow{0}{*}[5.5cm]{\includegraphics[width=0.25\columnwidth]{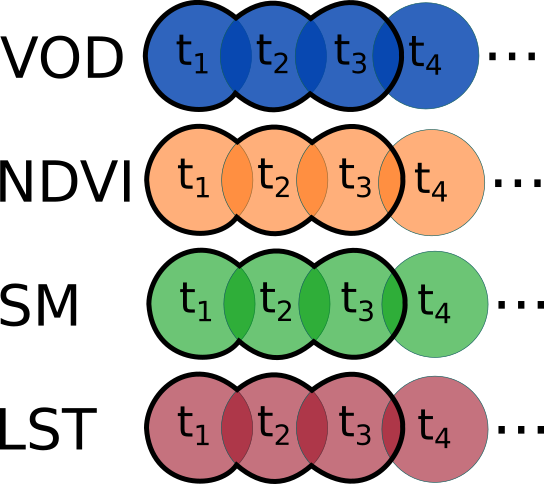}}\\
(b) \\
\includegraphics[height=5.5cm]{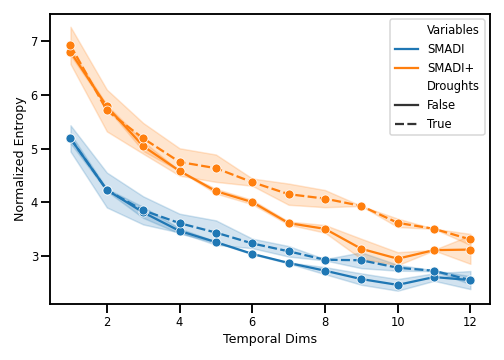} & 
\includegraphics[height=5.5cm]{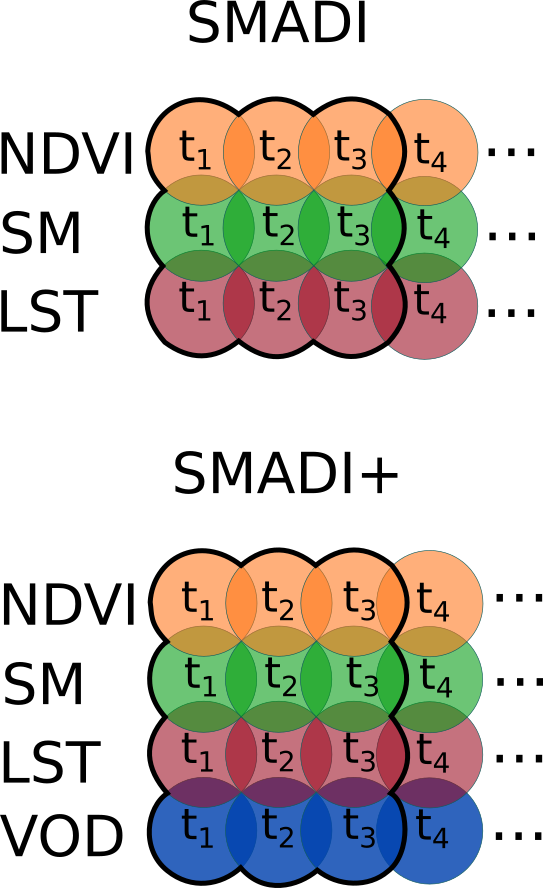}\\
\end{tabular}
\vspace{+0.15cm}
\caption{(a) Compares the Entropy for VOD, LST, NDVI and SM individually against the number of temporal dimensions considered. (b) Compares the contribution in entropy of VOD on the joint multidimensional variables integrated in SMADI [LST, NDVI, SM] and SMADI+ [LST, NDVI, SM, VOD] and how it changes as we include more temporal dimensions. The solid lines are mean estimates and the shaded regions are the variance estimates for the non-drought years (2010-2011) and the drought years (2014-2015). \textcolor{black}{Next to each graphic we show a Venn diagram to illustrate the measured information for 3 temporal dimensions as an example, following the same criteria as in Fig.\ref{fig:it_measures3}.}} 
\label{fig:droughtH}
\end{center}
\end{figure}

%Figure 6(a) shows the amount of expected information (H)for each of the four variables and how it changes as we includemore temporal dimensions. We see that VOD has the highestentropy compared to the other other variables whereas SM asthe  lowest  entropy  of  all  of  the  variables.  Figure  6(b)  showsthat VOD increases the amount of expected information whenadded to the SMADI variables ensemble.In figure 6(a), we see that the amount of entropy for VOD(blue) is higher than all other variables with different temporalfeatures. VOD and all other variables decrease in entropy as weadd more temporal features. NDVI saturates at around 1.5 bitswhereas the other variables have a steady smooth decline. Wecan also see that LST and VOD has a relatively large differencebetween  drought  and  non-drought  years  as  compared  to  theother variables NDVI and SM. This would indicate that witha  large  amount  of  temporal  features  could  be  more  useful  indetecting droughts.
\begin{wrapfigure}{r}{4.4cm}
\vspace{-0.5cm}
\begin{mdframed}[backgroundcolor=gray!20] 
How much information is adding a particular variable for drought monitoring? What temporal scales are the most informative? RBIG can answer these questions explicitly in bits 
\end{mdframed}
\vspace{-0.5cm}
\end{wrapfigure}
The amount of expected information $H$ for each of the four variables and how it changes as we include more temporal dimensions is analyzed in Figure~\ref{fig:droughtH} (a). Entropy will always increase with more features. So the entropy shown here has been normalized by the total amount of features present which allows us to quantify the amount of {\em entropy per feature}. It can be seen that the amount of entropy for VOD is the highest in all temporal settings, closely followed by LST. All variables decrease in entropy as we add more temporal features. NDVI saturates at around 1.5 bits whereas the other variables have a steady smooth decline. We can also see that LST and VOD show the largest difference between drought and non-drought years, and that the difference is largest as we increase the temporal dimension. This result suggests that LST and VOD observed during longer periods could be more useful in detecting droughts. %This would indicate that with a large amount of temporal features, adding VOD as a drought indicator could be more useful in detecting droughts.
Figure~\ref{fig:droughtH} (b) shows that VOD increases the amount of expected information when added to the SMADI variable ensemble in all the temporal settings considered, suggesting that it would be worthwhile to include VOD in agricultural drought studies. Results indicate operational settings of vegetation monitoring could benefit from synergistic approaches that allow including multi-sensor multi-dimensional variables, in particular under stress and disturbances such as agricultural droughts.  

\begin{figure}[h!]
\small
\begin{center}
\setlength{\tabcolsep}{-0pt}
\begin{tabular}{cc}
\includegraphics[height=5.5cm]{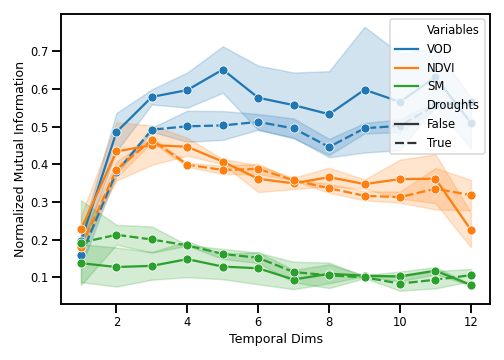} &
\multirow{0}{*}[5.5cm]{\includegraphics[width=0.25\columnwidth]{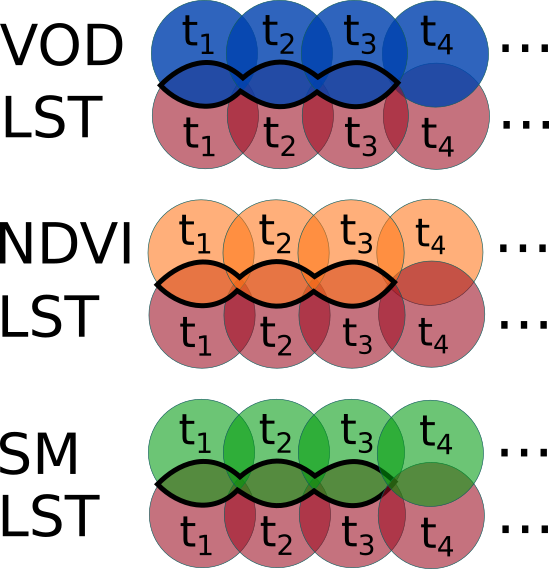}}\\
\end{tabular}
\vspace{-0.3cm}
\caption{Mutual information between pairs of multidimensional variables: LST-VOD, LST-NDVI, and LST-SM.  The solid lines are mean estimates and the shaded regions are the variance estimates for the non-drought years (2010-2011) and the drought years (2014-2015). \textcolor{black}{The Venn diagram illustrates the measured information for 3 temporal dimensions as an example, following the same criteria as in Fig.\ref{fig:it_measures3}.}}
%At the right we show a Venn diagram (similar to Fig.\ref{fig:it_measures3}) of the information measured for 3 temporal dimensions as an example, the information measured is the one contained inside the bold line.} 
\label{fig:droughtMI}
\end{center}
\end{figure}

The MI of every pair of multidimensional variables was analyzed to investigate the relation and redundancy between them as well as the optimal time scales to combine them. Note standard measures for pairwise comparison such as Pearson's correlation are restricted to one temporal dimension and hence do not allow exploring these scales. The MI scores obtained for LST relations are shown in Figure~\ref{fig:droughtMI}.
Interestingly, it shows that LST-NDVI and LST-VOD show an increase in mutual information up to about 2-4 temporal dimensions and then it saturates. This result suggests that a period of about 1-2 months is needed to capture the soil-plant status with the remotely-sensed variables analyzed in our study region. The curves are relatively similar irregardless of whether it is a drought year or not, and the spread of values for the drought years is considerably reduced for all variables and especially for VOD. This could be related to a reduced variability (limited range of values) under drought episodes, but further studies are needed to confirm this. We also observed that MI is consistently low between SM and all variables with any number of temporal dimensions, and is also low between NDVI and VOD, highlighting the value of combining optical and microwave variables for vegetation/land monitoring.

%The RV Coefficient (the multivariate extension of the Pearson's correlation coefficient)~\cite{Robert1976RV,Josse2016RV} was also calculated for all variables combinations (see results for LST in Fig. \ref{fig:droughtMI}). Correlation was very low in all cases, with a steady increase as more temporal features were added. The higher coupling of LST with NDVI and VOD at 2-4 temporal dimensions is not captured by RV. 
% Figure~ZZ(lag experiment) shows the mutual information between VOD and NDVI as a function of time lag, and reveals that the optimal information lag between NDVI and VOD variables is around 70 days, which could be used to derive models (autoregressive or based on differential equations) for their coupling and understanding.

%% file: sections/43_world_info.tex
\begin{figure*}[t!]
    \centering
    \small
\centerline{\includegraphics[width=15cm]{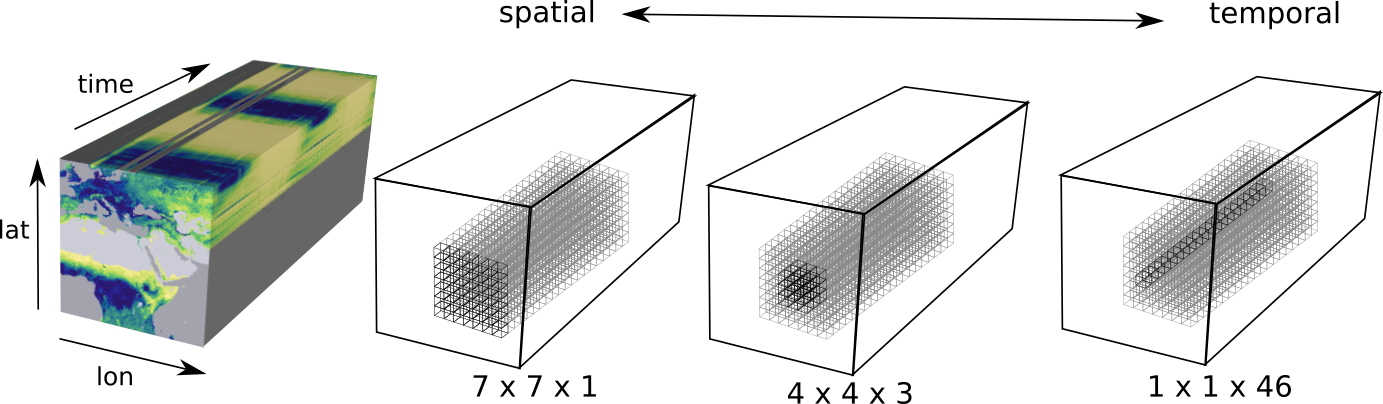}}
    \caption{\textcolor{black}{This figure illustrates an example decomposition of the Earth science data cube (ESDC)\cite{Mahecha19esdc} into different spatial-temporal configurations ranging from completely spatial to completely temporal. The $7\times 7 \times 1$ spatial configuration is all spatial pixels and no temporal pixels; this is very similar to spatial patches. The $1\times 1 \times 46$ configuration is all temporal pixels but no spatial pixels which is essentially a time series. The $4\times 4 \times 3$ configuration is a mixture of spatial and temporal pixels. Through-out this article we see different notions of spatial-temporal representation of the ESDC data.} }
    \label{fig:ESDC_data}
\end{figure*}

\begin{figure}[t]
    \centering
    \small
%\centerline{\includegraphics[width=10cm]{figures/Dibuix.png}}
\centerline{\includegraphics[width=15cm]{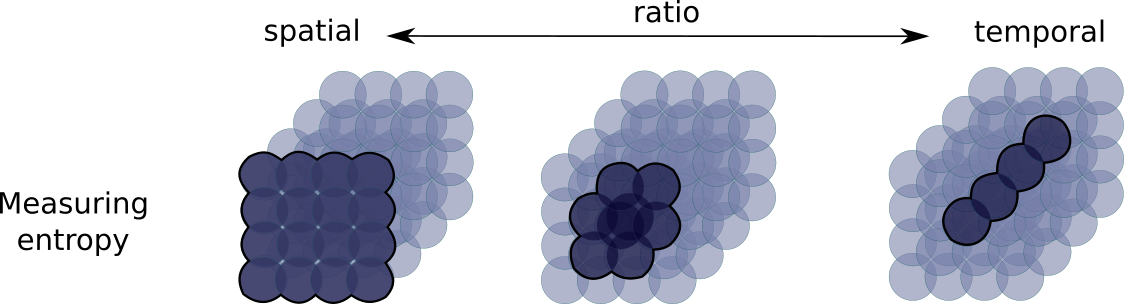}}
    \vspace{3mm}
    \begin{tabular}{cc}
% \multicolumn{2}{c}{Precipitation}  \\
    % \includegraphics[width=4cm]{figures/Dibuix.png} & 
\multicolumn{2}{c}{\includegraphics[width=7cm]{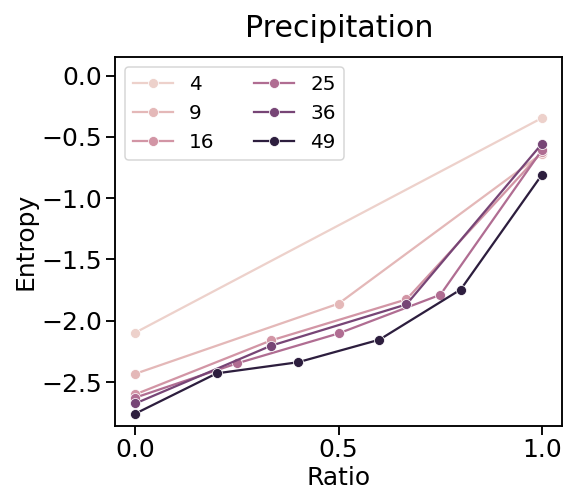}}\\
% Sensible Heat & Evaporation\\
\includegraphics[width=7cm]{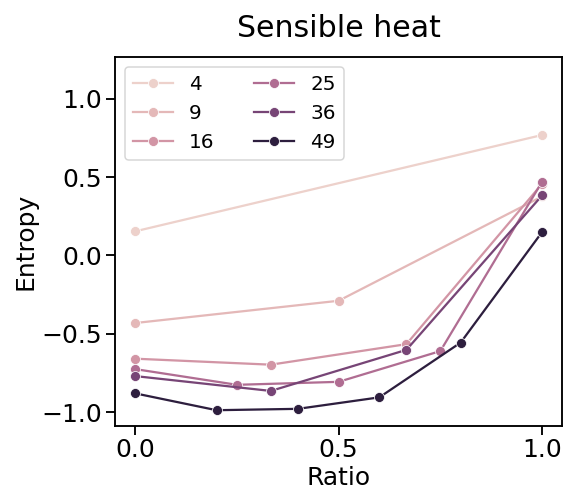} & \includegraphics[width=7cm]{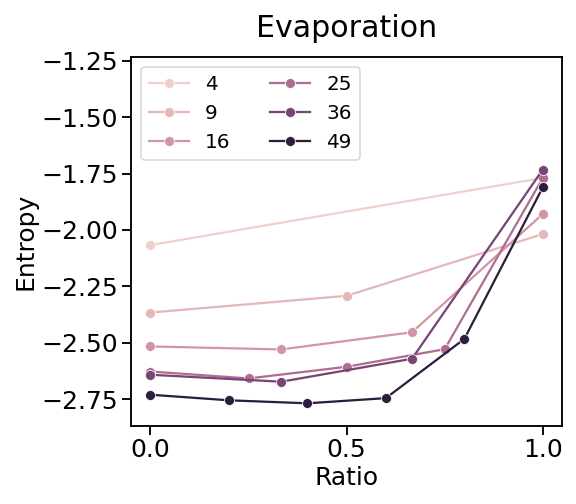}\\
\end{tabular}        
\caption{\textcolor{black}{Measuring entropy for different spatio-temporal configurations in the ESDC variables. The \textbf{top row} information theory Venn diagram representation of Fig. \ref{fig:ESDC_data} and how this relates to measuring entropy: the expected uncertainty. In the \textbf{middle and bottom rows}, we show how the measured entropy for precipitation, sensible heat and evaporation from the Earth science data cube \cite{Mahecha19esdc} changes with different spatial-temporal representations, ranging from fully spatial (ratio = 0) and fully temporal representation (ratio = 1)}.}
    \label{fig:entropypanel}
\end{figure}

%%%% THE MAPS
\begin{sidewaysfigure}[htb]
    \centering
    \small
    \setlength{\tabcolsep}{1pt}
    \begin{tabular}{c|ccc}
& Precipitation & Sensible Heat & Evaporation \\
\rotatebox{90}{\hspace{10mm} $7\times 7 \times 1$} &
    \includegraphics[width=7.5cm]{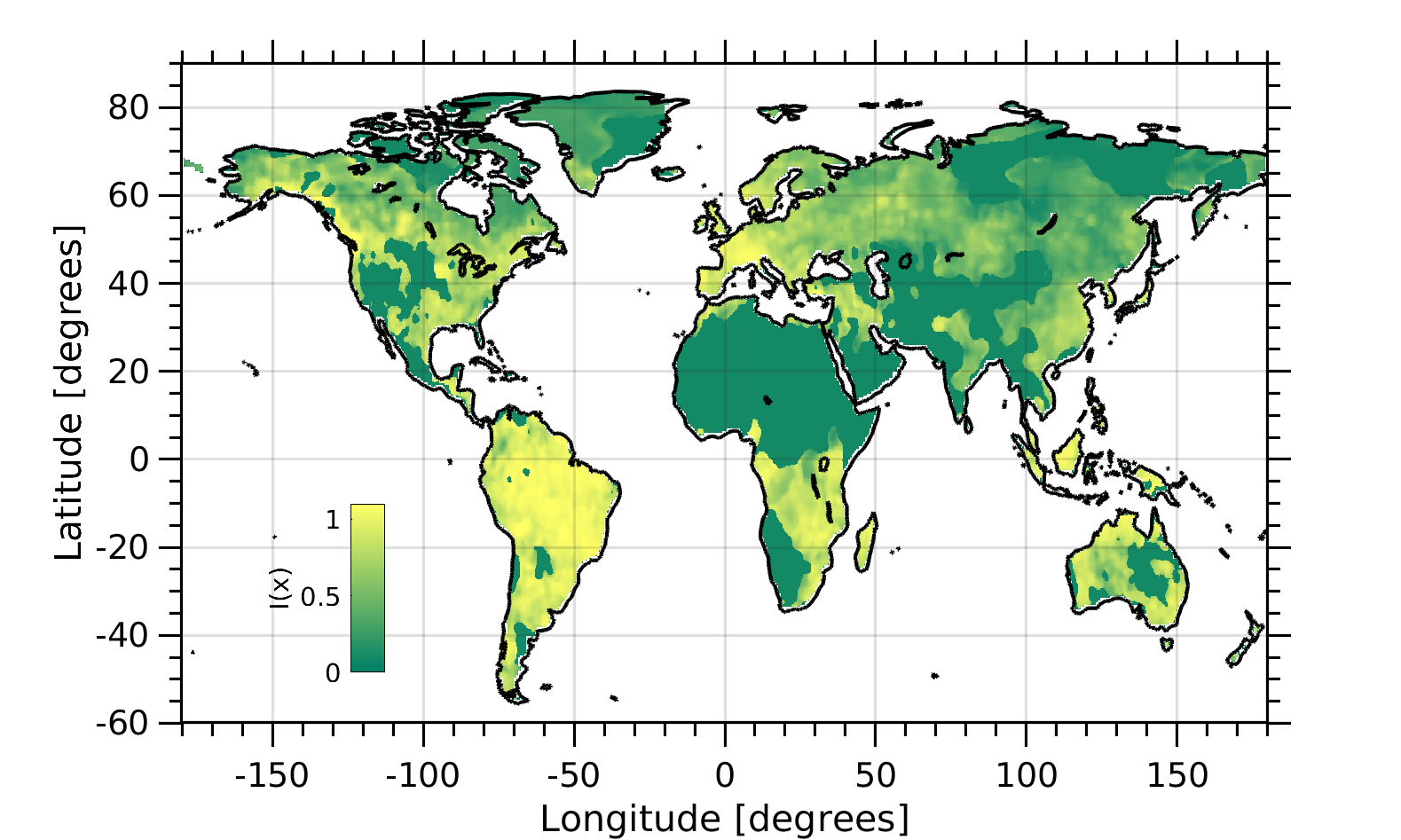}  &
    \includegraphics[width=7.5cm]{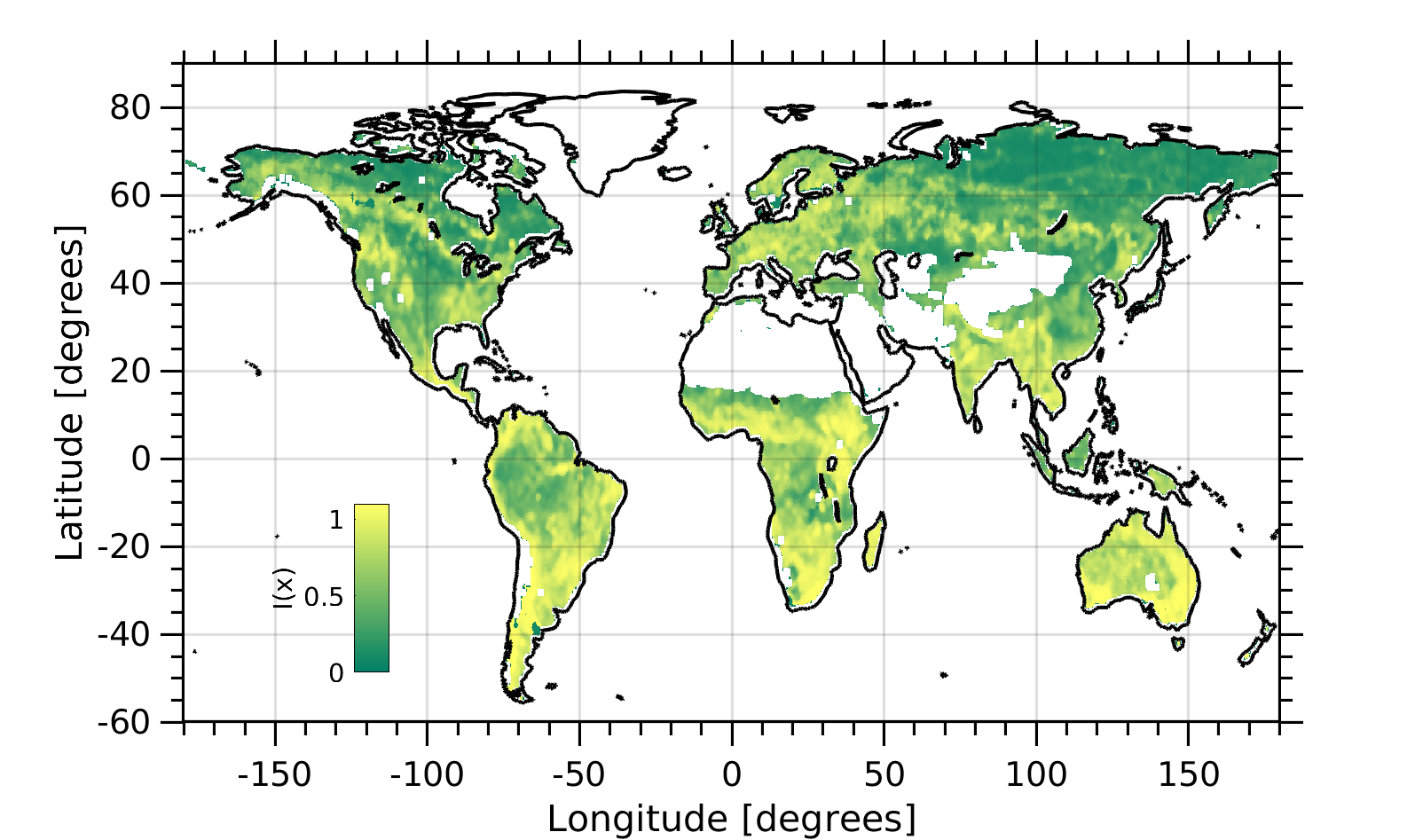}  &
    \includegraphics[width=7.5cm]{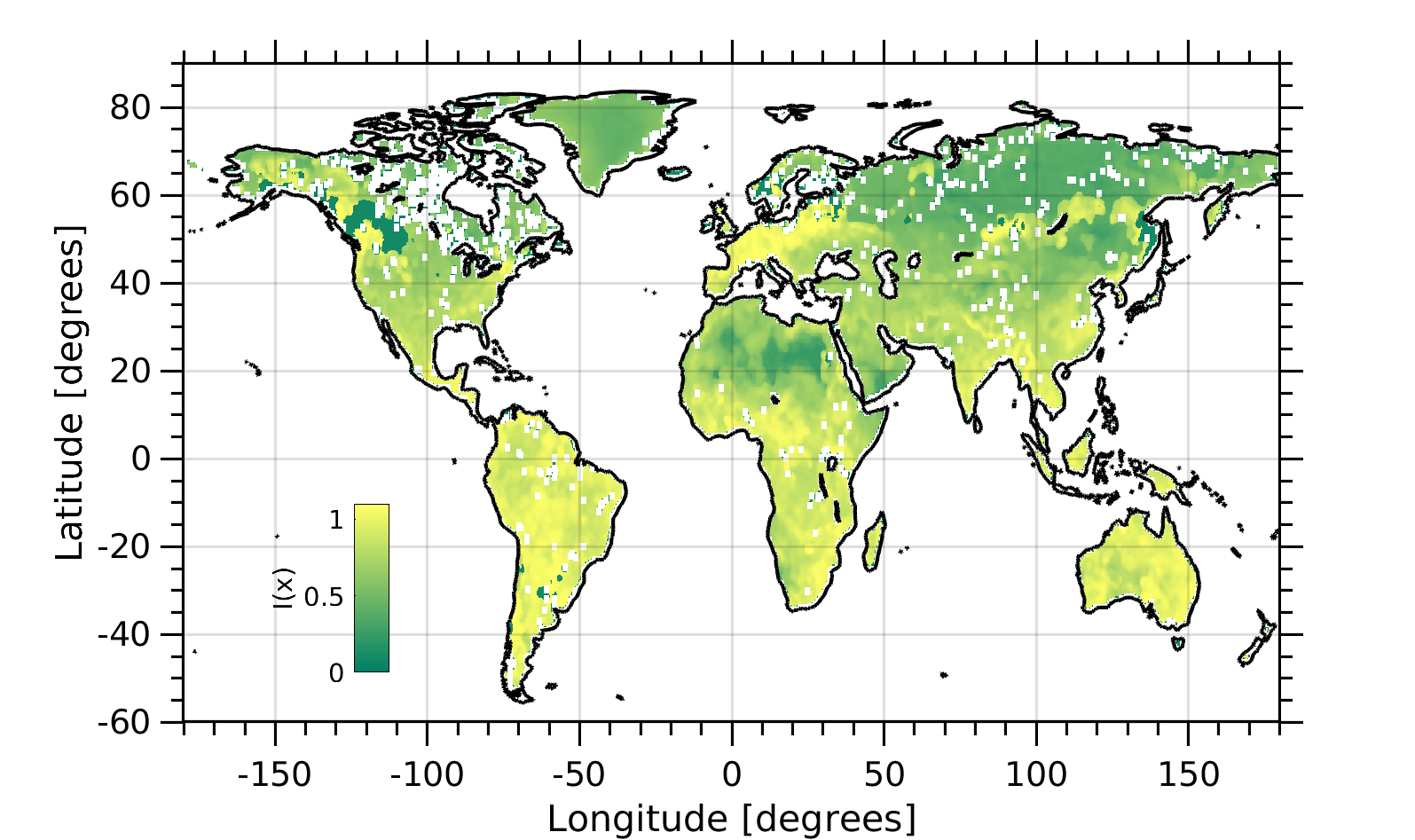}  \\
%    \rotatebox{90}{\hspace{5mm} $3\times 3 \times 6$} &
%    \includegraphics[width=6cm]{figures/experiments/spatial_temporal_2/var1_config2.png}  &
  %  \includegraphics[width=6cm]{figures/experiments/spatial_temporal_2/var3_config2.png}  &
 %   \includegraphics[width=6cm]{figures/experiments/spatial_temporal_2/var4_config2.png}  &
  %  \includegraphics[width=6cm]{figures/experiments/spatial_temporal_2/var5_config2.png}  \\
    \rotatebox{90}{\hspace{5mm} $1\times 1 \times 46$} &        \includegraphics[width=7.5cm]{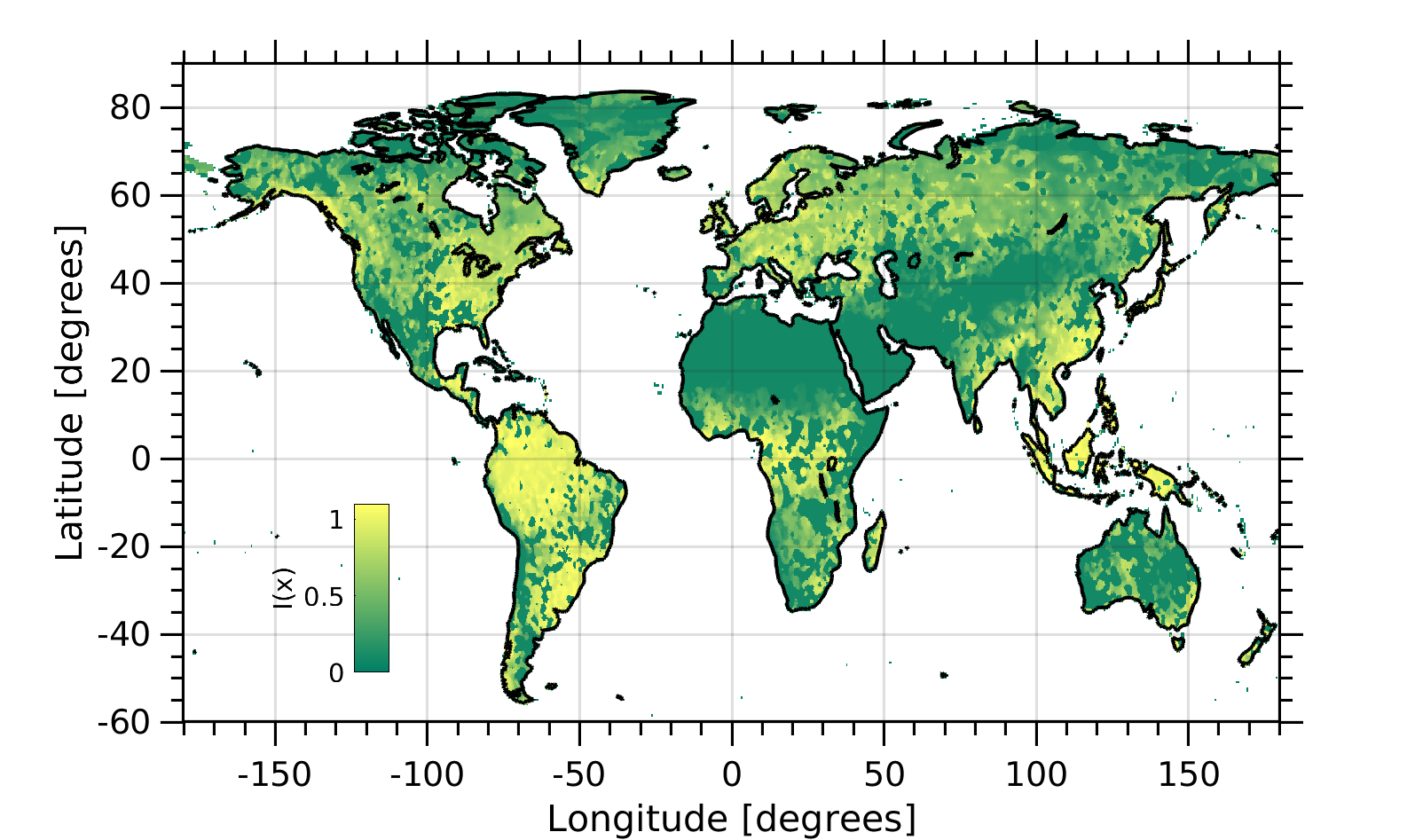}  &    %\includegraphics[width=6cm]{figures/experiments/spatial_temporal_2/var3_config3.png}  &
    \includegraphics[width=7.5cm]{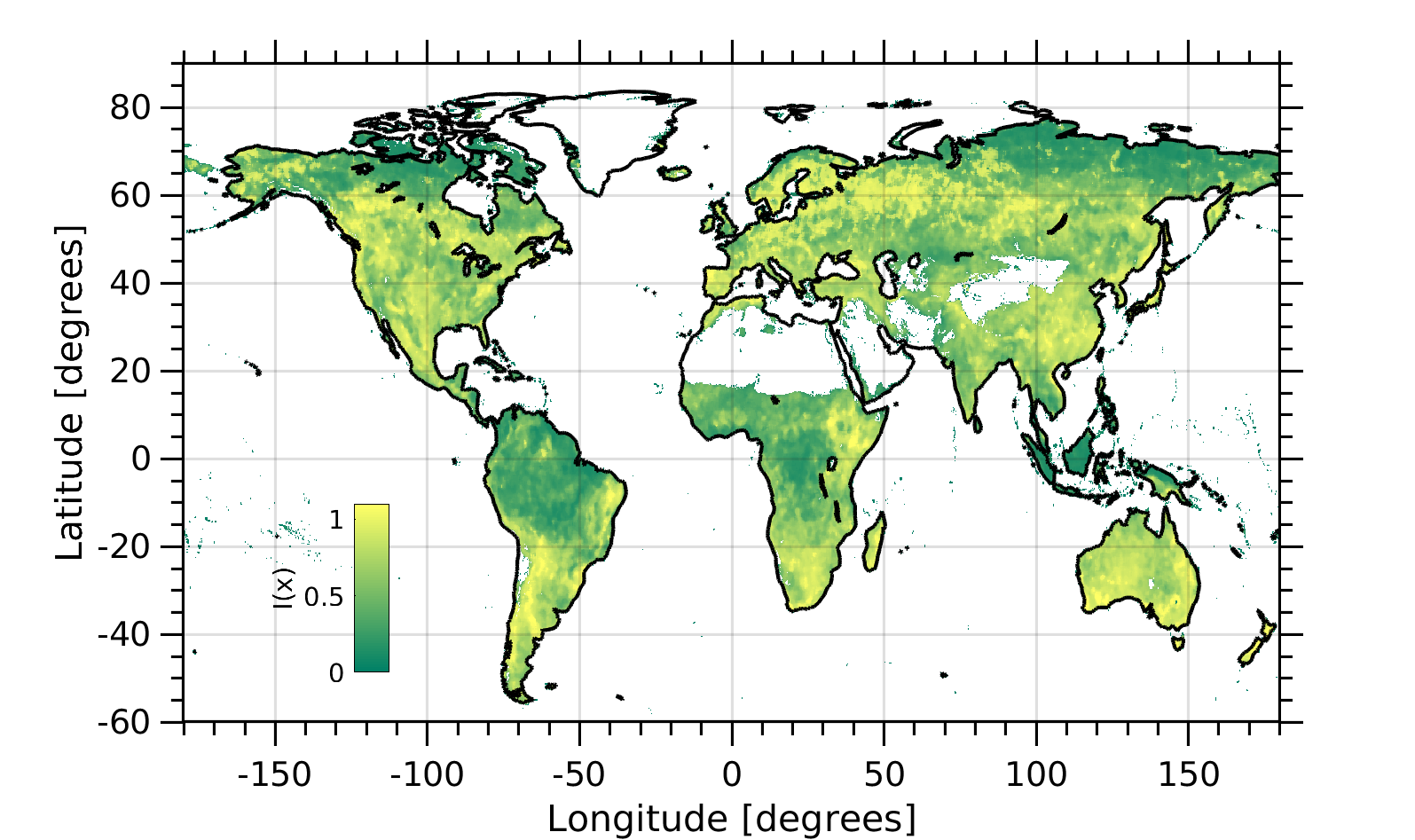}  &
    \includegraphics[width=7.5cm]{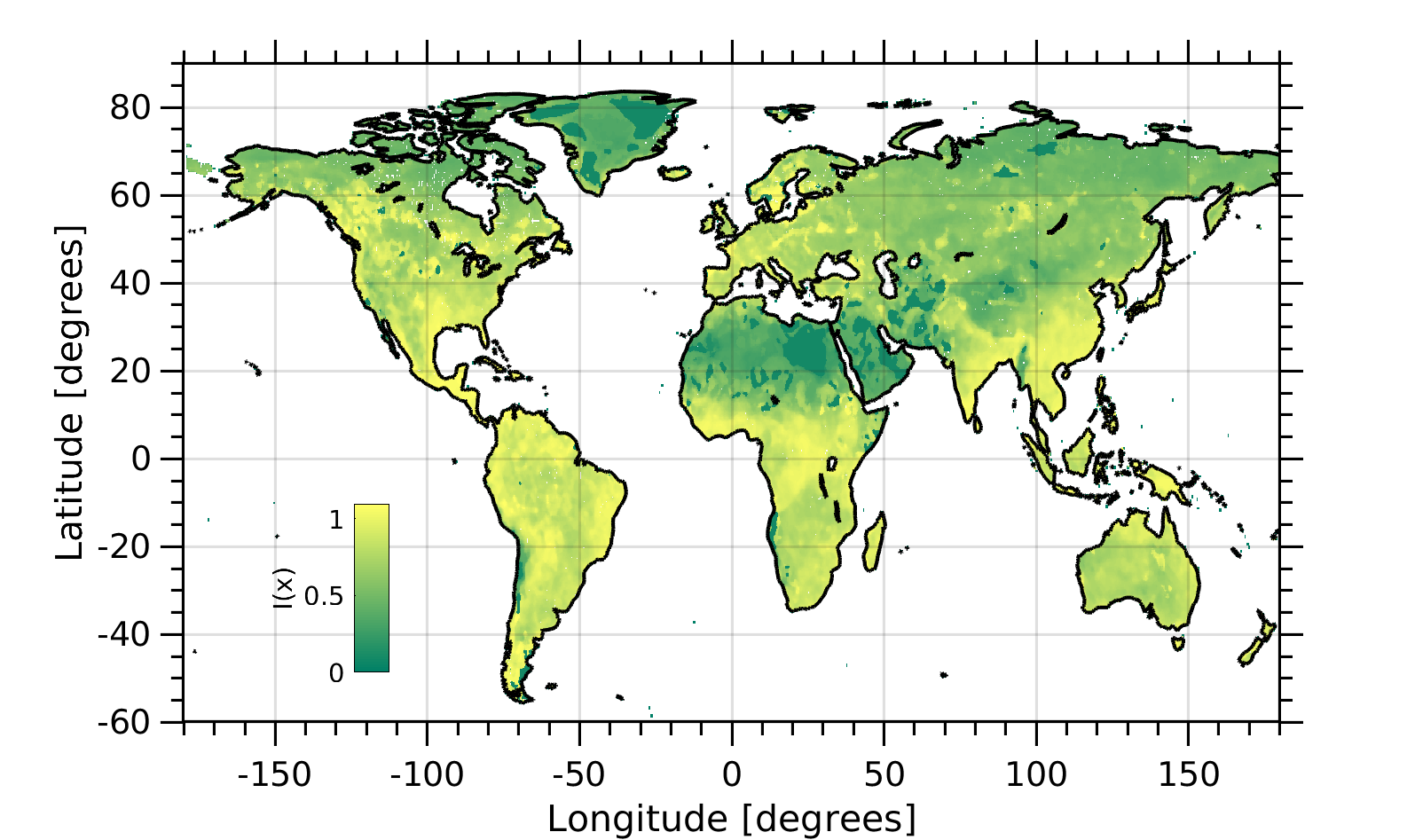}  \\
    \rotatebox{90}{\hspace{5mm}  Temp.-Spat. $I(x)$} &
    \includegraphics[width=7.5cm]{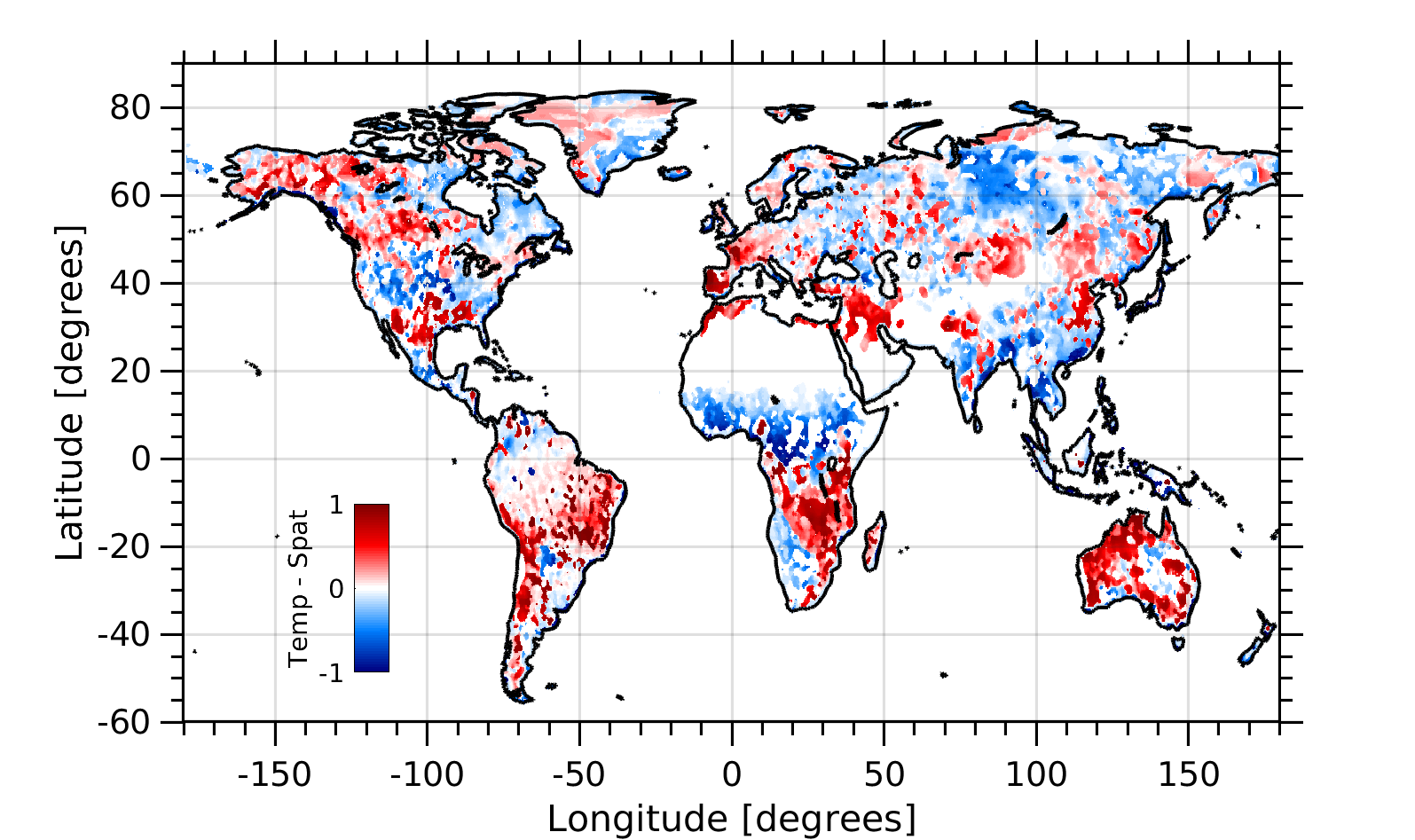}  &
    \includegraphics[width=7.5cm]{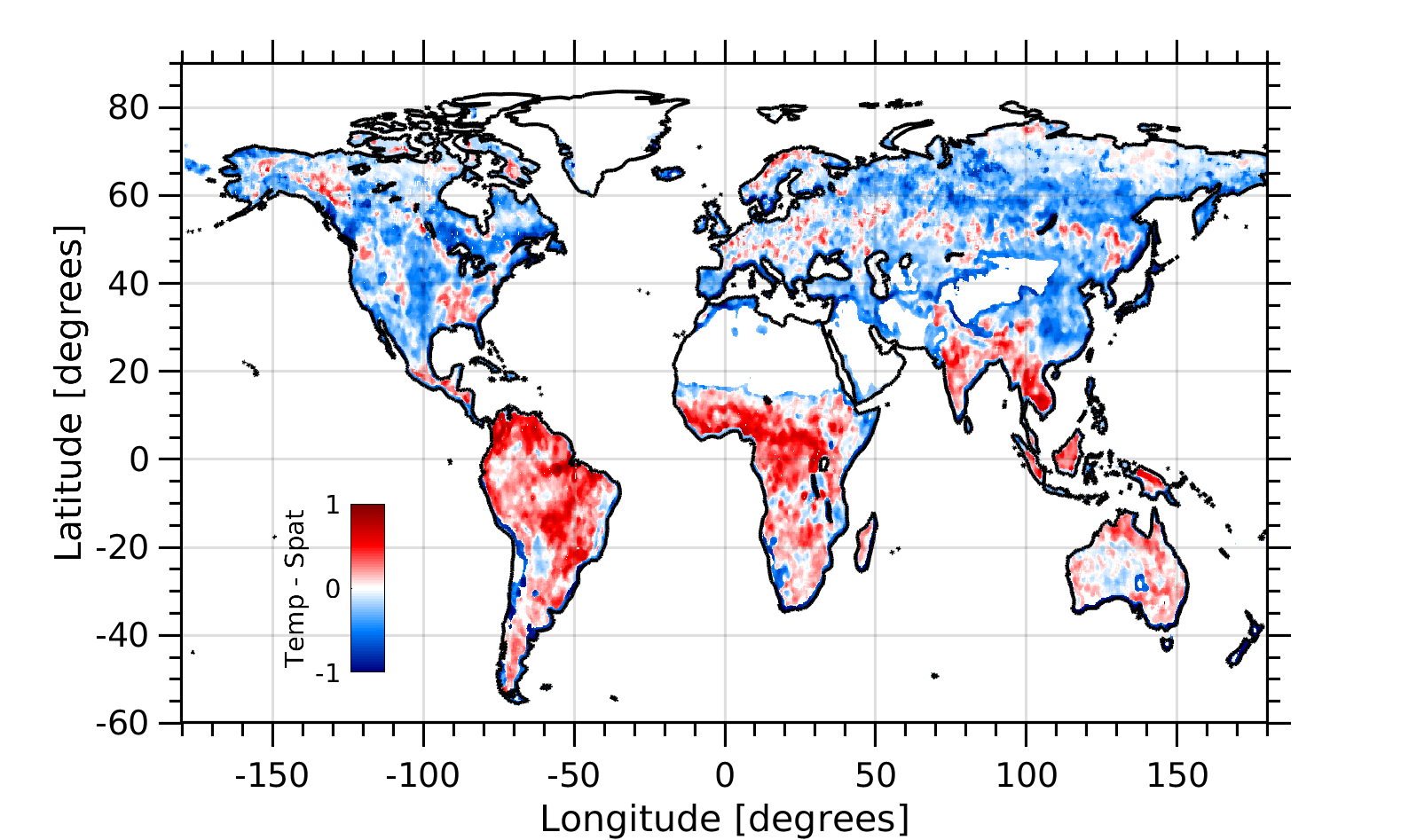}  &
    \includegraphics[width=7.5cm]{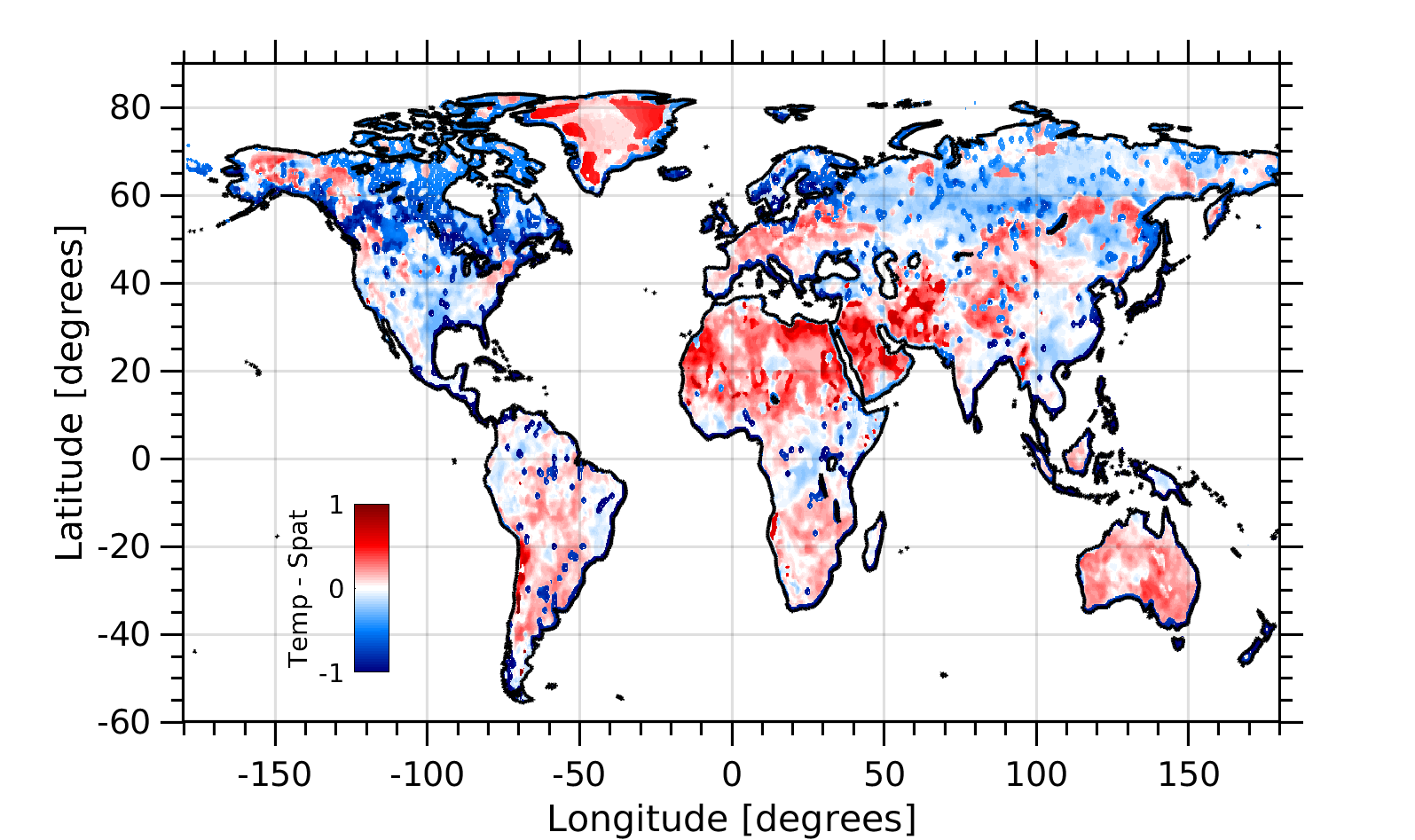}  \\
    \end{tabular}
    \caption{The two first rows show information content maps for precipitation, sensible heat, and evaporation using a fully spatial ($7\times 7$ spatial width, $1$ temporal length) and a fully temporal %, (b) $3\times 3$ spatial width, $6$ temporal length and (c) 
    ($1\times 1$ spatial width, $46$ temporal length) configuration. 
    The bottom row shows a divergent map of the trade-off (subtraction) between the information content of fully spatial and fully temporal per each variable.} % ($1\times 1\times 46$). }%(D) Information tradeoff per selected biomes.}
    \label{fig:spatiotemporalpanel}
\end{sidewaysfigure}

\subsection{Experiment 3: Information in Spatial-Temporal Earth data} 
%%% at some point we should read and cite this superrecent paper from Miguel: https://link.springer.com/article/10.1007/s10651-020-00446-4

\iffalse
\begin{itemize}
\item More data description + ESDL
\item Mega-maps of information (triangle + anomalies)
\item isolated areas, (zoomed in regions) + bars per biome
\item Entropy to summarize, H
\end{itemize}
\fi

\subsubsection{Data}

For our experiments we used observational and model simulated variables from the Earth Science Data Lab (ESDL)~\cite{Mahecha19esdc}\footnote{\url{https://www.earthsystemdatalab.net/}}, which is a platform that provides an opportunity for data-centric processing methodologies. %That coupled with cloud computing facilities enables us to prototype many approaches with a central environment for processing and sharing of results. 
The analysis-ready data-cube contains and harmonizes more than 40 variables relevant to monitor key processes of the terrestrial land-surface and atmosphere. %Notationally, the platform contains spatially and temporally gridded Earth data variables, $\cX=\x(u,v,t,z)$, where $u$ is the longitude, $v$ is the latitude, $t$ is the time and $z$ is the variable. Our temporal range is for the year 2001 with intervals every 8 days (46 time stamps for one year) and the spatial resolution is 0.05$^o$. 
Data exhibit clear spatial-temporal relations, which need to be taken into account to properly convey and quantify information. Figure~\ref{fig:ESDC_data} illustrates how we represent this spatial-temporal relations as inputs given a single variable. Here we focus on three key %four 
land-surface variables: % which exhibit nonlinear relations in space and tim: 
precipitation, %root-zone soil moisture, 
sensible heat and evaporation, which are outlined below:
\begin{itemize}
%%%%% GPP AND LAI JUST IN CASE VALERO'S RESULTS MAKE SENSE
%\item {\bf GPP} \red{GPP is the rate of fixation of carbon dioxide through the photosynthesis and one of the largest single flux in the global carbon cycle. However, the process is sensitive to climate variability. For instance, it has been shown that regional extreme events like droughts, heatwaves, and other types of disturbances may even influence the inter-annual variability of the globally integrated GPP  \cite{ZSCHEISCHLERetal2014}.  Hence, it is key to understand the spatial and temporal dynamics of GPP at regional and global scales. Here, we consider the GPP FLUXCOM (\url{http://www.fluxcom.org/}) product, computed as described in \cite{TRAMONTANAetal2016,bg-17-1343-2020}.}    
% \item {\bf LAI}
\item {\em Precipitation, $Precip$.} This is a fundamental variable in land-atmosphere processes. The collected data comprises the period 1980--2015, and comes from the Global Precipitation Climatology Project (GPCP)~\cite{ADLERetal2003,HUFFMANetal2009}.

%\item {\em Root-zone soil moisture, RZSM.} Soil moisture plays a fundamental role for the environment and climate system, as it influences hydrological and agricultural processes, runoff generation and drought development processes, and land-atmospheric feedbacks. We used root-zone soil moisture (RSM)~\cite{MARTENSetal2017,MIRALLESetal2011} in \href{https://www.gleam.eu/}{GLEAM}, which covers 2001--2011, that is a more sensitive variable to monitor water stress and droughts in vegetation. 

\item {\em Sensible heat, $SH$.} These data comprises 2001--2012, and was generated by training an ensemble of machine learning algorithms with eddy covariance data from FLUXNET and satellite observations in a cross-validation approach, regressions from these observations to different kinds of carbon and energy fluxes were established and used to generate data sets with a spatial resolution of 5 arc-minutes and a temporal resolution of 8 days. The H resembles the sensible heat flux from the surface and is expressed in [W~m$^{-2}$]~\cite{TRAMONTANAetal2016}. 

\item {\em Evaporation, $E$.} These data covers 2001--2011, and builds on the Global Land Evaporation Amsterdam Model (GLEAM), which consists on a set of algorithms that separately estimate the different components of land evaporation using input forcing data sets from reanalyses, optical and microwave satellites and other merged sources. The model itself consists of four modules: potential evaporation (Priestley and Taylor equation), interception (Gash analytical model), soil (multilayer soil model plus data assimilation) and stress (semi-empirical). The data are sampled on a grid of 0.25$^\circ$ and have a daily temporal coverage~\cite{MARTENSetal2017,MIRALLESetal2011}. 
\end{itemize}
The data is organized in a 4-dimensional data cube ${\bf x}(u,v,t,k)$ involving (latitude, longitude) spatial coordinates $(u,v)$, time sampling $t$, and the variable $k$. The available data is provided at two spatial resolutions (0.083$^o$ and 0.25$^o$) and at a temporal resolution of 8 days, spanning the years 2001-2011. In our experiments, we focus on the lower resolution products and on the period 2008-2010.  

\subsubsection{Spatial-Temporal Analysis}

The considered variables (precipitation, sensible heat, evaporation) are fully coupled. 
Moisture and precipitation interactions are vastly modulated by both land-atmosphere exchanges and large-scale atmospheric circulation. %While precipitation directly increases soil moisture, the feedback of soil moisture to precipitation remains uncertain in terms of sign, strength and location. Actually, 
%Moisture can affect precipitation through its regulation of the water and energy exchanges between the land and the atmosphere, being the most direct pathway the SM regulation of evapotranspiration. 
Nevertheless, before understanding variable relations, it is important to {\em identify when and where} individual variables are expressive. This may help in assessing the coupling %, and eventual causal 
mechanisms between variables and improve Earth system models. 

\begin{wrapfigure}{r}{4.4cm}
\vspace{-0.5cm}
\begin{mdframed}[backgroundcolor=gray!20] 
What are the optimal spatial and temporal scales to exploit each variable in the coupled land-atmosphere system? RBIG sheds light on uncertainty quantification and information in multidimensional spatio-temporal Earth data cubes
\end{mdframed}
\vspace{-0.5cm}
\end{wrapfigure}
The question we want to address in this experiment is what are the optimal (in information terms) spatial and temporal scales to exploit each variable's information. Using RBIG, we show here that the ratio of spatial-temporal neighbouring pixels giving the most amount of information can be explicitly calculated. We used RBIG to calculate the entropy $H$ for the aforementioned variables under different spatial-temporal configurations (fully temporal, spatio-temporal and fully spatial) as well as the corresponding information $I(\x)$ for each time pixel and variable. % in $\x$. %From the density we can compute the entropy as well as the information content.

%% Entropy first
Figure~\ref{fig:entropypanel} shows the entropy for the different variables and configurations following the same procedure as in \cite{Laparra2015} (and used in experiment \S\ref{sec:ucmerced} too in the spatial domain only).  
Essentially we formed cubes with the same dimensionality but different spatio-temporal configuration and computed the entropy values for each of them. %Ratio 0 means fully spatial while ratio 1 means fully temporal. %We display maps to show the amount of information for each pixel globally. Then, we calculated at the entropy values for each of the minicubes and show how they change depending on each configuration (). 
We chose several configurations ranging from a ratio of purely spatial (ratio=0) up to purely temporal (ratio=1). We also looked at different configurations for the amount of spatial-temporal dimensions used, e.g. a maximum of 4 dimensions up to a maximum of 49 dimensions (temporally, this is approximately 1 year). %We chose three variables; precipitation, root zone soil moisture and water vapour. %; and show how the information map and the entropy values change depending on minicube configuration chosen. 
Notice how each variable has a different spatial-temporal relationship with entropy, but in general temporal configurations (ratio=1) convey more information than purely spatial (ratio=0) for all the considered variables. % for precipitation, sensible heat and evaporation than does for root zone soil moisture. 
Trends are clear in particular for precipitation, where incorporating temporal information for any amount of dimensions has higher expected information. %, in contrast to root moisture where incorporating more spatial context (pixels) gives more expected information. 
For sensible heat and evaporation the entropy paths are similar, and reveal a fast increase in entropy for particular spatio-temporal configurations (ratio$\sim$0.8). %Water Vapour has higher information at the extremes with the lowest in the middle. 
These results suggest different {\em optimal} (in information terms) time and space scales for different variables, which may have implications in further analyses and applications.

Using the same data configurations we have computed the information content of each sample following the procedure described in section \ref{sec:shannon_info}. This help us to visualize the regions with more and less information. We show in Fig. \ref{fig:spatiotemporalpanel} the results of a spatio-temporal analysis of the information content of all three variables. 
%Looking at the general spatial patterns of all variables, some conclusions can be extracted. 
In regions where we expect pronounced seasonal patterns, the information (complexity) is apparently high in fully temporal configurations as the seasonal cycle controls ecosystem dynamics. 
Actually, seasonal (temporal) modes are of lower informative content in the spatial domain, as they are mainly driven by solar forcing. %, and vice versa. 
The information values tend to be higher in tropical regions, whereas arid regions show low-complexity (low-information) patterns. Let us now look in deeper detail at the different spatio-temporal configurations and their information patterns.

%% Spatial Patterns for the 3 configurations

%%%%%%%%%%%%%%%%%%%%%%%%%%%%%%%%%%%%%%%%%%%%%%%%%%%%%%%%%
%%%%%%%%%%%%%%%  http://www.waterandclimatechange.eu/  
%%%%%%%%%%%%%%%%%%%%%%%%%%%%%%%%%%%%%%%%%%%%%%%%%%%%%%%%%

Global patterns in rainfall are traditionally related to a strong seasonality, dominated by the position of the Inter-Tropical Convergence Zone (ITCZ) in the tropics, and the El Ni\~no-La Ni\~na cycles, which occur irregularly at intervals of 2-7 years. Spatial information generally dominates with high probability in the Amazonia and the tropics and with low information in desertic areas (e.g. California, Arabian peninsula and central Australia). As we quantify information in spatio-temporal configurations, more clear patterns of low information (e.g. Australia) and high probability (e.g. east-west US gradient) emerge \cite{tuttle2016empirical}. Studying precipitation in the fully temporal configuration translates into a clear ruling of the winter season in Amazonia, Indonesia, as well as northern Europe. Yet, a comparison of temporal vs. spatial information in Fig.~\ref{fig:spatiotemporalpanel}[bottom row] reveals that spatial information dominates in desertic areas (e.g. Australia, Iberian peninsula, Sahara, Mexico) which are reasonably independent of time, and temporal information dominates in Sahel (Savanna), northern latitudes and SW China, which are generally characterized by high rain factors, seasons and moisture. 

%\red{*** RZSM * SPAT. not very informative map, all regions convey the same (and high, saturated) info. The variable does not vary much, neither the info. actually the curves are pretty flat compared to the others... * TEMP. same. * ANOMALYPLOT. fully temp scale dominates vastly, except in very humid and northern latitudes with high SM persistence. Maria!!!???? Remove it?}

Transfer of sensible heat $SH$ into the air is dependent on the temperature gradient between the surface and the air above. Patterns of the information of sensible heat $SH$ stand out clearly. While the (fully) spatial information dominates in the Northern hemisphere, the (fully) temporal information patterns appear in the tropics where rainfall is present over larger regions and seasons. The global spatial distribution of $SH$ information shows the largest values in subtropical dry regions where available energy is preferentially partitioned to sensible heat rather than latent heat~\cite{jung2011global}, and seem to be anti-correlated with the amplitude of the mean seasonal cycle. 
These results reveal a maximum information of $SH$ in the tropical and subtropical deserts, where the high surface temperature conducts much heat into the air above, and the lowest near the poles where the surface temperatures are much lower. 
Information is mainly concentrated in the tropics too, and show similar patterns to precipitation with the exception of clear spatial information in the Indian peninsula. 
Evaporation maps of information reveal that the spatial information dominates in deserts and dry regions where evaporation is limited, while temporal information (more interannual variability) resides in Northern latitudes. This is mainly due to the low temperatures and radiation which relates to little evaporation all year round. The temperate areas show increased evaporation information in both purely spatial and temporal configurations, coinciding with increasing temperatures over ground moistened by winter rains. Cooler winter temperatures in Southern hemisphere reduce evaporation, which is also captured in the spatial-vs-temporal divergent maps. Note that in very dry regions information is higher (lower evaporative fraction), conversely for very humid regions, in agreement with~\cite{jung2011global}.  %, LE/(LE+H)

%% file: sections/6_conclusions.tex
This paper introduces a Gaussianization method and illustrates how to use it for multivariate density estimation in the context of Earth system science. The problem is highly relevant with the advent of all kinds of Earth data, both remotely sensed and in situ observations, novel products and model simulations. Density estimation is a long-standing unresolved problem in statistics and machine learning, mainly because of the curse of dimensionality. Besides, the data in remote sensing and geosciences pose additional challenging problems for PDF estimation: high dimensional data, nonlinear feature relations, many noise sources, and distinct spatial-temporal structures. 

%We introduced the framework of multivariate Gaussianization for density estimation. 
The Gaussianization method allows to simplify the problem by learning an invertible transformation of the data distribution to a multivariate Gaussian domain where features are independent. This not only makes the PDF estimation well-posed, but also allows us to estimate key information theoretic measures on multivariate datasets: information, entropy, total correlation  % Kullback-Leibler divergence 
and mutual information. We showed that the methodology can deal with high dimensionality and a high volume of data, and is simple to use and apply in spatio-temporal domains. We provide source code for the interested reader. 

We showed empirical evidence of performance in several Earth system data analysis problems, using a wide diversity of data (multispectral, hyperspectral, SAR, as well as global products from both satellites and Earth system models), and addressed the key problems of information estimation, redundancy, % divergences, 
and synthesis. Results confirmed the validity of the method, for which we anticipate a wide use and adoption. 

The framework enables us to tackle all applications involving a PDF estimation; from data classification to denoising and coding, which were not treated in this paper. The methodology also allows to compute other interesting IT measures, such as Kullback-Leibler divergence and conditional independence, % and other forms of divergences 
which will be a subject of future research.